\documentclass[useAMS,usenatbib]{mn2e}

\usepackage{epsfig}
\usepackage{amsmath}
\usepackage{amssymb}
\usepackage{graphicx}
\usepackage{deluxetable}
\usepackage{multirow}
\usepackage{array}
\usepackage{times}
\usepackage{color}
\usepackage[bookmarks]{hyperref}
\usepackage{csquotes}

\newcommand{\nat}{Nature}
\newcommand{\apj}{ApJ}
\newcommand{\apjl}{ApJ Letters}
\newcommand{\mnras}{MNRAS}

\newcommand{\aap}{A \& A}
\newcommand{\araa}{Annual Review of Astronomy and Astrophysics}

\newcommand{\pasa}{Publications of the Astronomical Society of Australia}
\newcommand{\apjs}{ApJ Supplement Series}

\title[AGN feedback, quiescence and CGM metal enrichment in early-type galaxies]
{AGN feedback, quiescence and CGM metal enrichment in early-type galaxies}
\author[M. Eisenreich et al.]{Maximilian Eisenreich,$^1$\footnotemark\, Thorsten Naab,$^1$ Ena Choi,$^2$ Jeremiah P. Ostriker,$^{3,4}$ \newauthor
and Eric Emsellem$^5$ \\
$^1$Max-Planck Institut f\"ur Astrophysik, Karl-Schwarzschild-Str. 1, 85741 Garching, Germany \\
$^2$Department of Physics and Astronomy, Rutgers, The State University of New Jersey, Piscataway, NJ 08854, USA \\
$^3$Department of Astrophysical Sciences, Princeton University, Princeton, NJ 08544, USA \\
$^4$Department of Astronomy, Columbia University, New York, NY 10027, USA \\
$^5$European Southern Observatory, Karl-Schwarzschild-Str. 2, 85741 Garching, Germany}

\date{Accepted ???. Received ???; in original form ???}

\pagerange{\pageref{firstpage}--\pageref{lastpage}} \pubyear{2017}

\begin{document}

\maketitle
\label{firstpage}

\begin{abstract}
We present three-dimensional hydrodynamical simulations showing the effect of kinetic and radiative AGN feedback on a model galaxy representing a massive quiescent low-redshift early-type galaxy of $M_* = 8.41\times 10^{10} M_\odot$, harbouring a $M_\mathrm{BH} = 4\times 10^8 M_\odot $ black hole surrounded by a cooling gaseous halo.
We show that, for a total baryon fraction of $\sim 20\%$ of the cosmological value, feedback from the AGN can keep the galaxy quiescent for about 4.35 Gyr and with properties consistent with black hole mass and X-ray luminosity scaling relations.
However, this can only be achieved if the AGN feedback model includes both kinetic and radiative feedback modes.
The simulation with only kinetic feedback fails to keep the model galaxy fully quiescent, while one with only radiative feedback leads to excessive black-hole growth.
For higher baryon fractions (e.g. 50\% of the cosmological value), the X-ray luminosities exceed observed values by at least one order of magnitude, and rapid cooling results in a star-forming galaxy.
The AGN plays a major role in keeping the circumgalactic gas at observed metallicities of $Z/Z_\odot \gtrsim 0.3 $ within the central $\sim 30$ kpc by venting nuclear gas enriched with metals from residual star formation activity.
As indicated by previous cosmological simulations, our results are consistent with a model for which the black hole mass and the total baryon fraction are set at higher redshifts $z > 1$ and the AGN alone can keep the model galaxy on observed scaling relations.
Models without AGN feedback violate both the quiescence criterion as well as CGM metallicity constraints.

\end{abstract}


\footnotetext{E-mail: meisenr@mpa-garching.mpg.de}

\section{Introduction}
\label{intro}
Massive early-type galaxies (ETGs) in the local universe are in their vast majority old, quiescent stellar systems which formed almost all of their stars $\sim 10$ Gyr ago, and show little to no signs of ongoing star formation \citep[e.g.][]{2003MNRAS.341...33K}.
This quiescence is a puzzle in two ways: First, in how it came to be, i.e. what mechanism quenched the star formation in these systems at high redshifts ($z\sim 2$), leaving them quiescent since then.
Theoretical arguments and numerical simulations both point towards powerful feedback from the galaxies' supermassive black holes (SMBHs) ejecting and unbinding much of their gas (the fuel for their star formation) from their haloes as the cause for quenching \citep[e.g.][]{1998A&A...331L...1S, 2005Natur.433..604D, 2006MNRAS.365...11C, 2006MNRAS.370..645B, 2008ApJS..175..390H, 2010MNRAS.406..822M, 2010MNRAS.409..985D, 2013MNRAS.433.3297D, 2016MNRAS.463.3948D, 2012MNRAS.420.2859M, 2014MNRAS.441.1270L, 2014MNRAS.444.1518V, 2015MNRAS.452..575S, 2015MNRAS.446..521S}, see also the relevant sections in recent reviews on galaxy formation \citep{2015ARA&A..53...51S, 2016arXiv161206891N}.
Observations clearly indicate that most active galactic nuclei (AGN) in the local universe live in ETGs \citep{2003MNRAS.346.1055K}.
The direct connection between AGN activity and quenching is less clear:
While some measurements in individual galaxies show clear correlations between fast, AGN-driven outflows and star formation suppression \citep[e.g.][]{2012A&A...537L...8C, 2015A&A...578A..11B, 2016A&A...591A..28C}, some show the opposite \citep[enhanced star formation in connection with AGN winds, e.g.][]{2015A&A...582A..63C, 2015ApJ...799...82C}.
A similar division is seen in statistical studies: Some find evidence for star formation suppression through AGN winds \citep[e.g.][]{2012Natur.485..213P}, others do not \citep[e.g.][]{2015MNRAS.449..373D}.
Furthermore, some theoretical and numerical works also show enhanced star formation associated with AGN feedback \citep[e.g.][]{2012MNRAS.425..438G, 2013ApJ...772..112S, 2013MNRAS.433.3079Z}.
Nevertheless, it is observationally clear that feedback from AGN has significant impact on their host galaxies and the gas within them \citep[see e.g.][for a review of the corresponding evidence]{2012ARA&A..50..455F}.
Alternative quenching mechanisms usually involve interactions of galaxies with their environment \citep[e.g.][]{2010ApJ...721..193P, 2012MNRAS.419.3167S}, e.g. through the effects of merging with another galaxy, and it is very possible that various processes aid each other in the quenching (e.g. galaxy mergers causing favourable conditions for efficient AGN feedback, see e.g. \citet{2005ApJ...630..705H, 2008ApJS..175..390H}).

While AGN feedback is currently the most fashionable -- and indeed probably the dominant -- process for quenching galaxies, the physical explanations for quenching in massive systems have varied over time, and a thoughtful analysis would indicate that several other processes have contributed to comparable degrees:
\begin{itemize}
\item Firstly, there is an intrinsic physical effect, pointed out by three papers in 1977 \citep{1977MNRAS.179..541R, 1977ApJ...211..638S, 1977ApJ...215..483B} that tends to make any heating process more effective for these systems than for lower mass galaxies:
Cooling rates for a hydrogen-helium plasma are reduced significantly for gas at the higher virial temperatures of massive ETGs, and these papers pointed out that there is a critical mass above which the cooling time is longer than the dynamical time.
This mass is important in setting the upper mass scale for galaxies even though more modern calculations, which include the cooling effects of metal lines, have significantly altered the simple, original cooling criterion.
\item Secondly, we have stripping by ambient gas in clusters and groups.
The typical ETG lives in a dense cluster of galaxies, moving through the hot gas envelope seen in thermal X-ray emission and indicative of the hot, relatively dense environment within which these systems live.
\item Thirdly, type I supernovae are effective in blowing out the processed gas from the outer parts of normal ETGs.
Papers by \citet{1993ApJ...419...52R} and others have indicated that this is a major effect – but of course it could not prevent \textit{central} cooling flows and central starbursts.
\item Finally, in the cosmological context, gravitational heating from infall can be important, and \citet{2009ApJ...697L..38J} showed that this can add nearly $10^{60}$ erg over cosmic epochs.
This heating can balance a significant fraction of the normally expected thermal gas cooling and greatly retard star formation.
\end{itemize}
While each of these processes would significantly reduce the star formation which might otherwise occur, recent simulations (see above) have conclusively shown that a central SFR of several solar masses per year would still occur absent effective suppression of central cooling flows and star bursts.
It is the properly implemented AGN feedback that was found to be the dominant quenching mechanism in this domain.

In any case, galactic evolution leads to a population of elliptical galaxies that, at low redshifts, spend their time mostly quiescent, forming very few stars, while still containing significant amounts of hot gas \citep[e.g.][for some X-ray observations of the hot gas around ETGs]{1979ApJ...234L..27F, 2006ApJ...646..899H, 2007ApJ...668..150D, 2010ApJ...715L...1M, 2012MNRAS.422..494D, 2013ApJ...776..116K}.
Here, an important question is exactly how much hot gas these massive, local ETGs actually contain, as the gas mass (and its density, which is related), can have a significant impact on the efficiency of different feedback processes.
The hot gas mass, through its Bremsstrahlung and metal line emission, can be measured using X-ray telescopes, which by now quite a few groups have done, both for galaxies and larger-scale systems.
They correlate either the X-ray luminosity of the gas directly, or related quantities like the system's baryon fraction (its mass of stars and gas divided by its total mass including dark matter), to a variety of other quantities of the system, e.g. its stellar luminosity or mass, stellar velocity dispersion, total mass, or gas temperature \citep[e.g.][]{2007ApJ...668..150D, 2010ApJ...715L...1M, 2010ApJ...719..119D, 2011ApJ...729...12B, 2013A&A...555A..66L, 2013ApJ...776..116K, 2013MNRAS.432.1845S, 2015ApJ...812..127K, 2015MNRAS.449.3806A}.
These correlations show that more massive systems, i.e. systems in deeper potential wells, contain more gas, with the most massive---galaxy clusters---having a baryon fraction about equal to that measured from the cosmic microwave background (CMB, the so-called ``cosmological" baryon fraction).

On the smaller scales of individual galaxies, the most commonly measured relation was between the X-ray luminosity of the hot gas and tracers of the stellar luminosity ($L_\mathrm{B}$ or $L_\mathrm{K}$).
These relations indicate that quiescent ETGs contain less gas compared to their total masses than groups and clusters, but they also show an enormous scatter of up to several orders of magnitude \citep[e.g.][]{2011ApJ...729...12B, 2013MNRAS.432.1845S}.
Only recently, large galaxy surveys like the Sloan Digital Sky Survey (SDSS) made it possible to put tighter constraints on the gas mass in the less massive haloes on the scales of individual, massive galaxies.
\citet{2015MNRAS.449.3806A} used stacked SDSS data to measure the X-ray luminosity--mass relation down to stellar masses of $M_*=10^{10.8}M_\odot$, while the \citet{2013A&A...557A..52P} used the same data to constrain the hot gas mass via the Sunyaev-Zeldovich (SZ) effect.
Together, these observations indicate that massive galaxies contain enough hot gas within their virial radii to match the cosmological baryon fraction, but that their hot haloes have to be much less concentrated than in more massive systems.
\citet{2015MNRAS.449.3806A} also show that the X-ray luminosity--mass relation follows an unbroken power law down to the scales of massive galaxies, indicating that any influence of the AGN should be through a gentle, self-regulated feedback mode, and not through much more disruptive episodes of powerful ``thermal blast" feedback events.

The existence of massive hot gaseous haloes around ETGs leads to the second part of the puzzle: How do low-redshift ETGs \textit{stay} quiescent?
Without some energetic process stopping it, the hot gas in an ETG would flow towards the centre triggered by efficient gas cooling.
The gas would become dense enough to form new stars continuously, breaking the galaxy's quiescence, which is not observed \citep[e.g.][]{2003ARA&A..41..191M}.
Essentially the same issue exists in larger-scale structures, i.e. galaxy groups and clusters, where it is commonly known as the ``cooling-flow problem".
Numerical simulations on all scales from clusters down to individual ETGs show exactly the described behaviour when no feedback processes, or only feedback from stars (i.e. supernovae and stellar winds) are included, but they also show that feedback from the central SMBH can prevent a cooling catastrophe \citep[e.g.][]{1997ApJ...487L.105C, 2001ApJ...551..131C, 2007ApJ...665.1038C, 2006ApJ...643..120B, 2007ApJ...668....1N, 2010ApJ...717..708C, 2010MNRAS.406..822M, 2014MNRAS.441.1270L}.

In these simulations, the AGN feedback prevents a continuous cooling flow and keeps the system quiescent by heating the central gas and/or generating winds that decrease the central gas density.
Although, at these late times, these winds are not necessarily capable of ejecting the gas from the halo completely (the systems are generally too massive and their potential wells too deep), they can still prevent any further cooling and star formation.
If, in simulations, AGN-driven outflows keep ETGs quiescent, can we detect these outflows in the real universe?
Especially with the rise of integral-field spectroscopy over the recent years, there have indeed been various detections of likely AGN-driven outflows of gas in different phases (ionized, atomic, molecular) from the centres of ETGs \citep[e.g][]{2012A&A...540A..11K, 2012A&A...537L...8C, 2011ApJ...735...88A, 2015ApJ...798...31A, 2015ApJ...815...34D, 2015A&A...578A..11B, 2015A&A...582A..63C, 2015ApJ...799...82C, 2016A&A...588A..68G, 2016A&A...591A..28C, 2016arXiv160608442W}, which are often explained with interactions of the AGN-driven radio jet with the surrounding medium, but also with nuclear winds and radiation pressure \citep{2014A&A...562A..21C, 2015A&A...580A...1M, 2016Natur.533..504C}.

Even with modern supercomputers, it is impossible to simultaneously resolve the scales on which black hole accretion and feedback actually take place (i.e. the ~AU scales of accretion discs), and include a whole galaxy (or even larger scales) into a simulation.
Therefore, numerical implementations of AGN feedback are always sub-resolution models designed to catch the overall effect of the unresolved feedback processes on larger scales.
In the past, various groups tried several different such implementations:
One of the easiest ways to model the AGN feedback is probably in the form of pure thermal feedback as first implemented by \citet{2005MNRAS.361..776S}.
There, a fixed fraction $\epsilon_\mathrm{r}$ of the rest-mass energy accreted by the SMBH is assumed to be converted into radiation which then couples with the surrounding gas with a constant efficiency of $\epsilon_\mathrm{f}$, leading to isotropic heating of the interstellar medium (ISM) in the neighbourhood of the black hole.
The thermal energy input of this heating is strictly proportional to the black hole's mass accretion rate: $\dot{E}_\mathrm{FB} = \epsilon_\mathrm{f}\epsilon_\mathrm{r}\dot{M}_\mathrm{BH}c^2$.

Already this rather simple AGN feedback model leads to a regulation of the SMBH growth and the star formation that produces galaxies, groups, and clusters much more in line with observations than simulations without any AGN feedback, but it comes with some limitations.
For example, this simple thermal feedback model has a resolution-dependant, numerical ``overcooling" problem:
The surrounding gas can be heated to a temperature at which it cools rapidly, immediately radiating away the feedback energy, and making the feedback inefficient.
To avoid this problem, \citet{2009MNRAS.398...53B} modified the model to let the feedback energy accumulate until it is enough to heat a certain amount of gas by a set temperature such that losses through radiative cooling are reduced.
This model was also used in the large-scale cosmological simulations cosmo-OWLS \citep{2014MNRAS.441.1270L} and EAGLE \citep[e.g.][]{2015MNRAS.446..521S}.

Furthermore, some authors argue that---following both theoretical and observational indications---the AGN feedback should be composed of two different modes: a radiative, ``quasar" mode at high accretion rates, which corresponds to the radiative-thermal feedback described above, and a ``radio" mode relating to the creation of large X-ray cavities of hot, under-dense gas by radio jets \citep[see e.g.][]{2005MNRAS.363L..91C}.
\citet{2007MNRAS.380..877S} implemented such a two-mode feedback model by combining the \citet{2005MNRAS.361..776S} thermal feedback at high accretion rates above a set threshold with the creation of large, hot ``bubbles" (with properties dependent on the accretion rate) in the gas for accretion rates below that threshold.
However, this model has led to too powerful AGN-driven ejection of gas in massive haloes in the large-scale cosmological Illustris simulation \citep{2014MNRAS.445..175G}, leaving the haloes almost devoid of gas, in contradiction to observations.

\citet{2015MNRAS.448.1504S} instead chose to mimic the transition between quasar and radio mode by using the sum of two feedback efficiencies (one for each mode) with different dependencies on the accretion rate to calculate the total feedback energy, which was then distributed thermally to the surrounding gas.
In a different approach, \cite{2012MNRAS.420.2662D} used a combination of pure thermal feedback in the ``quasar" mode at high accretion rates, and a jet-like feedback model in the ``radio" mode at lower accretion rates.
The jet feedback is modelled by distributing mass, momentum and energy in a small cylinder with the momentum aligned to the cylinder axis.
This model was used in the large-scale cosmological simulation suite Horizon-AGN \citep{2016MNRAS.463.3948D}.
More recently, \cite{2017MNRAS.465.3291W} combined the usual purely thermal energy injection for the ``quasar mode" with a purely kinetic energy and momentum injection in random directions for the ``radio" mode.
Yet another approach \citep[e.g.][]{2000A&A...356..788C, 2000ApJ...534L.135M} assumes that the feedback energy is injected in the form of relativistic particles filling radio-bright bubbles.
In this model, the amount of injected momentum is small, and most of the energy is first stored as the enthalpy of the buoyantly rising bubbles.
As they cross several pressure scale-heights, the bubbles gradually release their energy into the surrounding gas, thereby heating it \citep[][]{2001ApJ...554..261C, 2002MNRAS.332..729C}.

Another problem common to purely thermal models of AGN feedback is that, when they are strong enough to drive outflows and affect the gas on large scales, they often heat the interstellar, circumgalactic, and intergalactic medium (ISM, CGM, and IGM, respectively) to overly high temperatures compared to what is observed \citep[e.g.][]{2014MNRAS.442..440C, 2015MNRAS.449.4105C}.
If, instead, the feedback is implemented in a more physically motivated way, by directly modelling the interaction of the observed radiation and winds of the AGN with its surrounding gas, and specifically taking into account the momentum transfer that takes place in these interactions, this problem can be solved.
Following this approach, extensive work has been done in a series of papers:
Starting with first simple \citep{1997ApJ...487L.105C, 2001ApJ...551..131C}, then much more sophisticated \citep{2005MNRAS.358..168S, 2007ApJ...665.1038C} models of radiative feedback (Compton and photoionization heating plus the corresponding radiation pressure), later also including kinetic momentum feedback modelling the broad-line region winds observed in AGN \citep{2009ApJ...699...89C, 2010ApJ...717..708C, 2010ApJ...711..268S}, these authors investigated the influence of AGN feedback on the evolution of elliptical galaxies in one-dimensional (1D) simulations.
Comparing both types of feedback, they found that both together are necessary to match the observed properties of ETGs, as pure radiative feedback leads to much too high SMBH masses, while pure momentum feedback is incapable of keeping the galaxy quiescent without expelling too much gas.
More recently, they expanded their studies to two-dimensional simulations \citep{2010ApJ...722..642O, 2011ApJ...737...26N, 2014ApJ...789..150G}, finding much more stochastic and less efficient AGN feedback than in 1D simulations, and also (as for their 1D work) testing the effects of different (constant and accretion-dependent) feedback efficiencies.

Similar models have also been developed and explored by other authors: 
\citet{2011MNRAS.412.1341D} implemented AGN feedback in the form of pure momentum input through radiation pressure, and \citet{2012MNRAS.420.2221D} added a kinetic feedback model describing broad-line region winds.
\citet{2011ApJ...738...54K} used a combination of radiative feedback (using a radiative transfer technique) and mechanical feedback in the form of bipolar jet-like winds.
More recently, \citet{2017MNRAS.464.1854B} investigated the impact of winds driven by radiative AGN feedback, using radiative-hydrodynamical simulations to determine how the AGN radiation couples to the ISM.
Finally, \citet{2016MNRAS.458..816H} studied the effects of both broad-line region wind feedback and radiative heating (a model similar the one used in this paper) in small-scale simulations of the central region of a massive galaxy.

\citet{2012ApJ...754..125C} implemented the momentum and radiative feedback models from Ciotti, Ostriker et. al (see above) into the fully three-dimensional hydrodynamical code \textsc{gadget}3 \citep{2005MNRAS.364.1105S}, and compared it to the ``standard" thermal feedback model by \citet{2005MNRAS.361..776S}, first in simulations of isolated spiral galaxies and mergers of such \citep{2014MNRAS.442..440C}, and later in cosmological zoom-in simulations of ETGs \citep{2015MNRAS.449.4105C}.
They find that their momentum feedback implementation is much more successful at driving outflows and preventing recent star formation in local ETGs than the thermal feedback model, while also producing galaxies with X-ray (i.e. hot gas) luminosities similar to those observed, while the thermal feedback results in much higher X-ray luminosities.

In this work, we use an improved black-hole feedback model of \citet{2012ApJ...754..125C} to investigate the influence of AGN feedback on the late evolution of massive, quiescent elliptical galaxies, as they are observed in the local universe.
To this purpose, we run controlled hydrodynamical simulations of an isolated spherical galaxy set up to closely resemble a typical observed local, massive ETG in its major properties.
We examine the effects of the different parts of the AGN feedback (i.e. momentum and radiative feedback) on the star formation history of the galaxy (especially with regard to maintaining its quiescence), and on its ISM and CGM properties (especially the generation of large-scale outflows, their effect on the gaseous metal distribution, and the X-ray characteristics of the galactic hot gas; see e.g. \citet{2007MNRAS.380..877S, 2010MNRAS.406..822M} on these feedback effects on the scales of galaxy clusters and groups), and compare the results to observational constraints.

This paper is organized as follows:
In section \ref{sim}, we briefly describe the numerical code and the physical sub-resolution models used for our simulations, present the initial conditions of our simulated galaxy, and give an overview of the simulation runs.
Afterwards, we present and discuss our simulation results by comparing the star formation history and black-hole growth (section \ref{sfh}), the ISM evolution (section \ref{cenISM}), the metal enrichment of the CGM through gas flows from and to the centre (section \ref{CGM}), and the X-ray properties of the galaxy (section \ref{Xray}) between the different runs and with observational constraints.
Finally, in section \ref{summ}, we summarize our work and the conclusions we draw from its results.

\section{Simulations}
\label{sim}

\subsection{Numerical code and subgrid models}
For our simulations, we use the N-body smoothed particle hydrodynamics (SPH) code SPHGal \citep{2014MNRAS.443.1173H}, an improved version of \textsc{gadget}3 \citep[see][for \textsc{gadget}2, the last public version of this code]{2005MNRAS.364.1105S}.
The improvements introduced in SPHGal include amongst others the use of a pressure-entropy formulation, a Wendland $C^4$ kernel with $N_\mathrm{ngb} = 100$ neighbours, and a new implementation of artificial viscosity.
These changes significantly reduce the numerical artefacts present in the fluid mixing of the original \textsc{gadget} version, and also improve the convergence rate noticeably.

To include physical processes beyond gravity and hydrodynamics to our simulations, the code is supplemented by subgrid models for metallicity-dependent gas cooling, star formation, energy and momentum feedback from stars and the central SMBH, as well as metal production and diffusion.

The model for gas cooling, star formation, stellar feedback and metal enrichment was originally implemented by \cite{2005MNRAS.364..552S,2006MNRAS.371.1125S}, and further improved and extended by \cite{2013MNRAS.434.3142A}.
Each gas particle that falls below a temperature threshold set to $12,000$ K and above a density threshold of $1.94\times 10^{-23} \text{ g }\text{cm}^{-3}$ has a probability of $1 - \mathrm{e}^{-p_\text{SF}}$ to turn into a star particle in the current time-step of size $\Delta t$, where $p_\text{SF}$ is defined as
\begin{equation}
p_\text{SF} = \epsilon_{\text{SFR}} \sqrt{4\pi G \rho} \Delta t
\end{equation}
which is larger for higher gas densities $\rho$.
The star-formation-rate efficiency $\epsilon_{\text{SFR}}$ is a free parameter that is necessary because the actual scales of star formation are not resolved in the simulation.
It is set to $\epsilon_{\text{SFR}} = 0.02$.
Compare also \citet{2003MNRAS.339..289S} for the ultimate origin and motivation of this star-formation model.

The gas cools with a rate dependent on their current temperature, density and metal abundances.
The chemistry is taken into account by tracking the abundances of Hydrogen, Helium, and the nine metals most important for the cooling rate (C, N, O, Ne, Mg, Si, S, Ca and Fe) in each gas (and star) particle.
Each chemical element contributes separately to the cooling rate of the gas (see \cite{2013MNRAS.434.3142A} for details).
The abundances change over time in the gas due to enrichment from stellar feedback, as well as diffusion between gas particles.

The stellar feedback implemented in the code has three major effects: It enriches the gas with metals, accelerates it (kinetic feedback), and heats it up (thermal feedback).
If a star particle is tagged as giving feedback (either through supernova type Ia or II explosions, or through winds from asymptotic giant branch (AGB) stars) it distributes mass (in a mix of all tracked elements depending on the particle's metallicity and the feedback type) to its 10 closest neighbouring gas particles.
The fraction of mass that each gas particle gains is weighted with the SPH smoothing kernel of the feedback-giving star particle, and therefore depends on the distance between the two.

Following the simplifying assumption that all stars in a single population exploding in supernovae type II (SNII) do so at the same age $\tau_\text{SNII}$, each newly created star particle undergoes a SNII feedback event exactly once at a time $\tau_\text{SNII}$ after its creation, where $\tau_\text{SNII}$ is only dependent on the particle's metallicity.
While this is a valid approximation for SNII feedback because the SNII progenitor stars all have very short lifetimes, the supernova type Ia (SNIa) progenitors are old white dwarfs which can have vastly different ages at the time of their explosion.
Therefore, the SNIa feedback is modelled quasi-continuously: Each star particle older than 50 Myr undergoes SNIa feedback events repeatedly every 50 Myr until a maximum age of 10 Gyr, after which the feedback stops.
The feedback effects of these SNIa events (i.e. the released mass, energy and momentum) decline with the age $\tau$ of the particle as $\tau^{-1}$, thereby following the delay time distribution of the SNIa rate presented in \cite{2012PASA...29..447M}.
The AGB feedback follows the same procedure as the SNIa feedback, but with smaller yields and a different elemental distribution in the released mass.

Through the stellar feedback, energy and momentum are injected into the surrounding gas in the following way, using a model implemented by \citet{2017arXiv170101082N}.
A supernova event (of each type) is assumed to eject mass in an outflow with a velocity $v_{\text{out,SN}} = 4000 \text{ km}/\text{s}$, corresponding to an energy of
\begin{equation}
E_{\text{SN}} = \frac{1}{2} m_{\text{ejected}} v_{\text{out,SN}}^2
\label{eq:snfb}
\end{equation}
where $m_{\text{ejected}}$ is the total mass ejected by the star particle undergoing a supernova.
It depends on the age (for SNIa) the metal composition and the total mass of the star particle, following tabulated mass yields from \citet{1995ApJS..101..181W} for SNII and \citet{1999ApJS..125..439I} for SNIa.
Depending on the distance between the supernova-undergoing star particle (SN particle) and the affected gas particle, this outflow is then assumed to be in one of three characteristic phases of interaction with the ambient gas.
This supernova-ejecta phase then determines which fraction of $E_{\text{SN}}$ is injected into the gas particle as kinetic and thermal energy, respectively.
These phases are (in order of increasing distance to the SN particle) the momentum-conserving free-expansion phase, the Sedov-Taylor phase where 30\% of $E_{\text{SN}}$ are imparted as kinetic and 70\% as thermal energy \citep{1959sdmm.book.....S}, and the snow-plough phase where radiative cooling becomes important and reduces the total injected energy \citep[see][for details]{2017arXiv170101082N}.

For the AGB feedback, the feedback energy and momentum are always imparted to the neighbouring gas particles like they are in the free-expansion phase of the supernova feedback.
Here, the feedback energy is calculated according to equation \ref{eq:snfb}, but with $v_{\text{out,SN}} = 4000 \text{ km}/\text{s}$ replaced by $v_{\text{out,AGB}} = 25 \text{ km}/\text{s}$, and therefore much lower.

For the growth of the central SMBH of the simulated galaxy by accretion of gas and the resulting feedback we use a (slightly modified) model by \cite{2012ApJ...754..125C}, whose accretion model is in turn based on the model by \citet{2005MNRAS.361..776S}.
The black hole is represented by a single collisionless \enquote{sink} particle that grows by absorbing nearby gas particles.
The accretion rate follows the Bondi-Hoyle-Lyttleton formalism \citep{1939PCPS...35..405H,1944MNRAS.104..273B,1952MNRAS.112..195B} with several adjustments to its SPH implementation (the soft-Bondi criterion, the free-fall modification and an alternative averaging method) added by \citet{2012ApJ...754..125C}.
The inflow rate of gas onto the black hole is then given by
\begin{equation}
\label{eq:minf}
\dot{M}_\text{BHL} = \left\langle \frac{4\pi G^2 M_\text{BH}^2 \rho}{(c_\text{s}^2 + v^2)^{3/2}} \right\rangle\text{,}
\end{equation}
where $M_\text{BH}$ is the mass of the black hole, $\rho$ the surrounding gas density, $c_\text{s}$ the sound speed in the gas, $v$ the relative speed between the black hole and the gas, and the angle brackets stand for the SPH kernel averaging.
The probability of a gas particle to be part of the inflow onto the black hole in a given time step is then dependent on its kernel weight, the length of the time step, the inflow rate given above, and modifying factors corresponding to the soft-Bondi criterion and the free-fall time.

These modifiers are the following:
For the soft-Bondi criterion, the accretion probability is multiplied by a factor rising linearly from 0 at $r = r_\mathrm{B} + h$ to 1 (if $r_\mathrm{B} \geq h$) or to $\left(r_\mathrm{B}/h\right)^3$ (if $r_\mathrm{B} < h$) at $r = r_\mathrm{B} - h$, where $r$ is the distance between the black hole and the gas particle, $r_\mathrm{B}$ is the Bondi radius of the black hole, and $h$ is the SPH smoothing length of the gas particle.
This takes into account the limited resolution of the SPH simulation by only allowing gas particles to be accreted that are statistically within the Bondi radius.
The free-fall modification takes into account the relative free-fall times of the gas particles neighbouring the black hole to make it more likely for particles closer to the black hole (i.e. with shorter free-fall times) to be accreted than particles further away.

The inflow rate $\dot{M}_\text{inf}$ is limited by the Eddington rate \citep{1916MNRAS..77...16E}
\begin{equation}
\label{eq:medd}
\dot{M}_\text{edd} \equiv \frac{4\pi G M_\text{BH} m_\text{p}}{\epsilon_\text{r} \sigma_\text{T} c},
\end{equation}
which depends on the proton mass $m_\text{p}$, the Thompson cross-section $\sigma_\text{T}$ and the radiative efficiency of the AGN feedback $\epsilon_\text{r} = L_\text{r}/\dot{M}_\text{acc} c^2$ which is assumed to have a fixed value of 0.1, as is commonly done \citep[e.g.][]{1973A&A....24..337S}.
$L_\text{r}$ is the radiated luminosity of the accreting black hole, with $\dot{M}_\text{acc}$ being the rate of gas mass accreted by the black hole (i.e. incorporated into the BH mass).
The inflow rate is then $\dot{M}_\text{inf} = \min(\dot{M}_\text{BHL},\dot{M}_\text{edd})$.

The AGN feedback is a combination of kinetic-thermal wind feedback, and radiative feedback modelling the interaction of the SMBH's X-ray flux with the surrounding gas.
It works as follows:
We assume that only some of the inflowing gas $\dot{M}_\text{inf}$ is ultimately accreted onto the black hole while a large part of it is blown out again due to the feedback in the form of broad-line winds.
The rate with which this blown out gas is flowing out from the accretion region of the black hole is then simply $\dot{M}_\text{outf} = \dot{M}_\text{inf} - \dot{M}_\text{acc}$, and the energy and momentum fluxes of the AGN wind follow from conservation laws \citep{2010ApJ...722..642O}:
\begin{align}
\label{eq:ew}
\dot{E}_\text{w} \equiv & \epsilon_\text{w} \dot{M}_\text{acc} c^2 = \frac{1}{2} \dot{M}_\text{outf} v_\text{w}^2 \text{,} \\
\label{eq:pw}
\dot{p} = & \dot{M}_\text{outf} v_\text{w} \text{,}
\end{align}
where $\epsilon_\text{w}$ is the feedback efficiency (essentially a free parameter) and $v_\text{w}$ is the speed of the wind immediately after ejection by the AGN.
From equation \ref{eq:ew} follows:
\begin{equation}
\label{psi}
\frac{\dot{M}_\text{outf}}{\dot{M}_\text{acc}} = \frac{2 \epsilon_\text{w} c^2}{v_\text{w}^2} \text{,}
\end{equation}
i.e. the ratio between outflowing and accreted mass is only dependent on the wind speed $v_\text{w}$ and the feedback efficiency $\epsilon_\text{w}$.
Following \cite{2012ApJ...754..125C}, we fix the wind speed at $v_\text{w} = 10,000 \text{ km s}^{-1}$, a typical velocity for observed broad line winds \citep[e.g.][]{2003ARA&A..41..117C,2009ApJ...706..525M,2010ApJ...709..611D}.
The outflow-to-inflow ratio is now fully determined by choosing a feedback efficiency.
We choose $\epsilon_\text{w} = 0.005$ which leads to 90\% of the inflowing gas mass being ejected in the wind, while only 10\% ultimately contribute to the black-hole growth.
Numerically, this wind is implemented in such a way that gas particles that are part of the black-hole accretion region have a probability corresponding to the mass ratio $\dot{M}_\text{outf}/\dot{M}_\text{inf}$ to be ejected each time step.
Similarly, they are swallowed with a probability of $\dot{M}_\text{acc}/\dot{M}_\text{inf}$.
The broad line winds represented by this feedback collide with the gas in the immediate surroundings of the SMBH and create a larger, momentum-driven outflow.
As this happens on scales not resolved by the simulation, we mimic it by sharing the momentum of the ejected particle equally with two of its neighbouring gas particles.
The excess energy of the feedback is then distributed among the three particles as thermal energy.
The direction of the momentum is set to be either parallel or anti-parallel to the angular momentum of the corresponding particle relative to the black hole (with 50\% chance for each) before the ejection.
We choose this direction for the momentum, because the broad line outflows we mimic are stronger perpendicular to the accretion disc than parallel to it according to simulations by \citet{2004ApJ...616..688P}.
While the accretion disc is not resolved in our simulations, its angular momentum has to be that of the accreted gas particle.

In addition to the mechanical feedback described above, radiative feedback is used to both heat and accelerate the gas.
This part of the feedback has its source in the X-ray luminosity $L_\text{r} = \epsilon_\text{r}\dot{M}_\text{acc}c^2$ of the AGN which produces a flux $F_\text{r} = L_\text{r}/4\pi r^2$ at the position of each gas particle with the distance $r$ to the central black hole.
From this flux, the heating rate $\dot{E}_\text{X-ray}$ of the gas is calculated using formulae by \cite{2005MNRAS.358..168S} to describe the Compton and photo-ionization heating of the gas \citep[see][for details]{2012ApJ...754..125C}.
Other than the accretion luminosity and the distance to the AGN, the heating rate also depends on the temperature, the proton number density and the metallicity of the influenced gas particles.
The metallicity dependence was not in the original \cite{2005MNRAS.358..168S} formulae, but added in \citet{2016arXiv161009389C} to account for metal line absorption.
Note that, while $L_\text{r}$ is the bolometric luminosity of the AGN, only the effects of hard X-rays are taken into account.
Beside the heating, the X-ray flux from the black hole also creates a radiation pressure which is modelled as a momentum change of the gas particles
\begin{equation}
\dot{p}_\text{X-ray} = \frac{\dot{E}_\text{X-ray}}{c}
\end{equation}
radially away from the the black hole.
Last but not least, the actual Eddington force is also included in this feedback model.
It is based on Thompson scattering of the AGN's radiation with the surrounding electrons in the gas, and is implemented as a momentum change of the gas particles radially away from the black hole defined as
\begin{equation}
\dot{p}_\text{Edd} = \frac{F_\text{r} N_e \sigma_\text{T}}{c}
\end{equation}
where $N_e$ is the number of electrons in the considered gas particle.

\subsection{Initial conditions}

We simulate the evolution of a spherically symmetric, isolated, large and local early-type galaxy consisting of an old stellar population embedded in a hot gaseous halo, a dark matter halo, and containing a central supermassive black hole.
These four components of our model galaxy are represented by four different particle types, all of which but the gas particles are collisionless and only interact gravitationally.
The gas particles additionally interact via hydrodynamic forces.

The stellar component of the ETG has a mass of $M_* = 8.41\times 10^{10} M_\odot$, and is modelled with $N_* = 841,000$ particles with individual masses of $m_* = 10^5 M_\odot$ and gravitational softening lengths of $\epsilon_* = 20 \text{ pc}$ which trace a Hernquist density profile \citep{1990ApJ...356..359H} with a scale length of $a_* = 2.21$ kpc, corresponding to an effective (i.e. projected half-mass) radius of $R_\text{e} = 4.01$ kpc.
The mass and size of the stellar spheroid are scaled according to a relation by \citet{2010ApJ...713..738W}.
Each star particle has an initial age $\tau_\text{ini}$ and metallicity (tracked by multiple elemental abundances as described above).
Ages are randomly assigned to the stars in a log-normal distribution with a mean age of $\mu = 6$ Gyr and a logarithmic spread of $\sigma = 0.1$, corresponding to the predominantly very old stellar population found in most ETGs \citep[e.g.][esp. Fig. 10 and Fig. 14, respectively]{2005ApJ...621..673T, 2015MNRAS.448.3484M}.
Note that our simulations cover a time of about 4.5 Gyr, hence an initial stellar age of 6 Gyr corresponds to 10.5 Gyr at the end of the simulation, i.e. redshift 0.
The stellar metallicities follow the observations of \cite{2013ApJ...776...64G}: They are solar at the galactic centre and decline exponentially with increasing radius such that $[\text{Fe}/\text{H}] = -0.3$ at $r = 2R_e \approx 8 \text{ kpc}$, giving a linear slope for $[\text{Fe}/\text{H}]$ of $0.0375 \, \text{dex/kpc}$.

The mass of the central SMBH is set according to $M_\text{BH} = 4\times 10^8 M_\odot$, following the observed $M_\text{BH}-M_*$ relation by \cite{2013ARA&A..51..511K}.
Its softening length is identical to the stellar one: $\epsilon_\text{BH} = \epsilon_*$.
The virial dark matter mass (i.e. the dark matter mass within the virial radius) follows from the stellar mass via an abundance matching relation \citep{2013MNRAS.428.3121M}, yielding $M_\text{DM,vir} = 6.92\times 10^{12} M_\odot$.
This dark matter mass is distributed spherically symmetrically with a \cite{1990ApJ...356..359H} density profile where the scale length $a_\text{DM} = 74.7$ kpc is determined by associating it with a NFW profile \citep{1996ApJ...462..563N} of the same $M_\text{DM,vir}$ and a concentration factor of $c_\text{con} = 9$ \citep[see][]{2005MNRAS.361..776S}.
This leads to a virial radius $R_\text{vir} = 402$ kpc (where $R_\text{vir} = R_{200, \mathrm{crit}}$ i.e. the radius at which the density is 200 times the critical density).
With these choices for the stellar and dark matter distribution, the dark matter mass fraction within the half-mass radius of the galaxy is about 50\%, which is on the high end of, but still fully consistent with, the spread of observational values \citep[e.g.][]{2011MNRAS.415.2215B}.
The dark matter density profile is traced by $N_\text{DM} = 10^6$ dark matter particles of masses $m_\text{DM} = 9.71\times 10^6 M_\odot$ and softening lengths $\epsilon_\text{DM} = 200$ pc.
The angular momentum of the DM halo is given by the spin parameter $\lambda_\text{DM} = 0.033$.

The last component of the model ETG is the hot gas halo whose mass $M_\text{gas}$ is determined from the stellar and DM mass by setting the total baryon fraction of the galaxy within its virial radius $f_\text{b} = (M_\text{gas}+M_*)/M_\text{DM}$.
Due to the difficulty of detecting low-density gas at large distances from the galactic centre, the total baryon fraction of elliptical galaxies is not particularly well constrained \citep[e.g.][]{2013MNRAS.432.1845S}.
Still, there are several X-ray observations of hot gaseous haloes that show the existence of a scaling relation between the hot gas X-ray luminosity of a given ETG with some of its other properties (e.g. total mass, stellar luminosity), albeit mostly with a very large scatter \citep[see e.g.][]{2011ApJ...729...12B, 2013ApJ...776..116K, 2013MNRAS.432.1845S, 2015MNRAS.449.3806A}.
As $L_\text{X,gas}$ depends on the total hot gas mass, we use one of these relations to estimate the baryon fraction to be roughly 20\% of the cosmological value (see Fig. \ref{fig:Xray}), $f_\text{b,cosm} = 0.1864$ \citep{2014A&A...571A..16P}.
This corresponds to a hot gas mass of about $M_\text{gas,vir} = 1.74\times 10^{11} M_\odot$.
The hot halo is sampled with particles of the same mass and softening length as are used for the stellar component (i.e. $m_\text{gas} = 10^5 M_\odot$, $\epsilon_\text{gas} = 20$ pc), resulting in a gas particle number of $N_\text{gas} = 2.12\times 10^6$.
These particles trace the radial dependence of the gas density in form of a $\beta$-profile \citep{1976A&A....49..137C,1984ApJ...276...38J,1998ApJ...503..569E} with a slope parameter $\beta = 2/3$ \citep[following][]{1984ApJ...276...38J} and a core radius $r_c = 0.22 R_\text{S} = 9.8 \text{ kpc}$ \citep[following][]{1998ApJ...497..555M}, where $R_\text{S} = R_{200}/c_\text{con} \approx 44.7 \text{ kpc}$ is the scale radius of the dark matter profile.
The gas density profile is cut at a radius of $r_\text{cut} = 50r_c \approx 492 \text{ kpc}$, i.e. there are no gas particles beyond this radius.
From the density distribution, the temperature profile follows with the assumptions that the gas is in hydrostatic equilibrium and distributed isotropically \citep[see][]{2011MNRAS.415.3750M}.
Every gas particle is also given initial metal abundances, such that the radial metallicity profile follows that of the stars, but with a slightly lower metallicity peak of 93\% solar at the centre \citep[following][]{2014ApJ...783....8K}.
The hot gas halo has the same spin parameter as the dark matter halo ($\lambda_\text{gas} = 0.033$).

\begin{figure}
\centering
\includegraphics[width=.47\textwidth]{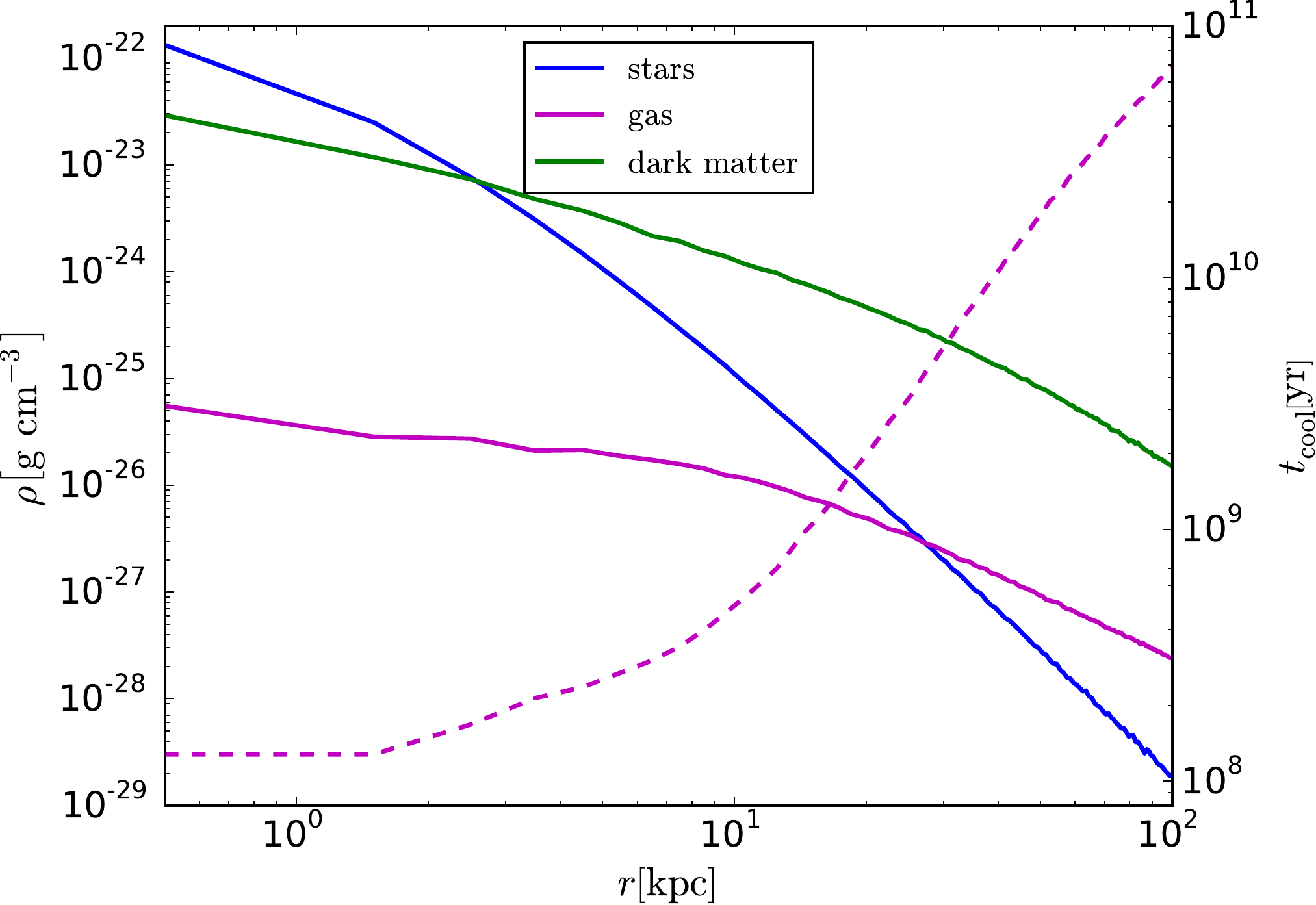}
\caption{Initial radial density profiles of the stars (blue), gas (magenta), and dark matter (green) in the simulated galaxy (solid lines, left y-axis), as well as the initial cooling time profile of the gas (dashed magenta line, right y-axis).
The cooling time is calculated using the same cooling rate that is used in the simulation itself, and is not taking into account gas heating due to feedback processes.}
\label{fig:inidens}
\end{figure}

With a gas particle mass of $m_\text{gas} = 10^5 M_\odot$, we are able to resolve the Jeans mass of all gas except the densest and coldest (compare Fig. \ref{fig:maps10}, bottom row), almost all of which is star-forming and hence turned into star particles after some time (if it is not affected by accretion or feedback processes).
Therefore, our resolution should be sufficient for investigating the impact of AGN feedback on the global properties of the galaxy, and the larger-scale behaviour of the hot gas.
An overview of the most important initial condition parameters is given in Table \ref{tab:IC}.
The initial density profiles of the components of the model galaxy (solid lines), as well as the cooling time of its gas (dashed line, calculated with the same cooling rate that is used in the simulation, and neglecting heating by feedback processes) are shown in Fig. \ref{fig:inidens}, on the left and right ordinate, respectively.
Within the central 5 kpc ($\sim$ the stellar half-mass radius), the cooling time varies between $\sim 100$ and $\sim 200$ Myr.
\begin{table*}
\centering
\caption{Overview of important parameters of the initial conditions.}
\begin{tabular}{c|cccc}
\hline
Parameter & Stars & Gas & Dark matter & Black hole \\
\hline
total mass $M_\text{tot}$ [$M_\odot$] & $8.41\times 10^{10}$ & $2.12\times 10^{11}$ & $9.71\times 10^{12}$ & $4\times 10^8$ \\
virial mass $M_\text{vir}$ [$M_\odot$] & $8.33\times 10^{10}$ & $1.74\times 10^{11}$ & $6.92\times 10^{12}$ & \\
particle mass $m$ [$M_\odot$] & $10^5$ & $10^5$ & $9.71\times 10^6$ & $4\times 10^8$ \\
number of particles $N$ & $8.41\times 10^5$ & $2.12\times 10^6$ & $10^6$ & $1$ \\
softening length $\epsilon$ [kpc] & $20$ & $20$ & $200$ & $20$ \\
\hline
\end{tabular}
\label{tab:IC}
\end{table*}

We compare three different simulation runs in this work, one without black-hole accretion and feedback (no-BH), one with BH accretion and the wind-feedback model described above, but without the radiative feedback (BH-W), and one with the full BH feedback model (radiative heating and radiation pressure in addition to the wind feedback, BH-WR).
We compare the no-BH model to the feedback models to investigate if BH feedback is necessary and/or sufficient to keep the ETG quiescent, and then compare the feedback models to probe the importance and effects of different feedback implementations on the galactic evolution.
Table \ref{tab:models} summarizes these simulation runs.
\begin{table}
\centering
\caption{Summary of simulations showing the run name and the active black-hole model components.}
\begin{tabular}{c|ccc}
\hline
Name of run & no-BH & BH-W & BH-WR \\
\hline
BH accretion &  & $\checkmark$ & $\checkmark$ \\
BH Eddington force &  & $\checkmark$ & $\checkmark$ \\
BH wind feedback &  & $\checkmark$ & $\checkmark$ \\
BH radiative feedback &  &  & $\checkmark$ \\
\hline
\end{tabular}
\label{tab:models}
\end{table}

To check how the mass of the hot halo affects the AGN feedback's ability to influence the galactic evolution, we include an additional run ($\text{BH-WR}_{50}$) using the BH-WR feedback model, but containing 50\% of the cosmological baryon fraction in the initial condition (raising the virial gas mass to $M_\text{gas,vir} = 5.61\times 10^{11} M_\odot$, and the number of gas particles to $N_\text{gas} = 6.86\times 10^6$) which is otherwise identical to the other runs.
This run lies an order of magnitude above the observed X-ray scaling relation already in the initial condition (see Fig. \ref{fig:Xray}, top panel).
It also completely fails to be quiescent (see Fig. \ref{fig:sSFR}), which shows that the amount of hot gas in ETGs has to be reduced to a low baryon fraction (or the gas density significantly reduced, compare \citet{2015MNRAS.449.3806A}) at higher redshifts for AGN feedback to be capable of regulating the late-time evolution of the galaxy.
In the further discussion in this paper, we will concentrate on the runs with 20\% of the cosmological baryon fraction (see Tab. \ref{tab:models}).

\section{Star formation history \& black hole growth}
\label{sfh}
\begin{figure}
\centering
\includegraphics[width=.47\textwidth]{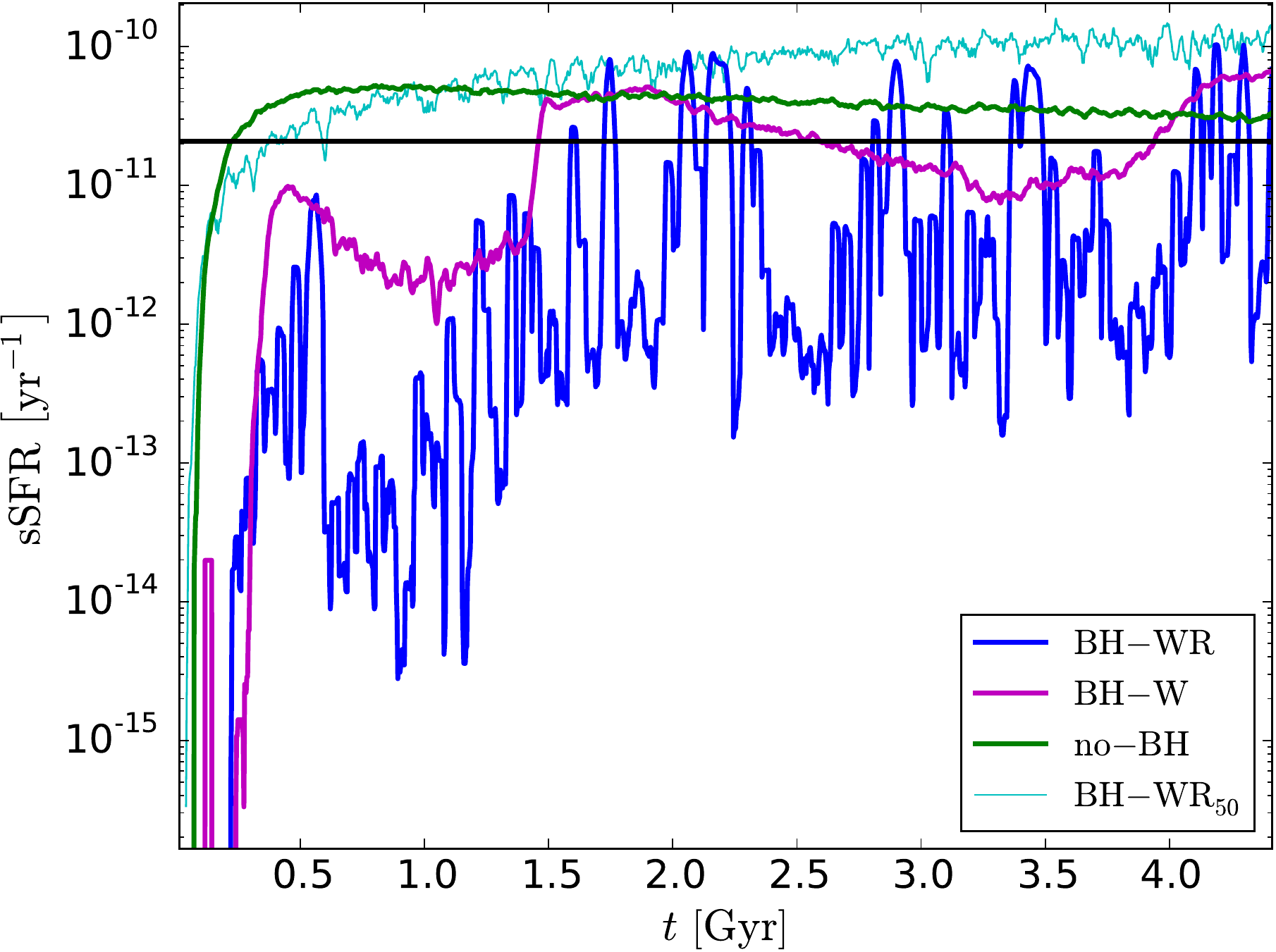}
\caption{Running mean with a step size of about 30 Myr (10 snapshots) of the specific star formation rate over time for the different models (see legend). The black horizontal line is the quiescence limit \citep{2008ApJ...688..770F}.
Without black-hole feedback, the galaxy is actively star-forming at almost all times; including the feedback reduces the overall star formation significantly.
With only wind feedback, there are alternating long periods of complete quiescence and star formation, while the full feedback model including radiation leads to fast oscillation between star-forming and quiescent states of the galaxy.
For 50\% of the cosmological baryon fraction, the AGN feedback fails to prevent star formation.}
\label{fig:sSFR}
\end{figure}

In Fig. \ref{fig:sSFR}, we show the evolution of the total specific star formation rate (sSFR) over time for all of our runs to investigate if the black-hole feedback can efficiently keep the galaxy quiescent.
As the sSFR oscillates quite strongly between individual snapshots, and is therefore noisy, we plot its running mean, averaged over 10 snapshots (about 30 Myr) each, to get a clearer picture of its time evolution.
The black, horizontal line in this figure is the quiescence limit according to \citet{2008ApJ...688..770F}, i.e. $\mathrm{sSFR} = 0.3\tau_\mathrm{Hub}^{-1}$ where $\tau_\mathrm{Hub} \equiv H_0^{-1}$ is the Hubble time.

Using this definition, the galaxy is actively star-forming for the whole simulation time except the first $\sim 200$ Myr (5\% of the total time) if we use the no-BH model, where the short quiescent period at the beginning of the simulation is just the time the initial hot gas needs to cool down in the centre of the galaxy (compare Fig. \ref{fig:inidens}).
However, if we include black-hole feedback, i.e. in the BH-W and BH-WR models, the overall star-formation rate of the galaxy is reduced significantly; the ETG is now quiescent for about 63\% and 87\% of the total simulation time in the BH-W and BH-WR models, respectively (Tab. \ref{tab:quiet}).
This shows that black-hole feedback is necessary to keep an isolated early-type galaxy quiescent at low redshifts.
\begin{table}
\centering
\caption{Quiescent time fractions: Total time above \& below the quiescence limit, as well as fraction of total time below the quiescence limit (quiescent fraction) for the different runs}
\begin{tabular}{c|cccc}
\hline
Name of run & no-BH & BH-W & BH-WR \\
\hline
time active (Gyr) & 4.12 & 1.63 & 0.60 \\
time quiescent (Gyr) & 0.23 & 2.80 & 3.83 \\
\textbf{quiescent fraction} & 0.05 & 0.63 & 0.87 \\
\hline
\end{tabular}
\label{tab:quiet}
\end{table}

The star-formation history of the galaxy differs distinctly in the two feedback models:
The additional radiative feedback in the BH-WR model not only reduces the overall star-formation of the galaxy compared to the BH-W model, and keeps it quiescent for a significantly longer fraction of the simulation time (compare Tab. \ref{tab:quiet}), it also increases the time-variability of the sSFR.
While in the BH-W model, the sSFR rises and falls moderately over long times, in the BH-WR model, it oscillates much faster and over a much higher range of values (with more than two orders of magnitude between the lowest and highest sSFR values in the 30 Myr running mean. Looking at individual snapshots without the running mean, the picture is largely the same: The sSFR repeatedly drops to zero in the BH-WR model, while its scatter is always less than one order of magnitude (usually below a factor of 2) in the BH-W model and only in the 10\% range in the no-BH model).

This difference is caused by the additional isotropic heating provided by the radiative feedback:
All the potentially star-forming dense gas in the galactic centre is heated to high temperatures by Compton scattering whenever a feedback event happens, strongly reducing the star-formation rate until the central gas can cool again.
In contrast, in the BH-W model, the wind feedback only accelerates and heats the gas directly surrounding the SMBH which then flows out of the central region only heating some of the cold, star-forming gas through shocks.
Hence, in the BH-W model, the sSFR changes less rapidly and not as strongly as in the BH-WR model.
\begin{figure}
\centering
\includegraphics[width=.47\textwidth]{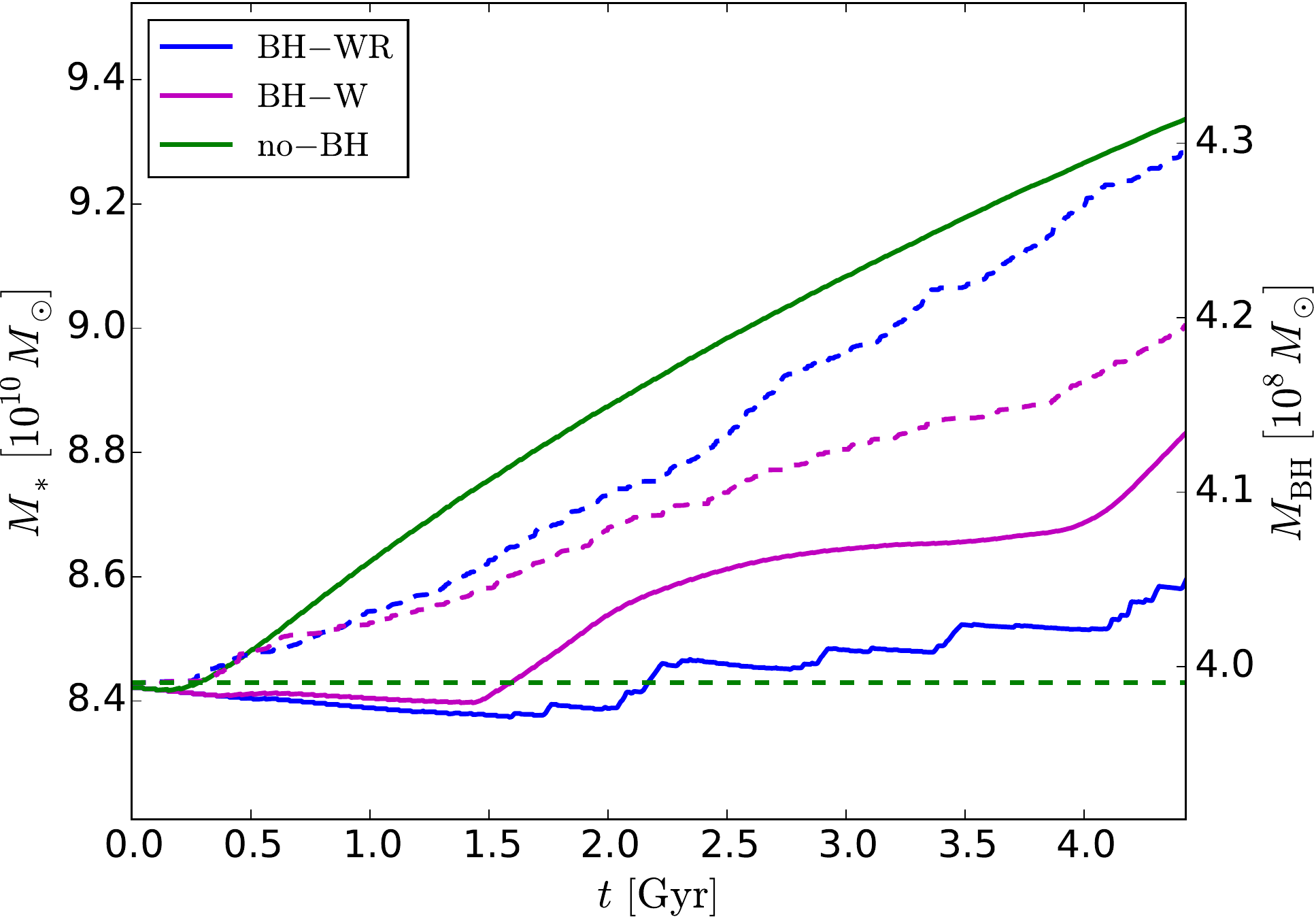}
\caption{Total stellar mass (solid lines, left y-axis) and black hole mass (dashed lines, right y-axis) of the galaxy over time for the same models.
Without black-hole feedback, the stellar mass grows by about 10\% over the whole simulation time; the feedback models reduce this growth to 4\% (BH-W) and 2\% (BH-WR) respectively.
In the BH-WR model, the black-hole mass grows more (about 7.5\%) than in the BH-W model (about 5\%), but in both feedback models the growth of the stellar and the black-hole mass is very small.}
\label{fig:MBH}
\end{figure}

Figure \ref{fig:MBH} shows the time-evolution of both the total stellar mass (solid lines, left ordinate) and the mass of the central black hole (dashed lines, right ordinate) for all models.
Without black-hole accretion and feedback (no-BH model), the stellar mass grows by about 10\%.
Both black-hole feedback models reduce the overall growth of the stellar mass significantly, compared to the no-BH model (to about 4\% in the BH-W, and about 2\% in the BH-WR model).
While the BH-WR model reduces the stellar mass growth more than the BH-W model, it leads to a slightly stronger growth of the black-hole mass (about 7.5\%, compared to about 5\% in the BH-W model), i.e. more gas accretion.
Overall, both feedback models keep the black-hole and the stellar mass growth small, and the galaxy quiescent for the majority of the time (though the BH-WR feedback is much more successful on that front).
\begin{figure}
\centering
\includegraphics[width=.47\textwidth]{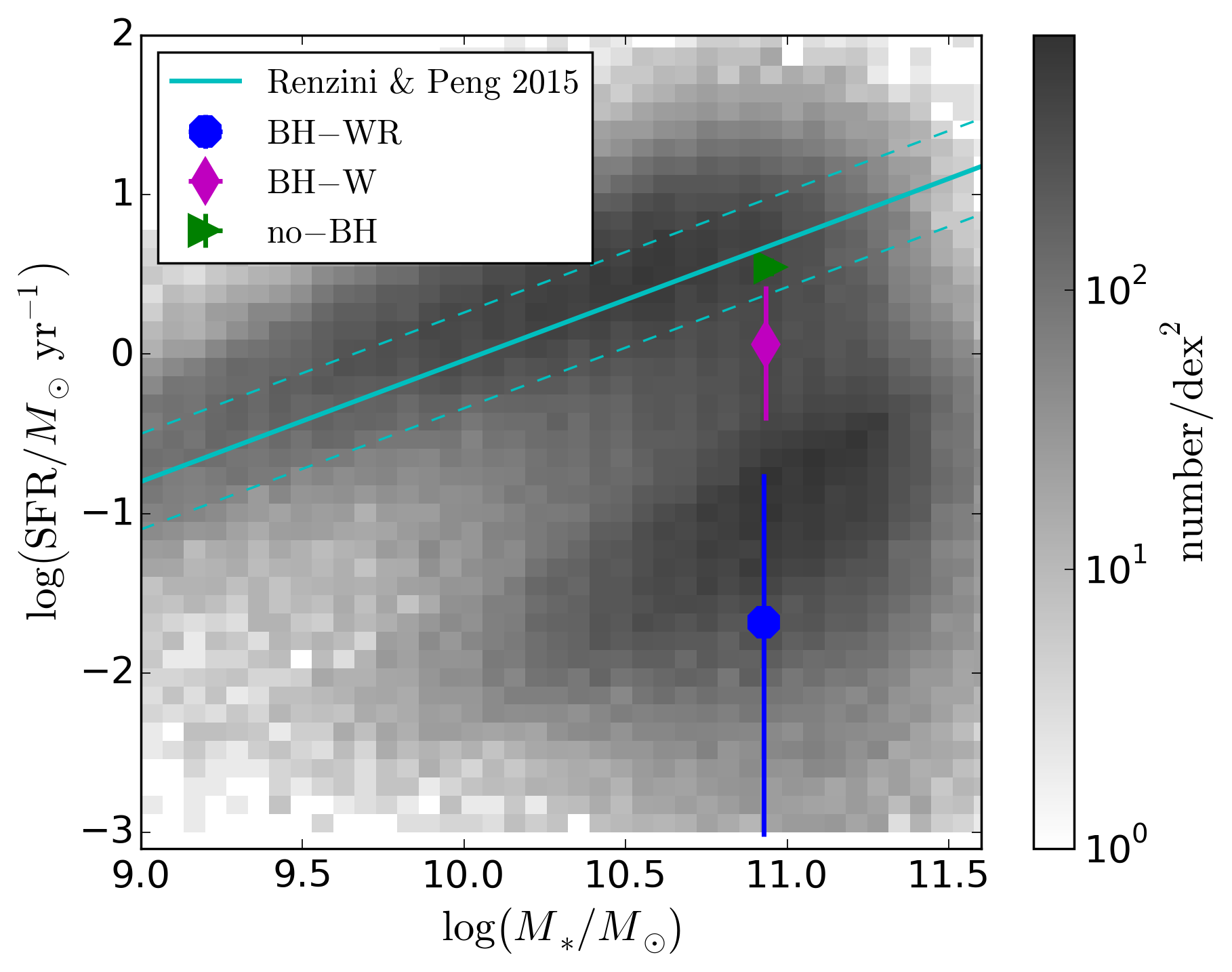}
\caption{SFR vs. stellar mass relation: Of the three models of this paper, the median values, as well as the scatter between the 25th and 75th percentile over the simulation time are plotted.
For comparison, observed data from \citet{2015ApJS..219....8C}---using the SDSS and WISE surveys---is shown as a grey-scale histogram depicting the number of galaxies in a given $M_*-\mathrm{SFR}$ bin, with a fit for the star-forming sequence by \citet{2015ApJ...801L..29R} (solid) and its 1$\sigma$ scatter (dashed) marked by the cyan lines.
To emphasise the bimodal split of the galaxy population into star-forming and quiescent galaxies, we plot the 1$\sigma$ upper limits for the observed star-formation rates and neglect $V_\mathrm{max}$ corrections.}
\label{fig:SFRMstar}
\end{figure}

In Fig. \ref{fig:SFRMstar}, we compare our simulations to the observed present day population of galaxies in a SFR -- stellar mass diagram.
For our simulations, we use the median star-formation rate and stellar mass over the whole simulation time (with the 1st and 3rd quartile of the spread in SFR given as error bars).
For the observations, we show data from \citet{2015ApJS..219....8C} \citep[see also][]{2015ApJ...801L..29R} as a grey-scale histogram of the abundance of observed galaxies in certain $M_*-\mathrm{SFR}$ bins. We use the 1$\sigma$ upper limit of the observed data without $V_\mathrm{max}$-correction to highlight the bimodal distribution between star-forming and quiescent galaxies.
To guide the eye, we also show the a fit for the star-forming sequence by \citet{2015ApJ...801L..29R} and its 1$\sigma$ scatter with cyan solid and dashed lines, respectively.

In the no-BH model, our simulated galaxy falls into the lower end of the star-forming sequence at all times during the simulation.
In the BH-W model, the galaxy has star-formation rates roughly corresponding to the ``green valley" between the blue and the red sequence of star-forming and quiescent galaxies.
Finally, in the BH-WR model, the galaxy's star-formation rate is firmly in the range of the quiescent galaxy population.
All in all, this confirms the conclusions made above: In our simulations, black-hole feedback is necessary to keep the galaxy quiescent, and including both wind and radiative feedback leads to significantly less overall star-formation than including only wind feedback.
\begin{figure}
\centering
\includegraphics[width=.47\textwidth]{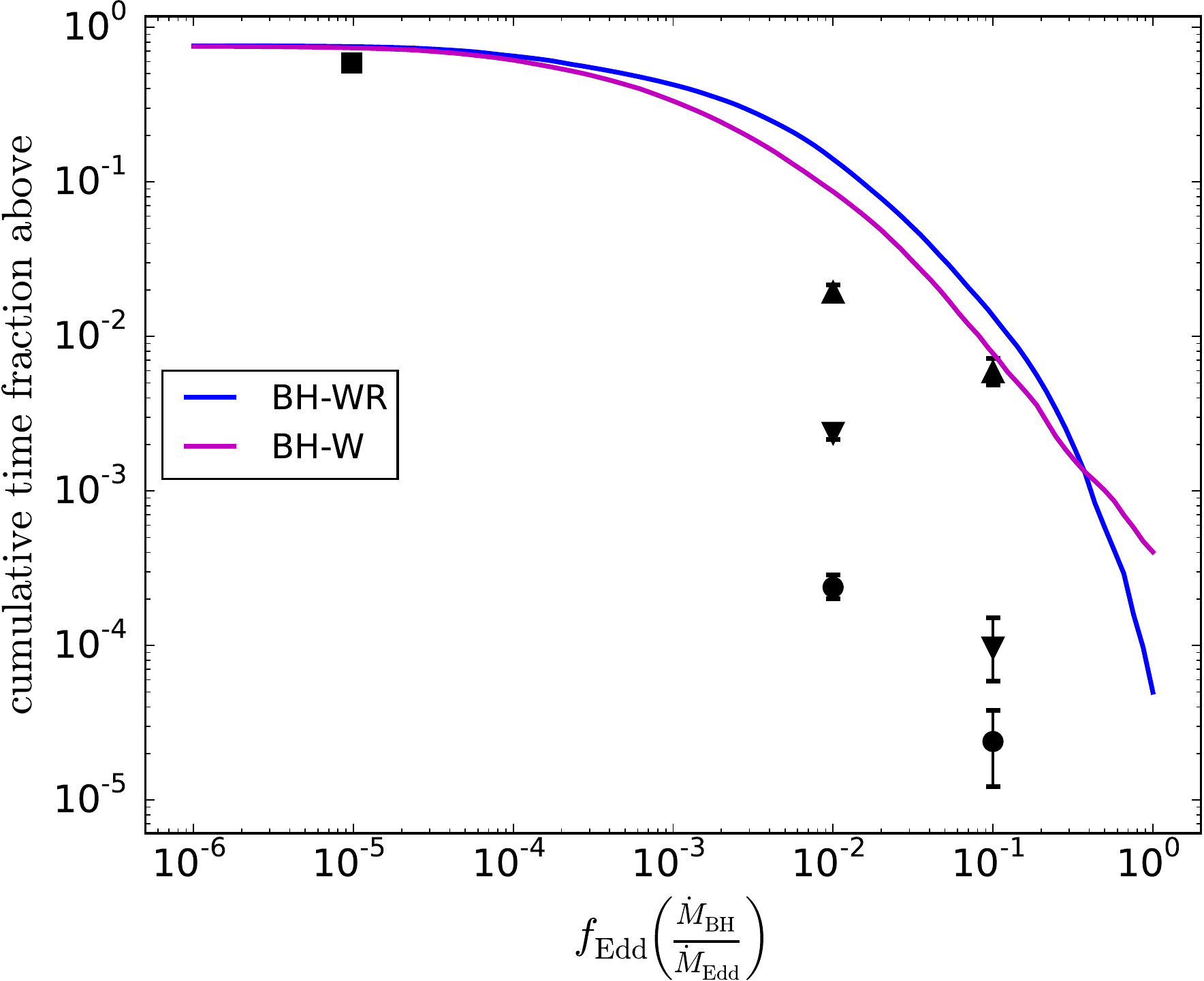}
\caption{Time distribution of different accretion rates in units of the Eddington limit (so called ``black-hole duty cycles") for the different models (see legend) compared to observations (square: \citet{2009ApJ...699..626H}, circles: \citet{2007ApJ...667..131G}, upward-pointing triangles: \citet{2009MNRAS.397..135K}, downward-pointing triangles: \citet{2004ApJ...613..109H}, all taken from \citet{2011ApJ...737...26N}).
Both simulated models fall above all of the observed values at Eddington ratios of 1\% and 10\%.}
\label{fig:BHgrowth}
\end{figure}

Lastly, we take a look at the accretion behaviour of the central SMBH in Fig. \ref{fig:BHgrowth}, which shows the Eddington ratio $f_\mathrm{Edd} = \dot{M}_\mathrm{BH}/\dot{M}_\mathrm{Edd}$ distribution of our two models including black-hole feedback compared with observational data from various sources \citep{2009ApJ...699..626H, 2007ApJ...667..131G, 2009MNRAS.397..135K, 2004ApJ...613..109H} which were compiled by \citet{2011ApJ...737...26N}.
We plot the cumulative fraction of simulation time during which the black hole accreted above a given $f_\mathrm{Edd}$ over the Eddington ratio.
Both models produce time fractions at high Eddington ratios that are too high compared to the observations, or only in agreement with the highest observational estimates, i.e. the SMBH in the simulations has a rather high duty cycle.
There is only little difference between the two models, though the BH-WR model has a slightly higher duty cycle than the onyWind model which matches the correspondingly higher SMBH mass growth (compare Fig. \ref{fig:MBH}).

\section{Black-hole-governed ISM evolution in the galactic centre}
\label{cenISM}
In the previous section, we have shown that feedback from the central supermassive black hole is both necessary to, and capable of keeping an isolated early-type galaxy quiescent.
The more physically complete feedback model, including both wind and radiative feedback, is more efficient at this than the pure wind-feedback model.
In this section and the next, we will now examine how the feedback achieves the quiescence through its influence on the galactic gas, and how it creates metal-rich, large-scale outflows that enrich the circumgalactic medium.
\begin{figure*}
\centering
\includegraphics[width=\textwidth]{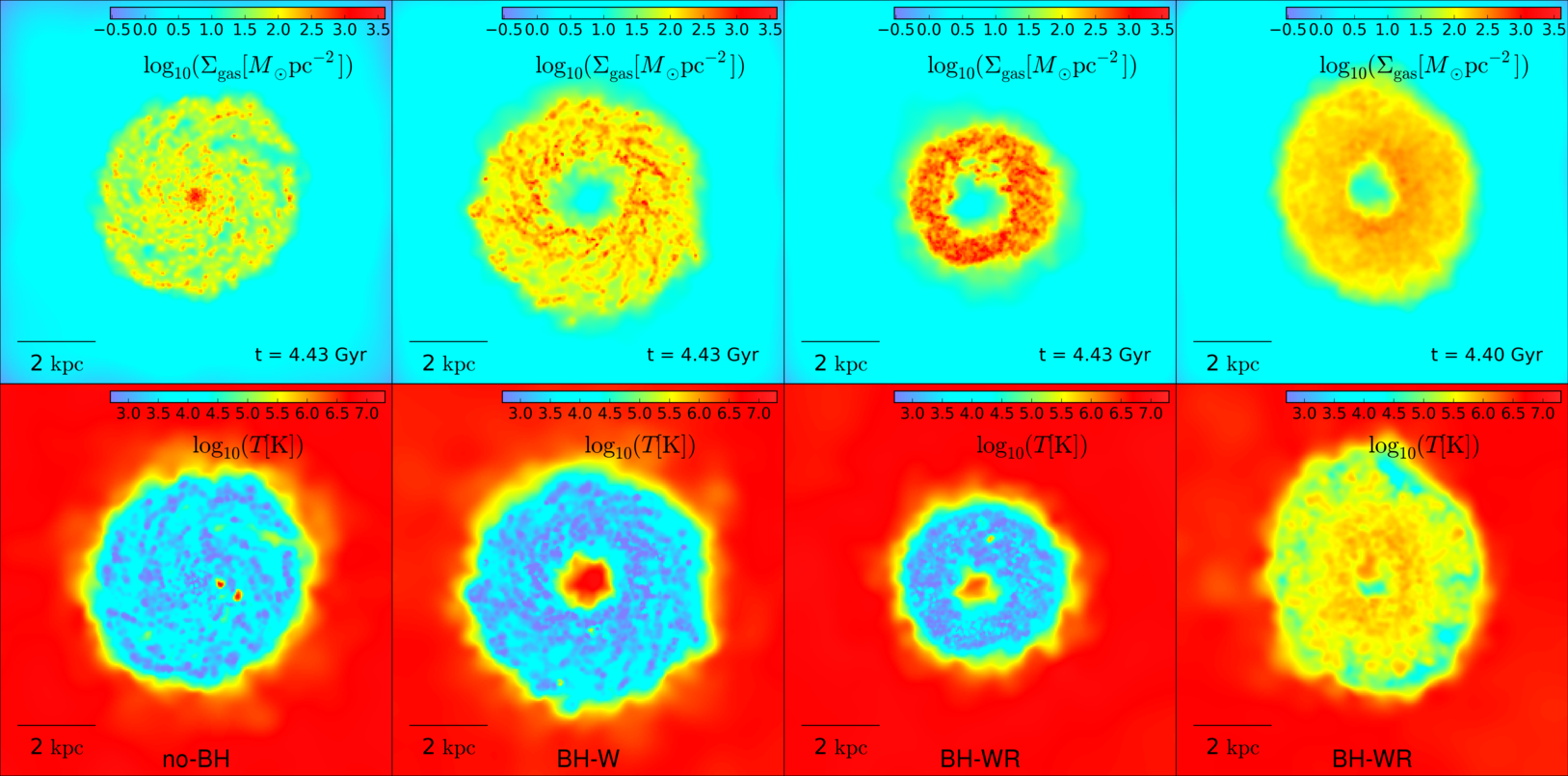}
\includegraphics[width=\textwidth]{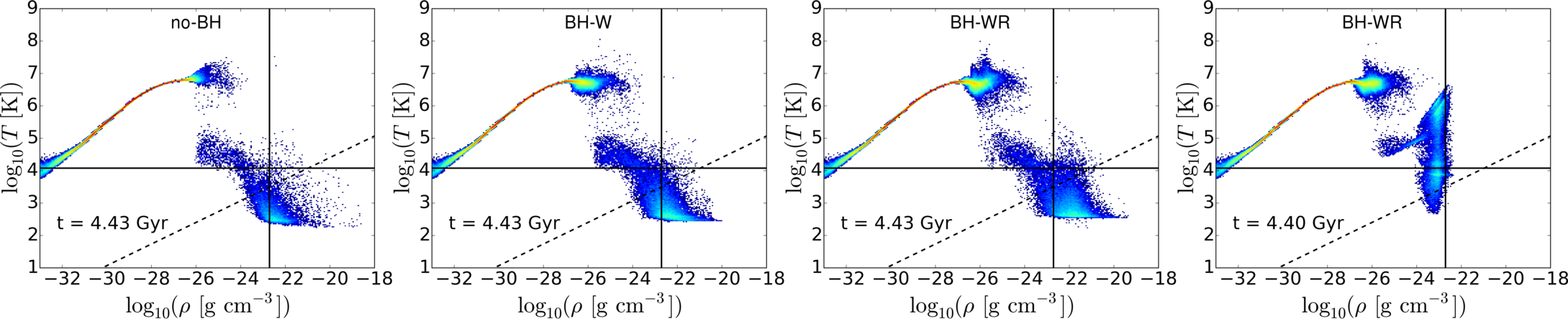}
\caption{Maps of the gas surface density (top row) and the density-weighted average temperature (central row) in a 10 kpc thick, 10x10 kpc wide region around the galactic centre, as well as temperature-density phase diagrams of the gas (bottom row).
The columns show, from left to right, the states at the end of the simulation time for the no-BH, BH-W and BH-WR models, as well as the state of the BH-WR model 30 Myr before the end of the simulation.
In the phase diagrams, the solid vertical and horizontal lines mark the density and temperature thresholds for star formation, respectively.
The dashed line marks the resolution limit of the simulation, i.e. the Jeans mass is resolved for all gas that lies to the top left of the dashed line in the phase space, and the colour shows the number density of gas particles in the temperature-density bins (rising number density from blue to red).
In all simulations, a central disc of cold, dense gas is formed.
While in the no-BH model, this disc includes an especially dense central core, in the other two models, the very centre is instead evacuated of gas by the wind feedback.
Additionally, in the BH-WR model, the whole disc is periodically heated and enlarged by the radiative feedback, before condensing and cooling down again.}
\label{fig:maps10}
\end{figure*}

We begin with Fig. \ref{fig:maps10}, which shows colour maps of the surface density (top row) and the density-weighted average temperature (central row) in a 10 kpc thick, 10x10 kpc wide region around the galactic centre (and therefore around the central SMBH), as well as the temperature-density phase diagrams of all the galactic gas (bottom row).
In the maps, the line of sight is chosen to be parallel to the initial angular momentum of the hot halo.
The solid vertical and horizontal lines in the phase diagrams mark the star formation thresholds in density and temperature, respectively, i.e. gas in the bottom right corner of the diagrams is star-forming.
The individual temperature-density bins are colour-coded for gas particle number density, which is rising from blue to red colours.
The dashed line marks the resolution of the simulation limit following the Jeans mass criterion:
Above the dashed line, the Jeans mass of the gas is resolved, i.e. it is larger than the kernel mass $M_\mathrm{ker} = N_\mathrm{ngb}m_\mathrm{gas} = 10^7 M_\odot$, while it is unresolved below the line, meaning that gas in this region could collapse under self-gravity due to numerical noise.
Almost all of the gas particles with unresolved Jeans mass are in the star-forming region of the phase space, and will be turned into star particles if they stay in this region.
Hence, while this resolution does not allow us to make predictions about the substructure of the dense, cool gas, it is high enough to investigate the global and larger-scale effects of AGN feedback that interest us in this paper.
The three left columns show the state of the three models (from left to right: no-BH, BH-W, BH-WR) at the end of the simulation time, while the last column to the right shows the state of the BH-WR model about 30 Myr before the end of the simulation.

In the no-BH model, we see a typical cooling-flow: The gas from the hot halo of the galaxy flows to the centre, cools down, and forms a disc of cold, dense gas in the central $\sim$6 kpc.
An especially dense core of up to a few times $10^{-19}\mathrm{g}/\mathrm{cm}^3$ forms at the very centre and leads to significant star-formation.
As the stellar feedback is too weak to affect the gas much, this central core and disc are stable over the whole simulation time, while some of the gas is constantly being turned into stars, and is in turn replenished by further cooling from the hot halo.

In both the BH-W and the BH-WR model, the gas cooling down into the centre forms a dense, ring-like disc with a hole of under-dense gas in the very centre of about 1 kpc size.
This central hole is produced by the wind feedback, which both heats the central gas and accelerates it parallel to the angular momentum of the gas disc, creating a fast, biconal outflow which evacuates the centre and thereby significantly reduces the star-formation rate.
In the BH-WR model, the star-formation rate is further reduced by the radiative heating:
While, in the BH-W model, the cold, dense disc around the centre is mostly undisturbed and forms stars constantly (though at a lower rate than the core in the no-BH model), in the BH-WR model, the whole disc is repeatedly heated and expanded (and thereby made less dense) by the radiative feedback (compare the two rightmost columns in Fig. \ref{fig:maps10}).
While the gas disc re-compresses and cools again quickly after being heated, the radiative feedback nevertheless reduces the galactic star-formation rate greatly, leading to the overall very quiescent state of the BH-WR model.

\section{Large-scale metal enrichment through feedback-driven winds}
\label{CGM}
\begin{figure*}
\centering
\includegraphics[width=\textwidth]{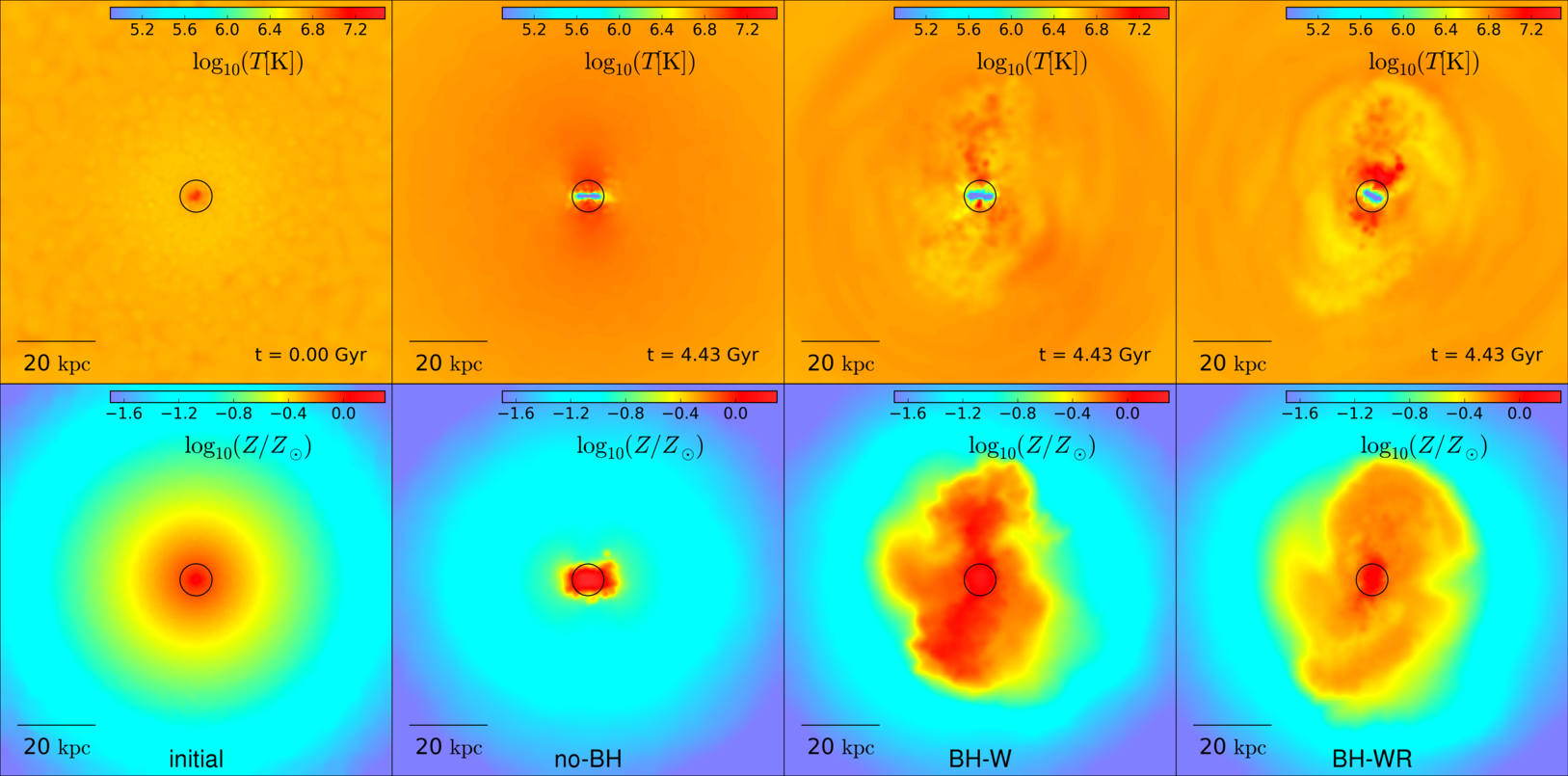}
\caption{Maps of the density-weighted average temperature (top row) and metallicity (bottom row) of the gas in a 3 kpc thick, 100x100 kpc wide region around the galactic centre.
The columns show, from left to right, the initial condition and the states at the end of the simulation time for the no-BH, BH-W and BH-WR models.
The black circle marks 1\% of the virial radius.
In the no-BH model, a very metal-rich and cold central disc is formed, surrounded by a hot accretion shock of infalling gas.
All gas outside this central region is depleted of metals.
In the other two models, the wind feedback drives large-scale, hot, metal-rich outflows, which keep the CGM of the galaxy metal-enriched out to ca. 30 kpc.}
\label{fig:maps}
\end{figure*}

In Fig. \ref{fig:maps}, we show the density-weighted temperature (top row) and metallicity (bottom row) in a 100x100 kpc wide, 3 kpc thick region around the galactic centre.
Unlike in Fig. \ref{fig:maps10}, the line of sight in this figure is perpendicular to initial angular momentum of the hot gas, allowing for the depiction of the biconal outflows produced by the black-hole feedback.
The left most column shows the initial condition; the other columns show, from left to right, the state of the no-BH, BH-W, and BH-WR models at the end of the simulation time, and the black circle marks $r = 0.01 R_\mathrm{vir}$.

In the no-BH model, we again see that a thin, cold disc is formed in the galactic centre.
This disc is surrounded above and below the plane by shock-heated gas that is being accreted onto it.
The cold disc and its surrounding accretion-shock region are the only part of the galaxy which remain metal-enriched by the end of the simulation:
All the metal-rich gas that is initially distributed throughout the galaxy cools down into the centre, where it is further enriched with metals produces by newly formed stars.
The lack of strong feedback (the stellar feedback being too weak to affect the central gas much beside the metal-enrichment) then leads to a highly metal-enriched central region of the galaxy and a total lack of any metals in the outskirts and the CGM.

The situation looks very different in the two models that include black-hole feedback, BH-W and BH-WR:
While we still see cold, central discs, in these models, the wind feedback produces biconal, hot, and metal-rich outflows perpendicular to the disc plane.
These outflows not only evacuate the galactic centre from gas that would otherwise form stars, they also transport metals produced in the centre up to ca. 30 kpc away from it, into the CGM.
Most of the outflowing, metal-rich gas falls back towards the centre within a few hundreds of Megayears, but the constantly renewed, black-hole feedback driven outflows from the centre still keep the CGM (up to ca. 30 kpc) enriched with metals.
\begin{figure}
\centering
\includegraphics[width=.47\textwidth]{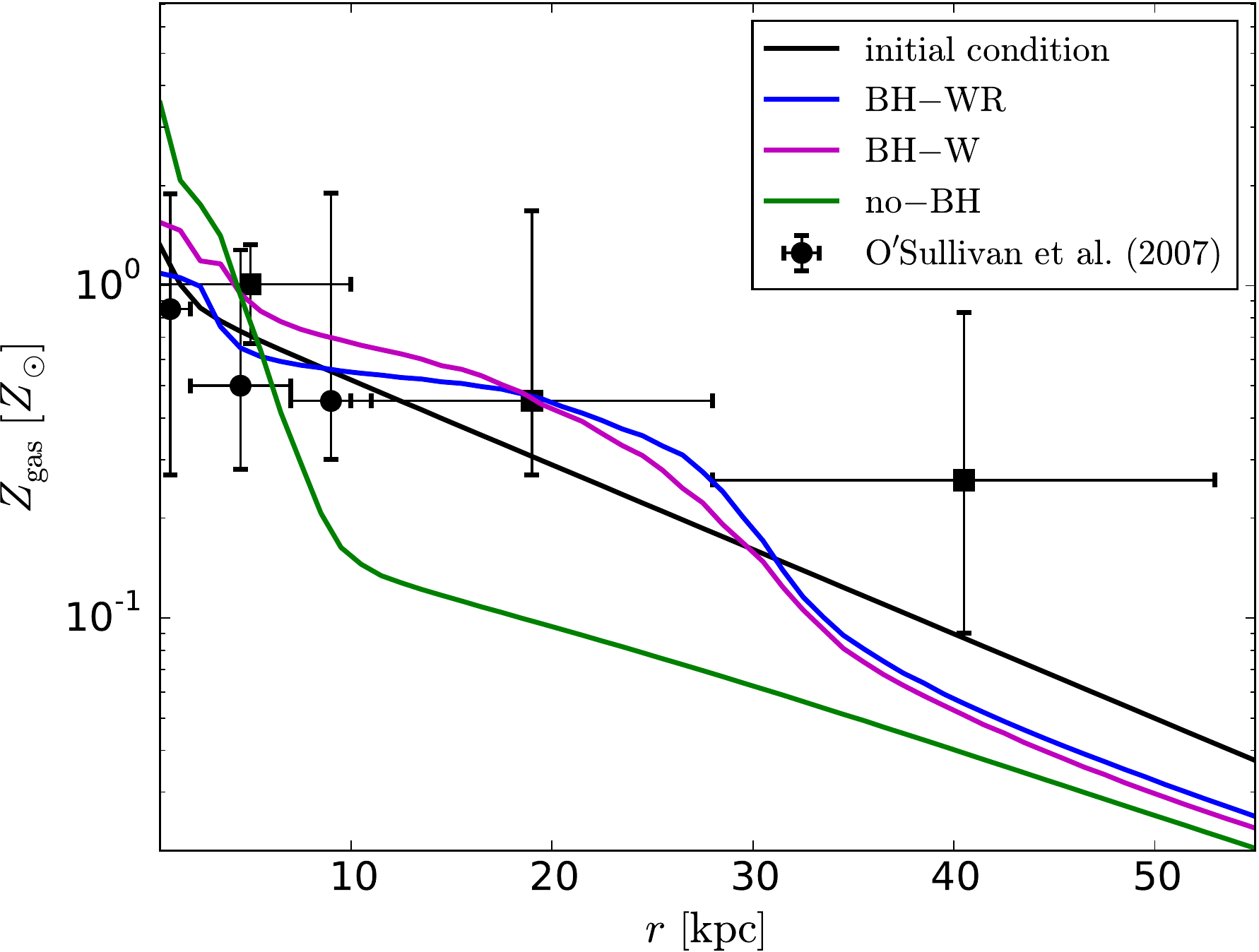}
\caption{Radial gas metallicity profiles at the end of the simulation, for the different models, as well as in the initial condition (see legend).
Also shown are observed values from \citet{2007MNRAS.380.1409O} for NGC 7796 (circles) and NGC 57 (squares).
In the no-BH model, only the centre of the galaxy stays metal enriched, while the volume further out than ca. 6 kpc becomes strongly depleted in metals.
In contrast, both in the BH-W and the BH-WR models, the gas stays enriched at more than 10\% solar metallicity out to about 30 kpc.}
\label{fig:metals}
\end{figure}

Figure \ref{fig:metals} demonstrates the same point by showing the 3D radial profiles of the gas metallicity at the end of the simulation for all three models, compared to the initial metallicity profile.
In the no-BH model, the star-formation and subsequent stellar feedback in the cold, dense disc and core raises the metallicity in the centre up to two to three times the solar value, but the constant inflow of gas from the outer parts of the galaxy, combined with the lack of any large-scale outflows, leads to a total depletion of metals in the CGM gas (with less than 10\% solar metallicity outside of a radius of 20 kpc).
On the other hand, with black-hole feedback (in the BH-W and BH-WR models), the central metallicity stays lower, close to the initial value, while the outer parts of the galactic gas become significantly enriched with metals out to a radius of about 30 kpc.
There is little difference between the two models that include black-hole feedback, as the wind feedback is the relevant part to create large-scale outflows capable of enriching the CGM.
Outside of a radius of ca. 30 kpc, the gas becomes depleted in metals even in the feedback-including models, as the AGN-driven outflows lose their momentum and stop progressing before they can reach further out.
The two models with black-hole feedback also agree much better with the observations of \citet{2007MNRAS.380.1409O}, at least out to $\sim 30$ kpc, but considering the observational uncertainties, as well as the slope of the initial metallicity profile in the simulations (which is set to follow the stellar profile, and therefore not necessarily realistic), on should be cautious before reading to much into this agreement.
\begin{figure}
\centering
\includegraphics[width=.47\textwidth]{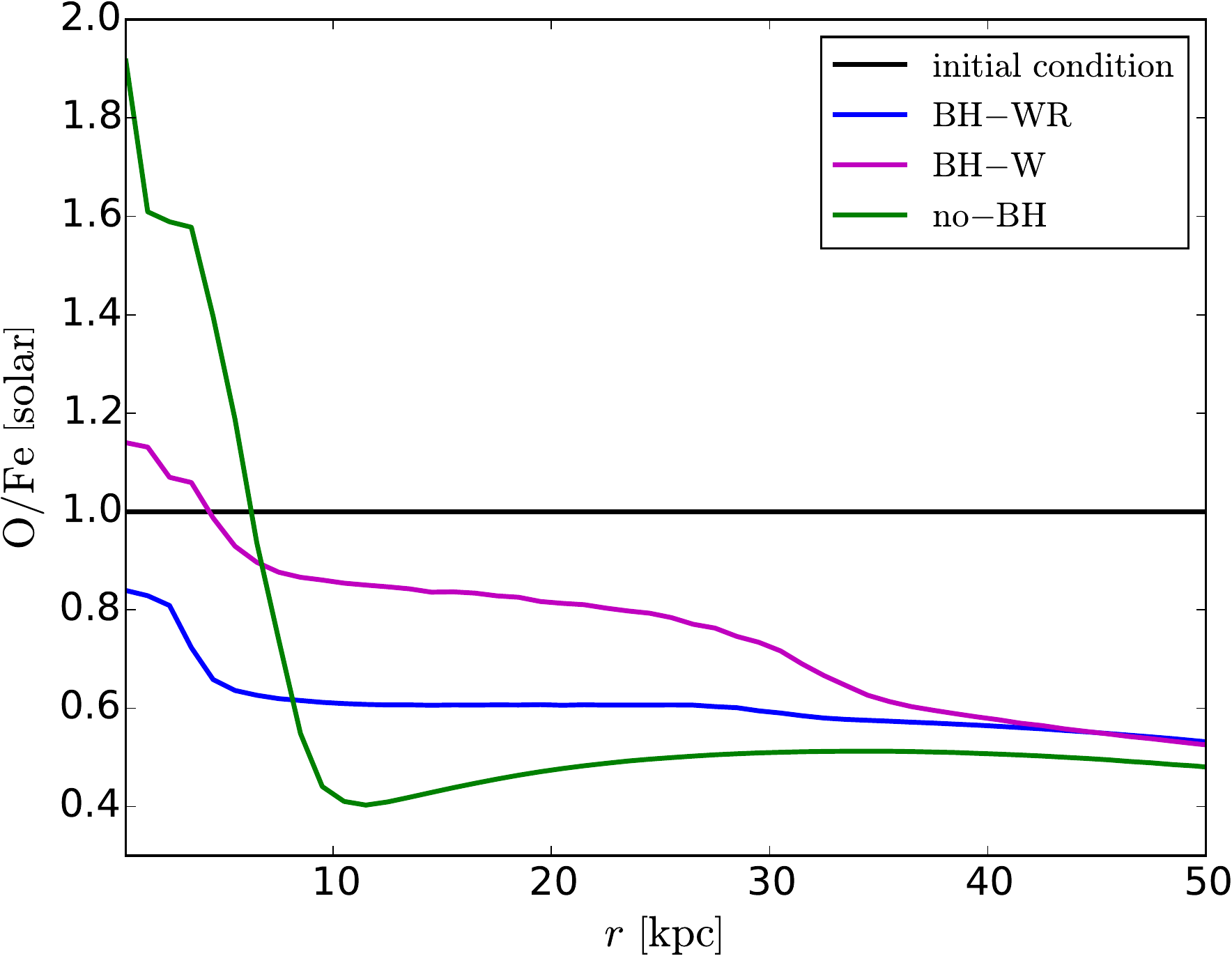}
\caption{Radial gas oxygen-to-iron ratio (O/Fe) profiles in solar units at the end of the simulation for the different models, as well as in the initial condition (see legend).
The no-BH model shows enhanced O/Fe in the star-forming galactic centre, and decreased O/Fe further out.
The BH-W model still has enough star-formation for a slightly enhanced O/Fe in the centre, and also shows higher O/Fe in the outskirts than the no-BH model due to its feedback-driven outflows.
The BH-WR model leads to decreased O/Fe at all radii due to the strong lack of star formation, but still shows higher O/Fe at large radii than the no-BH model due to large-scale outflows.}
\label{fig:ofe}
\end{figure}

In Fig. \ref{fig:ofe} we show the radial profiles of the oxygen-to-iron ratio in solar units (O/Fe) at the simulation's end for all three models, compared with the initial condition (which is set to the solar value, i.e. 1).
As oxygen is mainly produced in type-II supernovae of young, massive stars, while iron mostly originates from old white dwarfs undergoing type-Ia supernovae, and the initial stellar population in our simulations is made up exclusively of old stars, the radial O/Fe profile shows the combined effect of star-formation and feedback-driven metal distribution.

In the no-BH model, significant star-formation in the centre of the galaxy leads to an enhancement of O/Fe in the central regions up nearly twice the initial value.
None of the enriched gas from the star-forming centre is transported out beyond a few kiloparsecs, so no new, star-formation-dependent oxygen reaches the outskirts of the galaxy, while there are still old stars producing iron through SNIa in these regions.
Thereby, O/Fe decreases significantly beyond a radius of ca. 6 kpc, though not as drastically as the overall metallicity (compare Fig. \ref{fig:metals}).

In the BH-W model, the central star formation (and therefore oxygen production) is significantly lower than in the no-BH model, but still high enough to raise O/Fe slightly above the initial value in the galactic centre.
The wind-feedback driven outflow also transports some of the centrally-produced oxygen out into the CGM (up to about 30 kpc, the reach of the outflows), leading to O/Fe in these regions about twice as high as in the no-BH model, but still lower than in the initial condition.

In the BH-WR model, the star-formation rate is so low that the production of iron by old stars outweighs that of oxygen by newly formed ones everywhere in the galaxy: O/Fe is lower than initially at all radii.
Still, what little oxygen is produced in the slightly star-forming centre is distributed by the feedback-driven outflows into the CGM, leading to O/Fe in the region between about 6 kpc and 30 kpc lower than in the BH-W model, but still slightly higher than in the no-BH model.

Observed oxygen-to-iron ratios in massive elliptical galaxies mostly range from about 60\% to ca. 100\% of the solar value \citep[e.g.][and references therein]{2014A&A...572L...8P, 2011A&A...531A..15G}, and have roughly flat radial profiles \citep{2011A&A...531A..15G}.
The results of our BH-W and BH-WR runs are consistent with these observations, while those of the no-BH run fit less well to the data, especially in the centre, which is very oxygen-rich in the simulation.
\begin{figure}
\centering
\includegraphics[width=.47\textwidth]{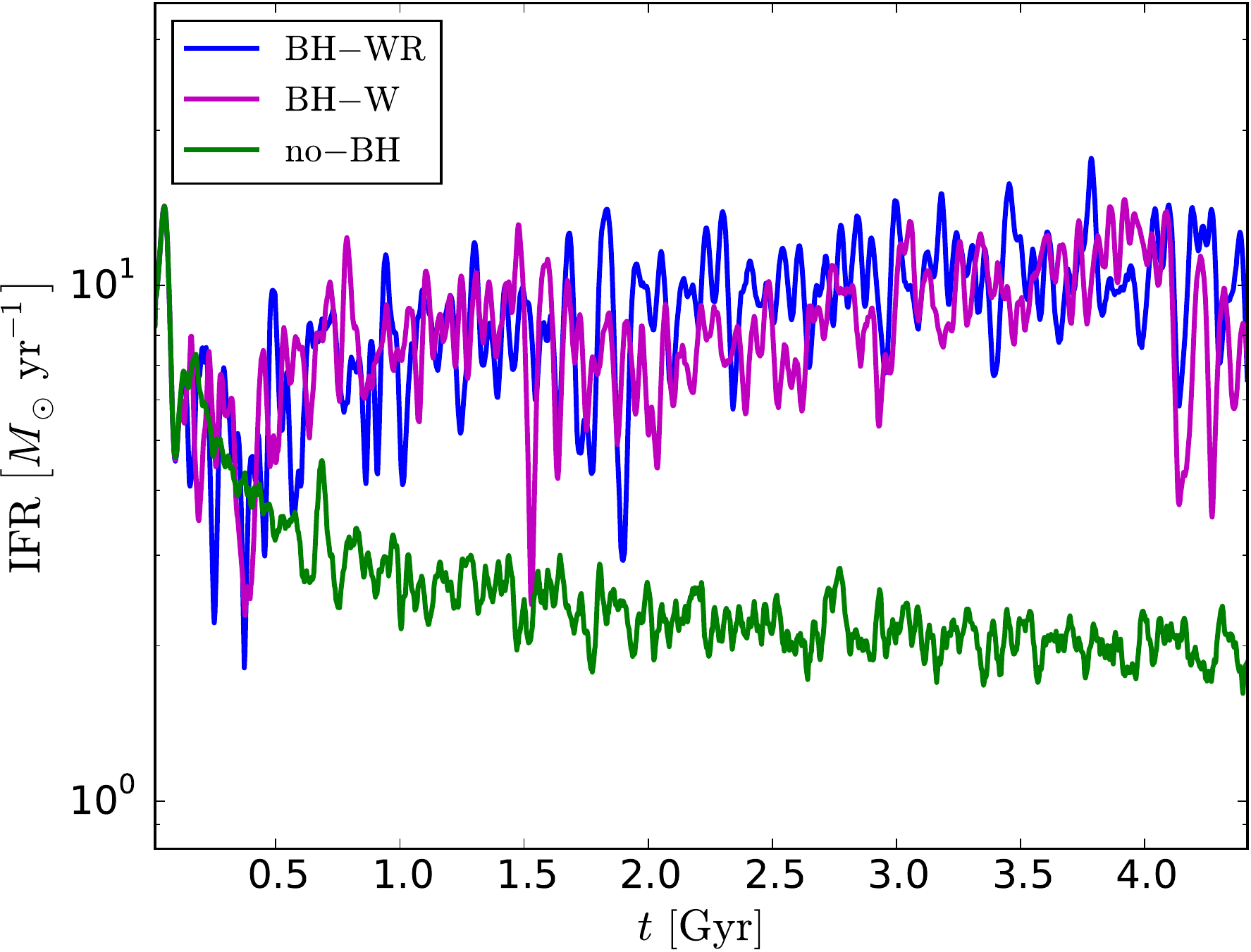}
\includegraphics[width=.47\textwidth]{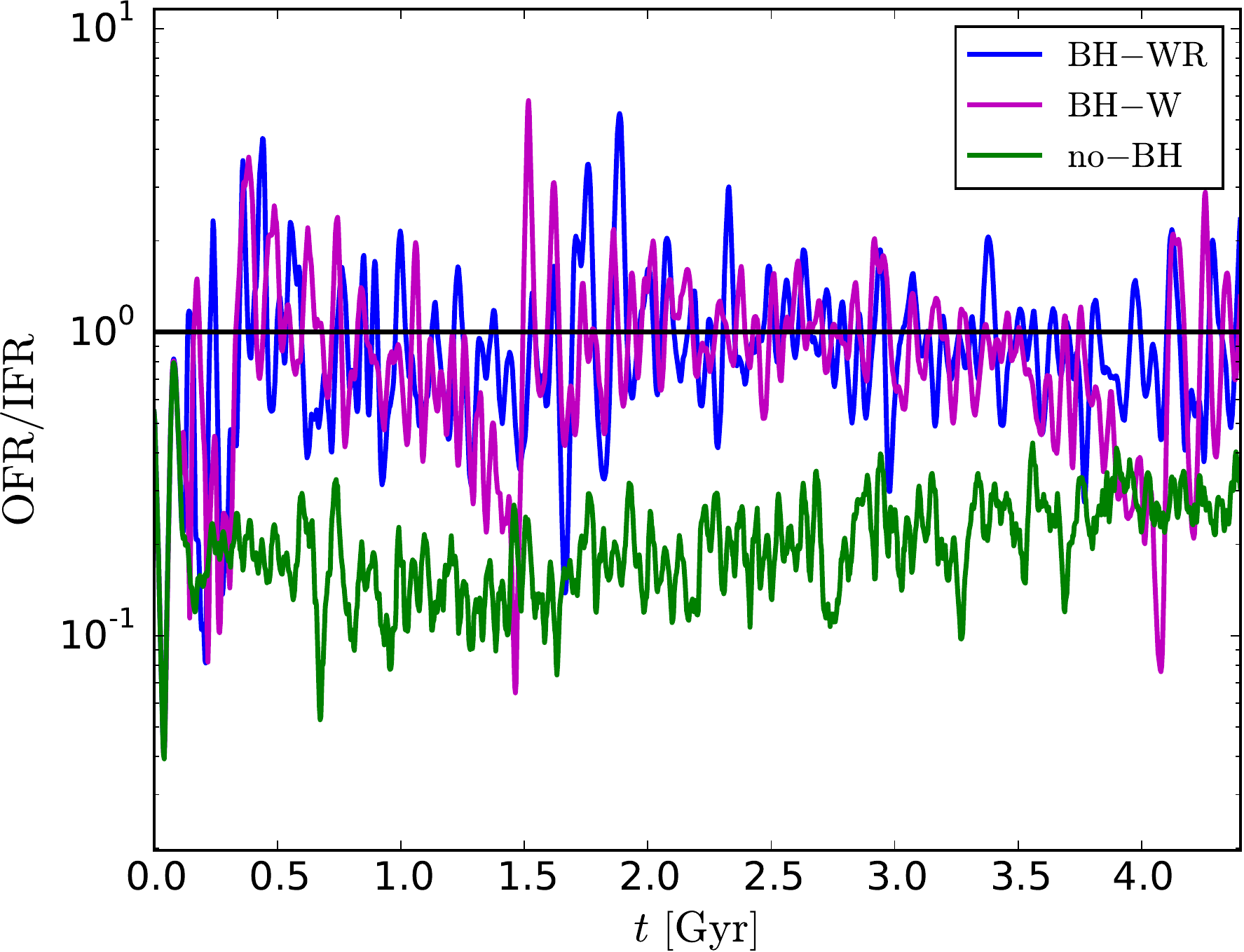}
\caption{Inflow (IFR, top panel), and the ratio of outflow to inflow rate (OFR/IFR, bottom panel) through a shell at 20 kpc distance from the centre over time for the different models (see legend).
For all flow rates, the running mean with a ca. 30 Myr (10 snapshot) step size is plotted.
The horizontal black line in the bottom panel shows a 1:1 ratio of OFR to IFR, i.e. zero net-flow of gas.
The BH-W and BH-WR models show very similar flow rates, with OFR/IFR oscillating around 1, while the no-BH model produces generally much lower flow-rates and constant OFR/IFR ratios far below unity.}
\label{fig:flows20}
\end{figure}

Using Fig. \ref{fig:flows20}, we examine the gas flow rates in the different models directly.
It shows, for all three models, the time evolutions of the gas mass inflow rate (IFR) through a shell at 20 kpc distance from the centre (top panel), and the ratio of outflow to inflow rate (OFR/IFR, bottom panel).
The flow rates are estimated taking the current positions and velocities of all gas particles on one side of the shell, and then determining which of them will be on the other side of the shell 3 Myr (about one snapshot-interval) later, assuming constant velocity during that time.
The total mass of all these gas particles is then the inflow or outflow rate, respectively.
Same as for the specific star-formation rate, we use a running mean over about 30 Myr for the flow rates to reduce the noise of the plots.

In the models that include black-hole feedback, BH-W and BH-WR, the flow rates are very similar.
Both inflow and outflow rate oscillate around about $8 M_\odot/\mathrm{yr}$, leading to an OFR/IFR ratio that fluctuates around 1.
Periods of net-outflow alternate with those of net-inflow: Gas is blown out by the black-hole-wind feedback, and falls back towards centre, leading to a quasi steady state (at least within the simulation time).
While the black-hole wind produces outflows, these are compensated for by gas flowing back into the galaxy, so the galactic centre is not completely depleted.
Instead, a so-called ``galactic fountain" is created.

In the no-BH model, there is overall much less movement of the gas (at this distance from the centre) than in the other models:
The inflow rate is lower by a factor of about 3-4, while the outflow rate is lower by more than an order of magnitude, leading to an OFR/IFR ratio that is constantly below 1, usually oscillating around $\sim 0.2$.
Without the black-hole feedback that drives gas out from the central regions which then slows down and starts flowing into the centre again, the only movement of the gas is an inflow to the centre where it is converted into stars.
The CGM gas is also not replenished by feedback-driven outflows, hence the total inflow rate is also significantly lower than in the feedback-including models.
As the hot gas halo and its metallicity are slowly depleted, the inflow rate shrinks further, though the effect is minor (from about $3 M_\odot/\mathrm{yr}$ to about $2 M_\odot/\mathrm{yr}$ in the last 4 Gyr).
\begin{figure}
\centering
\includegraphics[width=.47\textwidth]{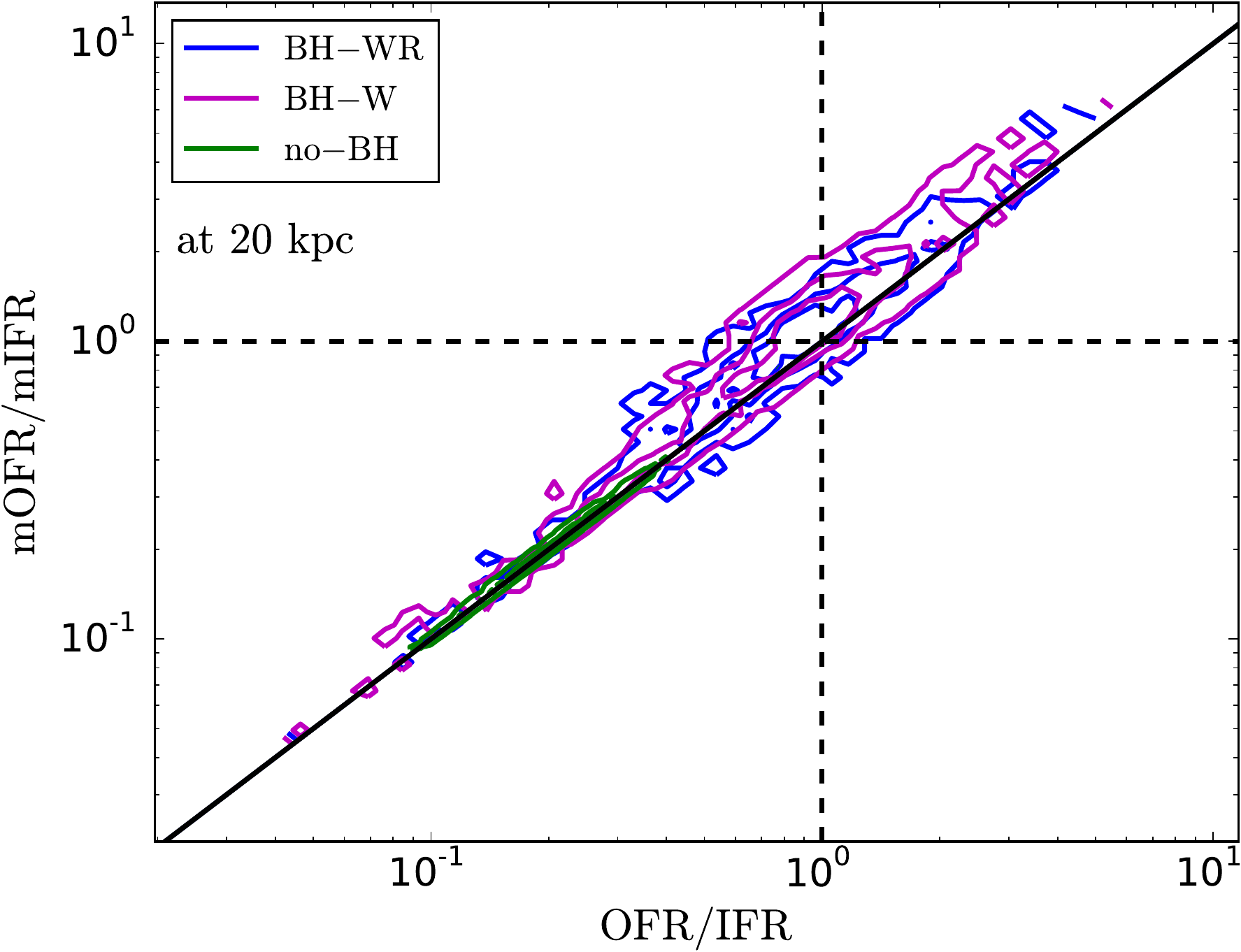}
\includegraphics[width=.47\textwidth]{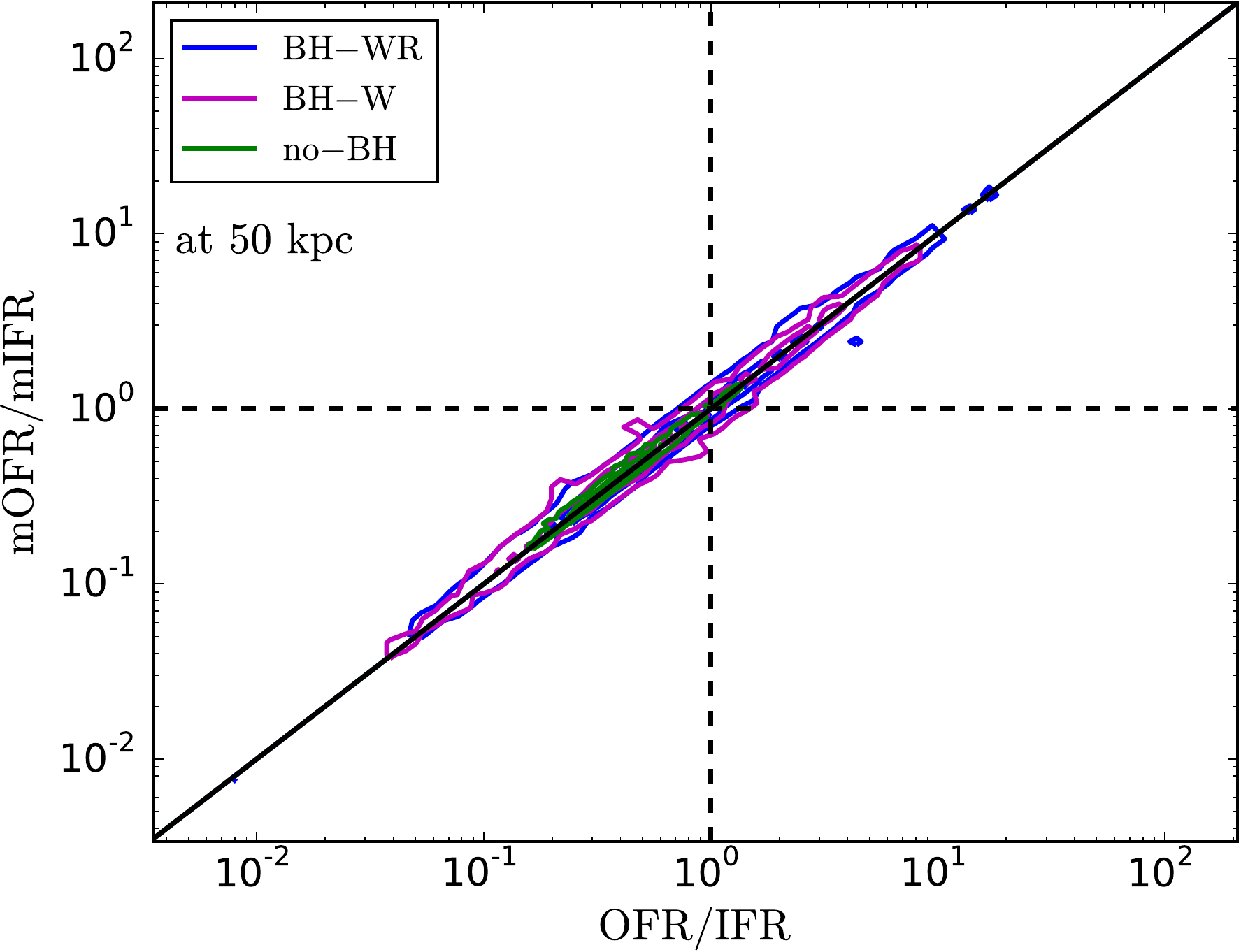}
\caption{Ratio of outflow over inflow rates of gaseous metals (mOFR/mIFR) vs. the same for all gas, metal or not (OFR/IFR), at 20 kpc (top) and 50 kpc (bottom) from the centre; the contours enclose 33\%, 67\%, and 95\% (from the inside out) of the distribution of values for all snapshots of all different models (see legend).
The solid black line marks a one-to-one relation between the two quantities, and the dashed black lines mark ratios of 1 (equal in- and outflow).}
\label{fig:mNFRvNFR}
\end{figure}

In the BH-W and BH-WR models, the total mass flow of the gas is in equilibrium.
The same is not true of the total flow of metals in the galaxy, as the AGN winds drive out gas that is on average more metal-rich than its inflowing counterpart.
To demonstrate this, we plot the distribution of ratios of metal mass outflow (mOFR) to inflow (mIFR) rate, i.e. the total mass of all elements except hydrogen and helium in the corresponding flows, against the distribution of the same ratio for the total gas mass (i.e. OFR/IFR).
This is shown in Fig. \ref{fig:mNFRvNFR} for flow rates through a shell at 20 kpc distance from the centre (top panel), and at 50 kpc distance (bottom panel).
The distribution is visualized with contours enclosing (from the inside out) 33\%, 67\%, and 95\% of the total number of values in all snapshots (i.e. values at different, evenly spaced simulation times).

In the no-BH model, the mOFR/mIFR ratio is about equal to the OFR/IFR at all times, both at 20 kpc and at 50 kpc.
This confirms that there is no outflow of centrally produced metals beyond 20 kpc in this model.
In contrast, in both the BH-W and the BH-WR model, there is a significantly higher ratio of outflow to inflow rate for metals than for the total gas mass at a distance of 20 kpc from the galactic centre, demonstrating that, although the the overall gas flow is balanced, the black-hole feedback still drives out a significant amount of metals into the CGM beyond 20 kpc.
This is different further away from the centre:
At a distance of 50 kpc, the two models that include black-hole feedback have mOFR/mIFR ratios close to the total OFR/IFR at most times.
The AGN-driven outflows have lost all of their outward momentum before reaching this far out from the galactic centre, and are therefore not able to enrich this region with metals.

\section{X-ray properties of the gas}
\label{Xray}
Finally, we take a look at the effects of the black-hole feedback on the X-ray luminosity of the galactic hot gas content.
The X-ray emission of the hot gas, produced by bremsstrahlung and metal-line cooling, is the main observable by which galactic hot haloes are detected.
It is also dependent on the temperature, on the metallicity, and (heavily) on the density of the gas, all of which are influenced by feedback from the central SMBH, potentially making it a good tool to investigate the effect of the feedback.

To estimate the X-ray luminosity of the gas in our simulations, we calculate $L_\mathrm{X, 0}(T, Z=0.4Z_\odot)$ (the normalized X-ray luminosity at 0.4 solar metallicity and temperature $T$), and $(\mathrm{d}L_\mathrm{X} / \mathrm{d}Z)(T)$ (the first-order metallicity dependence of the X-ray luminosity at temperature $T$) for a range of gas temperatures between $0.1$ and $60$ keV with XSPEC\footnote{see \url{http://heasarc.gsfc.nasa.gov/docs/xanadu/xspec/} for more information} \citep{1996ASPC..101...17A}.
Taking the tabulated temperature $T_i$ that is closest to the particles true temperature, we then use these pre-calculated tables to compute the X-ray luminosities of all gas particles $\left(L_{\mathrm{X, }i}\right)$ via:
\begin{align}
L_{\mathrm{X, }i} = & \mathcal{N}(\rho_i, n_{\mathrm{e, }i}) [L_\mathrm{X, 0}(T_i, Z=0.4Z_\odot) + \\
 & (\mathrm{d}L_\mathrm{X} / \mathrm{d}Z)(T_i)\cdot (Z_i - 0.4 Z_\odot)], 
\end{align}
where $\mathcal{N}(\rho_i, n_{\mathrm{e, }i})$ is a normalization factor dependent on the mass density and electron density of the gas particle.
With the luminosity of each particle known, we then compute the total X-ray luminosity of the galactic hot gas by summing over all particles, excluding those whose densities surpass the star-formation limit $\rho_i > 1.94\times 10^{-23} \text{ g }\text{cm}^{-3}$.
This very dense gas is assumed to reside in star-forming regions, where it is very likely to be obscured because of the high dust content.
\begin{figure}
\centering
\includegraphics[width=.47\textwidth]{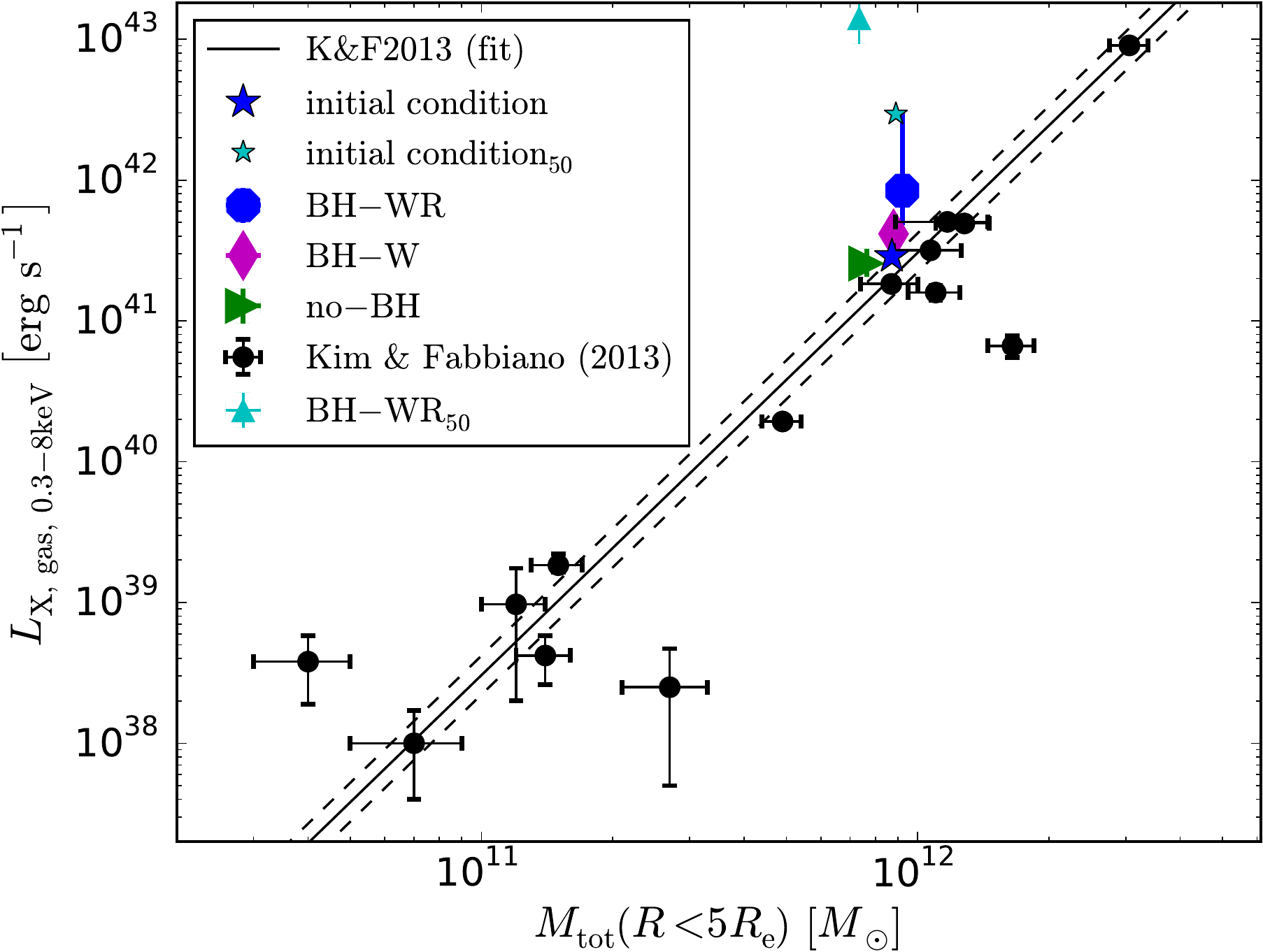}
\includegraphics[width=.47\textwidth]{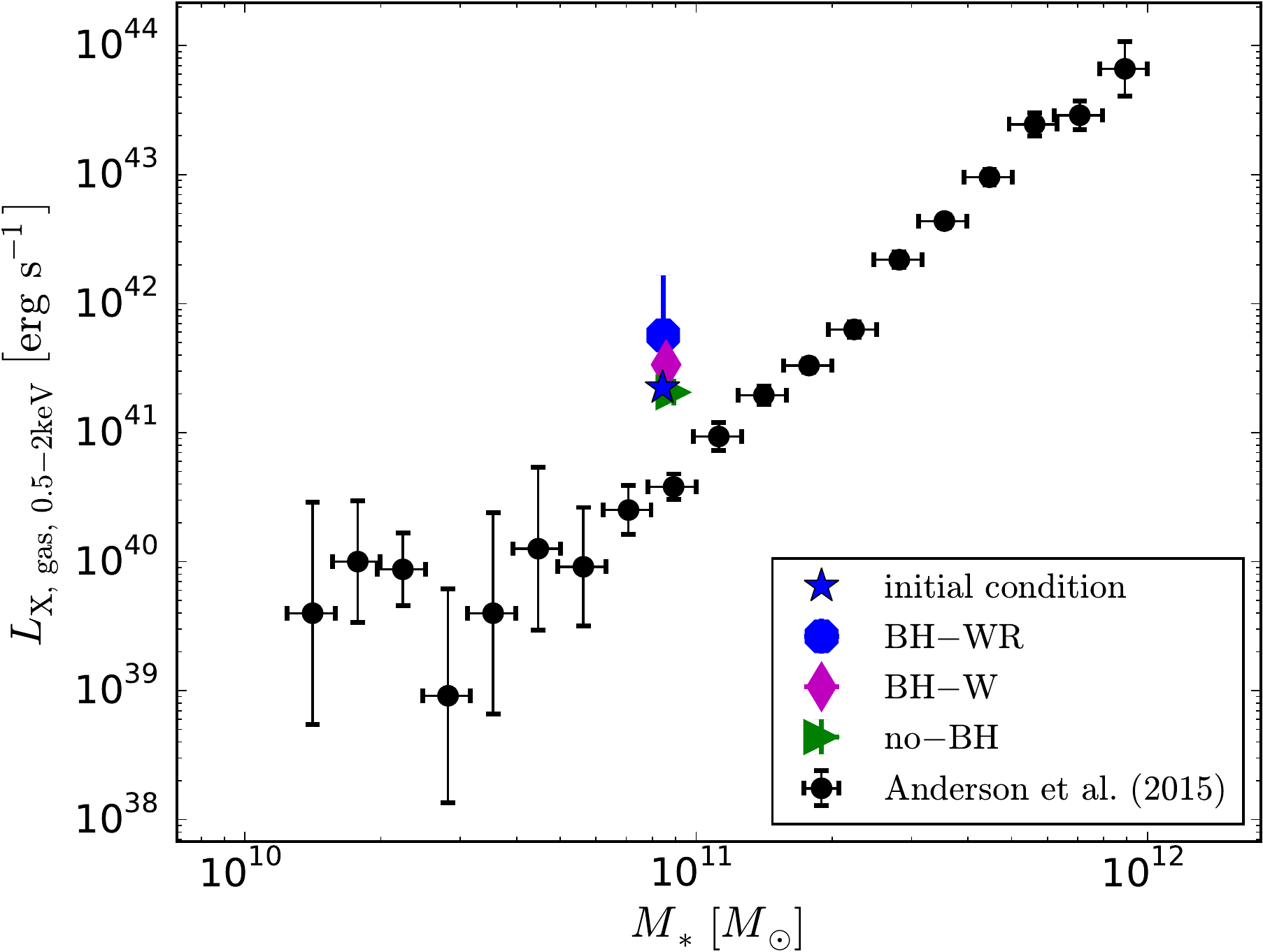}
\caption{Top: The X-ray luminosity of the hot gas in the 0.3-8 keV band over the total mass within five times the effective radius for the initial condition used in this work, the median value over the whole simulation time for all three models, and observations by \citet{2013ApJ...776..116K}.
Additionally, we show the initial and median values of the simulation with 50\% of the cosmological baryon fraction ($\text{BH-WR}_{50}$).
Also shown are the best fit (solid line) and its logarithmic scatter (dashed lines) for observational data.
The error bars on the simulation symbols represent the scatter between the 25th and 75th percentile over the whole simulation time.
See the text for how $L_\mathrm{X}$ was calculated.
Bottom: The X-ray luminosity of the hot gas in the 0.5-2 keV band (different band than in the top panel!) over the total stellar mass, with observations by \citet{2015MNRAS.449.3806A}.
There are only comparatively small differences between the median X-ray luminosities of the three different models with 20\% of $f_\mathrm{b,cosm}$, compared to the observational scatter.
The simulated values fit well to the observations in the $L_\mathrm{X}-M_\mathrm{tot}(R < 5 R_\mathrm{e})$ relation (top), while they are all too high in the $L_\mathrm{X}-M_*$ relation (bottom).
In the $\text{BH-WR}_{50}$ run, the X-ray luminosity is an order of magnitude too high for the observed $L_\mathrm{X}-M_\mathrm{tot}(R < 5 R_\mathrm{e})$ relation already in the initial condition, which increases to about 2 orders of magnitude for the median.}
\label{fig:Xray}
\end{figure}

\begin{figure}
\centering
\includegraphics[width=.47\textwidth]{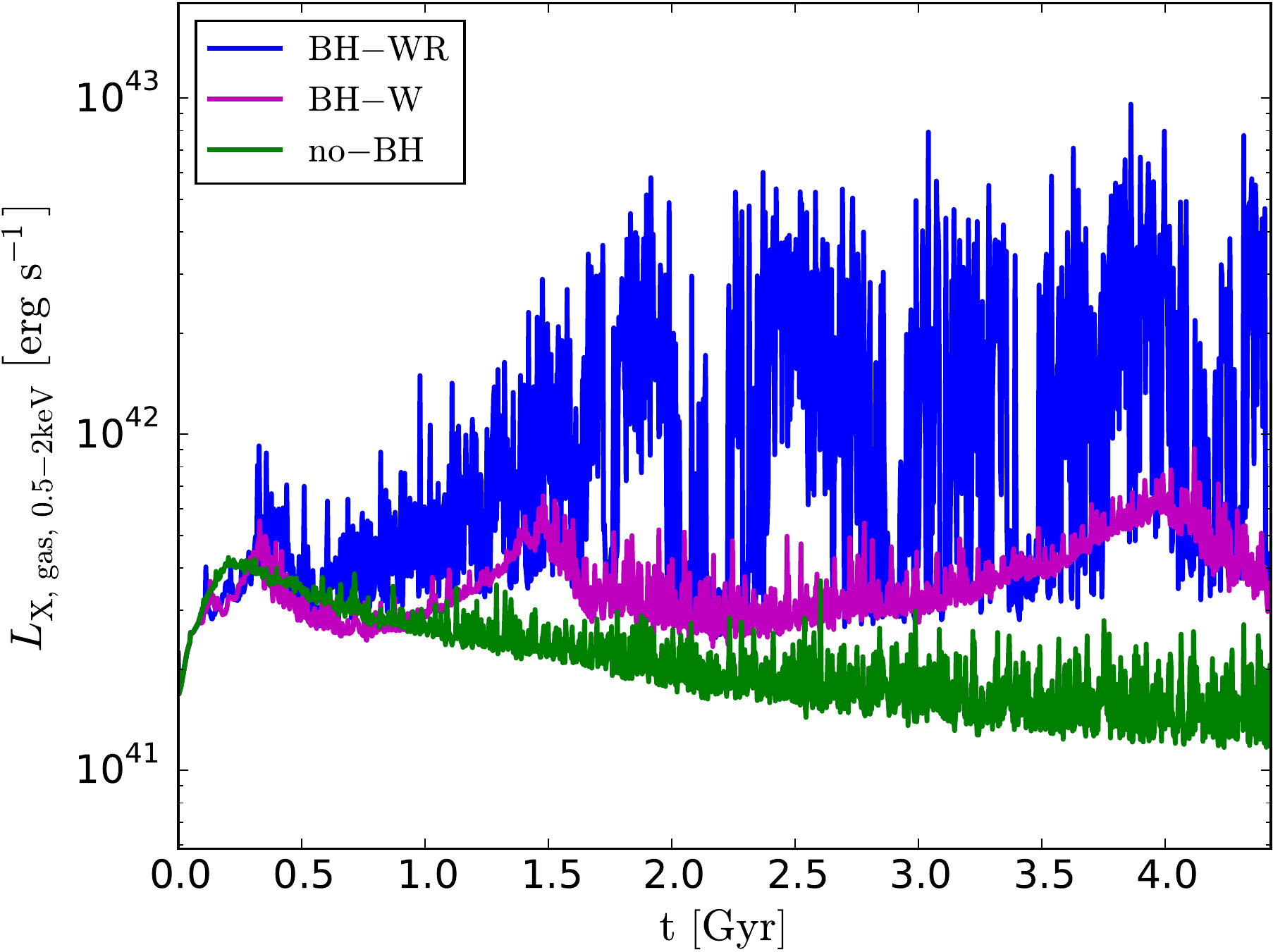}
\caption{X-ray luminosity in the 0.5-2 keV band over time for the different models (see legend).
In the BH-WR model, the heating of the central dense gas by the radiative AGN feedback causes a strong oscillation in the X-ray luminosity by more than order of magnitude, while $L_\mathrm{X}$ changes much less over time in the BH-W and no-BH models.}
\label{fig:lxtime}
\end{figure}

Using this estimate, we then compare the median total X-ray luminosity and its scatter in our three models with observations in two different scaling relations (Fig. \ref{fig:Xray}).
The upper panel shows the $L_\mathrm{X}-M_\mathrm{tot}(R < 5 R_\mathrm{e})$ relation with observations by \citet{2013ApJ...776..116K}, while the lower panel shows the $L_\mathrm{X}-M_*$ relation of \citet{2015MNRAS.449.3806A}.
The black-hole feedback does not influence the median X-ray luminosity of the hot gas much, compared to the already rather large scatter between individual observed galaxies.
The strongest effect can be seen in the BH-WR model: Here, the radiative heating of the central gas both raises the median luminosity by a factor of a few (compared to the initial condition), and also causes strong fluctuations in $L_\mathrm{X}$ over time (visible through the large scatter for the model, and in Fig. \ref{fig:lxtime}), as the heated gas quickly cools down again (compare Fig. \ref{fig:maps10}).
Nevertheless, the median values of all three models are broadly consistent with the observations by \citet{2013ApJ...776..116K} in the $L_\mathrm{X}-M_\mathrm{tot}(R < 5 R_\mathrm{e})$ relation.

On the other hand, they all fall significantly too high in the $L_\mathrm{X}-M_*$ relation, compared to the stacked observations of \citet{2015MNRAS.449.3806A}, as does the initial condition.
The reason for this contradiction might be that the stellar mass we assume in our initial condition is too low compared to the dark-matter (or total) mass.
We scaled our stellar mass to the dark-matter mass using the abundance matching relation of \citet{2013MNRAS.428.3121M}; had we used the relation of \citet{2014arXiv1401.7329K} instead, our stellar mass would be larger by a factor of $\sim 2.5$, enough to explain the difference between observed and simulated values, at least for the initial condition.
In the $L_\mathrm{X}-M_\mathrm{tot}(R < 5 R_\mathrm{e})$ relation, the no-BH model shows a significantly lower $M_\mathrm{tot}(R < 5 R_\mathrm{e})$ than the other models.
This is caused by a shrinking of the effective radius due to the comparatively strong central star-formation in this model.

In Fig. \ref{fig:lxtime}, we show the time evolution of $L_\mathrm{X}$ in the 0.5-2 keV band for all of our models.
The median luminosities only vary by a factor of a few between the different models (which is comparable to the scatter in $L_\mathrm{X}$ between observed galaxies), but in the BH-WR model, the X-ray luminosity has a much larger scatter than in the other two cases.
This is due to the radiative AGN feedback rapidly changing the density and temperature of the gas in the galactic centre, which is the densest part of the ISM and therefore contributes the most to the total $L_\mathrm{X}$.

In our simulations, the median X-ray luminosity of the gas fails as an observational diagnostic for the influence of the black-hole feedback on the gas, as the long-time feedback-effects on $L_\mathrm{X}$ are small compared to the effects of the total amount of gas available in the galaxy, and the total galactic mass that determines its overall density profile.
\citet{2017ApJ...835...15C} find a similar lack of long-term influence of the AGN feedback on the X-ray luminosity in their two-dimensional simulations of ETGs.
AGN feedback could have a larger impact on the total X-ray luminosity by permanently reducing the hot gas density, e.g. by expelling large quantities of gas to beyond the virial radius, into the intergalactic medium.
This is a possible mechanism to quench forming ETGs in the first place, but would need to happen at higher redshifts \citep[compare e.g.][]{2011MNRAS.412.1965M}.
For example, \citet{2015MNRAS.449.4105C}, who investigate the influence of different AGN feedback mechanisms on the formation of ETGs in cosmological simulations, find that the X-ray luminosity of their ETGs is $\sim 2-3$ orders of magnitude lower in simulations with the same mechanical-radiative AGN feedback model that was used in our BH-WR run than in  simulations without AGN feedback.
This is the case because the AGN wind feedback drives out large amounts of gas from the galaxies at high redshifts, strongly reducing the density and thereby the X-ray luminosity of the remaining hot halo gas.
Therefore, the total X-ray luminosity of a galaxy can be an observational constraint on the effect of AGN feedback on the galactic gas at high redshifts.

However, under the conditions investigated in our simulations, i.e. massive, local elliptical galaxies at low redshift, the feedback seems only capable to drive outflows up to $\sim 30$ kpc, beyond which they start cooling down into the centre again.
This, together with central heating, is enough to keep the galaxy quiescent, but it is not enough to permanently expel significant amounts of gas from the galaxy, and lower the overall X-ray luminosity which is mostly determined by the initial condition, i.e. the gas density distribution and gravitational potential of the galaxy about 4.5 Gyr before now.
\begin{figure*}
\centering
\includegraphics[width=\textwidth]{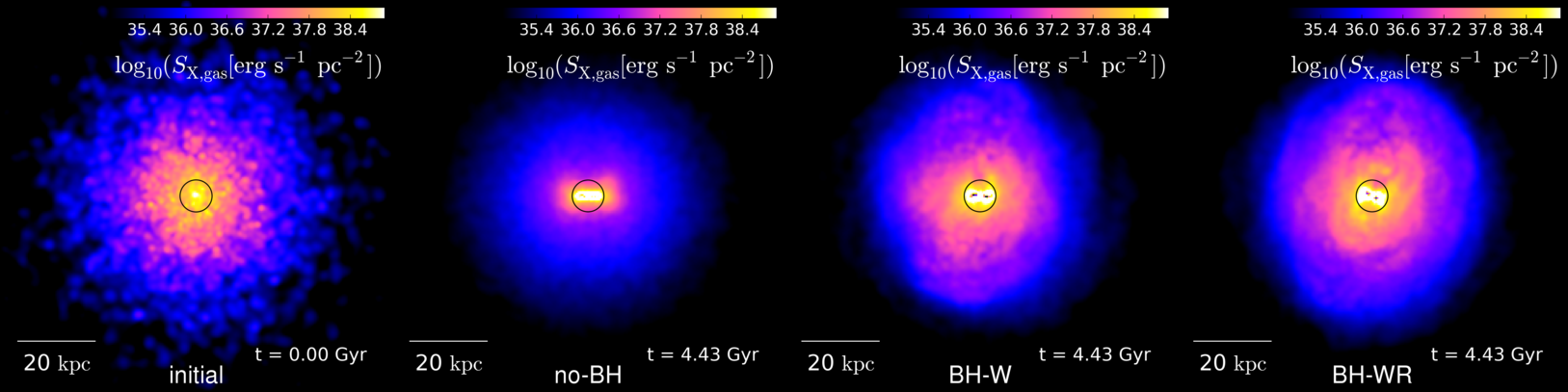}
\caption{Map of the X-ray surface brightness (0.5-2 keV band) of the gas in a 3 kpc thick, 100x100 kpc region around the galactic centre. 
The columns show, from left to right, the initial condition and the states at the end of the simulation time for the no-BH, BH-W and BH-WR models.
The black circle marks 1\% of the virial radius.
While the no-BH model produces a very bright core corresponding to the accretion region around the central star-forming disc, in the BH-W and BH-WR models, the X-ray bright gas is more spread out throughout the ISM and CGM, though the bulk of the luminosity still comes from the central core region.}
\label{fig:lxmap}
\end{figure*}

While the total X-ray luminosity fails to distinguish the models with and without feedback from one another, we can still see the effects of the feedback in the X-ray surface brightness distribution, which we show in Fig. \ref{fig:lxmap} for the initial condition and the final states of our three runs.
In all three runs, the formation of the central dense disc leads to a very bright core region of the X-ray emitting gas.
However, in the models including wind feedback (BH-W and BH-WR), the gas in the surrounding CGM (out to $\sim 30$ kpc) is kept about an order of magnitude brighter, as well as much more structured, than in the no-BH run.
\begin{figure}
\centering
\includegraphics[width=.47\textwidth]{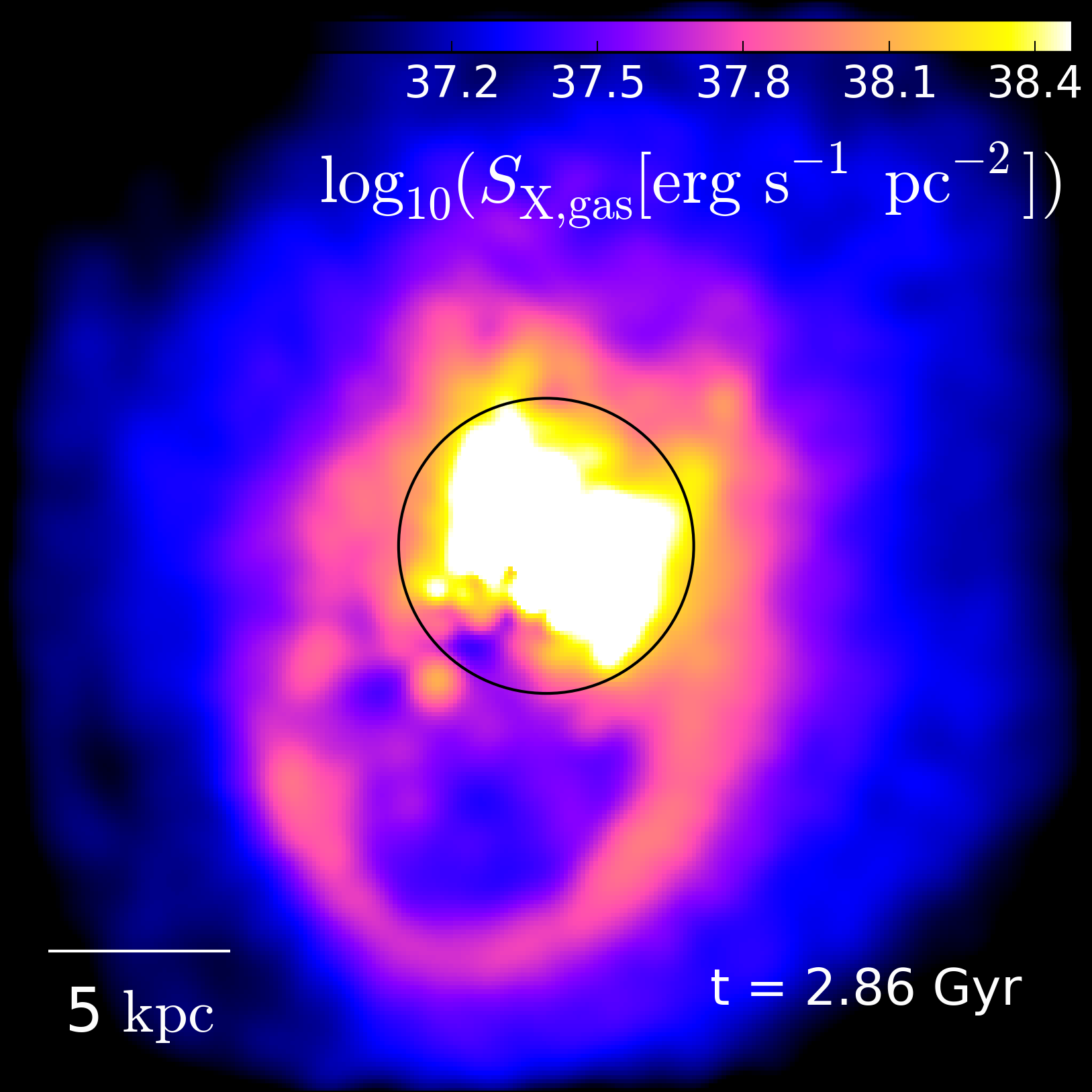}
\caption{Map of the X-ray surface brightness (0.5-2 keV band) of the gas in a 3 kpc thick, 30x30 kpc region around the galactic centre for the BH-WR model.
The black circle marks 1\% of the virial radius.
One can see the formation of an X-ray cavity in an AGN-driven outflow.}
\label{fig:Xcav}
\end{figure}

The hot outflows caused by the wind feedback in both the BH-W and BH-WR models result in several kpc large, outward moving cavities in the X-ray surface brightness.
Such X-ray cavities have been observed and associated with AGN activity in many galaxies, and their sizes are roughly comparable to those of the cavities found in our simulations \citep[e.g.][]{2000A&A...356..788C, 2004ApJ...607..800B, 2007ApJ...665.1057F}.
They are somewhat visible in the relevant panels of Fig. \ref{fig:lxmap}, but we also show one much clearer example of a forming X-ray cavity in the BH-WR model in Fig. \ref{fig:Xcav}.

\section{Limitations of the AGN feedback model}
\label{limits}
The AGN feedback model we use in this paper contains both a radiative and kinetic wind feedback mode, which together are efficient at limiting the SMBH growth, inhibiting star formation, and driving metal-enriched outflows up to about 30 kpc from the centre.
However, the model does not include every mode of AGN feedback that can be observationally motivated and might have an impact on the galactic gas content.
In particular, we neglect the potential effects of collimated radio jets on the evolution of the ISM and CGM.

It is difficult to directly model the highly relativistic synchrotron jets that are observed in many massive galaxies (e.g. M87) in galaxy-scale simulations as the ultra-high velocities in the jet lead to very small time steps.
Nonetheless, several groups have implemented jet-like AGN feedback in the form of non-relativistic (jet velocity $\sim 10000\, \mathrm{km}/\mathrm{s}$), strongly collimated winds mimicking outflows launched from the accretion disc around the SMBH into galaxy evolution simulations \citep[e.g.][]{2004MNRAS.348.1105O, 2010MNRAS.409..985D}, especially in the context of the radio mode in combined quasar/radio mode feedback models \citep[e.g.][]{2012MNRAS.420.2662D, 2013MNRAS.433.3297D}.

The simulations presented in this paper do not include a dedicated jet feedback mechanism, even though the SMBH spends most of its time at low accretion rates ($\sim 90\%$ of the total time at $f_\mathrm{Edd} < 1\%$, compare Fig. \ref{fig:BHgrowth}), i.e. in the radio-mode regime where such a feedback mechanism might be relevant.
The wind feedback included in our simulations is similar to the non-relativistic jet feedback models described and used in the papers cited above, so its effect would likely be qualitatively the same, i.e. driving out gas perpendicular to the central gas disc plane.

However, a jet feedback mode would be even more strongly collimated (with all gas in the jet being accelerated along the same axis) than our wind feedback (which would only be perfectly collimated if all gas flowing onto the SMBH had a uniform direction of angular momentum).
This might enable it to push the central gas farther away from the centre than the wind feedback, enriching the CGM beyond $\sim 30$ kpc with metals.
On the other hand, an additional jet feedback mode is unlikely to change the results of our simulations as far as the central region of the galaxy is concerned: The wind feedback is already able to evacuate the very centre, inhibiting the star formation there, and a jet would be just as incapable of affecting the surrounding dense gas disc (which is kept mostly quiescent by the radiative feedback) as the wind feedback due to its collimation.

\section{Summary \& Conclusions}
\label{summ}
We investigated the influence of different forms of AGN feedback on the evolution of isolated early-type galaxies at low redshift.
For this purpose, we used the SPH code SPHGal, an improved version of \textsc{gadget}3, to run simulations of an idealized, isolated, massive, local elliptical galaxy with three different models: one without any accretion onto or feedback from the central supermassive black hole (model no-BH), one where the black-hole feedback is implemented as a kinetic wind (model BH-W), and finally one where radiative X-ray heating is added to the wind feedback from the black hole (model BH-WR).
We compared the outcomes of these three simulations between each other, as well as to observations, with respect to their star-formation and black-hole growth history, the central density and temperature evolution of the ISM, the large-scale flow patterns and metal distribution in the gas, and the X-ray properties of the hot galactic gas.

Our results can be summarized as follows:
\begin{itemize}
\item Without black-hole feedback, a classical cooling-flow problem occurs in our simulation: Gas constantly flows into the central region and cools onto a cold, dense disc with a very dense core in the centre, resulting in permanent significant star formation.
Once gas had time to cool down from the initial condition, the galaxy is not quiescent anymore for the rest of the simulation time.

Including black-hole feedback reduces the star formation rate significantly:
Pure wind feedback (the BH-W model) only reduces the SFR moderately, resulting in a galaxy roughly in the ``green valley", changing between slightly quiescent and slightly active states.
This reduction of the SFR is achieved by bipolar outflow from the central cold, dense disc, that is caused by the wind feedback and evacuates the very centre of the galaxy, which would otherwise (as in the no-BH model) be the region with the highest SFR.

Adding radiative heating to the feedback as well (the BH-WR model) results in a very quiescent, ``red and dead", galaxy that shows only short bursts of significant star formation between longer periods of very little star formation.
Here, the dense, potentially star-forming gas around the wind-feedback-evacuated galactic centre is periodically heated and expanded by the radiative feedback, interrupting ongoing star formation, before it cools down again.
This feedback-driven alternation of heating \& expansion and cooling \& compression leads to a strongly oscillating, overall much lower SFR in the galaxy.
\item The growth of the black hole is quite limited in both feedback models, its mass increasing by only about 5\% and 7.5\% over the whole simulation time of $\sim 4.35$ Gyr in the BH-W and BH-WR models, respectively.
The duty cycle of the black hole is tendentially too high in both models.
\item Without black-hole feedback, the outer parts of the ISM and CGM are deprived of metals over time, as all metal-rich gas slowly flows into the galactic centre, leading to very low metallicity in the outskirts and very high, supersolar metallicity in the central region.
In contrast, black-hole-wind feedback leads to large-scale outflows, flattening the metallicity profile of the galactic gas out to about 30 kpc, while reducing it in the centre.
Additional radiative feedback does not affect this outcome.

The limited range ($\sim 30$ kpc distance from the centre) in which the AGN feedback has a significant effect on the CGM in our simulations is strongly influenced by the chosen initial condition, specifically the deep potential well of this massive, low-redshift ETG.
In cosmological simulations \citep[e.g.][]{2015MNRAS.449.4105C}, where the conditions are different (especially at higher redshifts), the AGN feedback can drive winds to much higher distances from the galaxy, or even eject gas from the galactic potential well completely.
Furthermore, a collimated jet mode of the AGN feedback, which is not included in our model, might also be able to move gas farther away from the centre.
\item The median total X-ray luminosity of the gas is only slightly affected by the different black-hole feedback models, compared to the observational scatter.
The radiative feedback has the largest effect, leading to strong oscillation of the X-ray luminosity as the central dense gas is constantly heated and cooling.
Overall though, all models fit reasonably well to the observed $L_\mathrm{X}-M_\mathrm{tot}(R < 5 R_\mathrm{e})$ relation, making the total gaseous X-ray luminosity a bad diagnostic to investigate black-hole feedback in low-redshift, massive ETGs.
At higher redshifts, where the AGN feedback can more efficiently reduce the overall hot gas density of the galaxy, its impact on the total X-ray luminosity can be much more significant \citep[compare the cosmological simulations of][]{2015MNRAS.449.4105C}, making $L_\mathrm{X}$ a more useful diagnostic at these earlier times.
Both the BH-W, and the BH-WR feedback models result in the formation of X-ray cavities in the feedback-induced hot outflows from the galactic centre, confirming the observational link between these cavities and AGN activity.
\end{itemize}

We conclude that feedback from the central supermassive black hole is necessary to keep massive, isolated elliptical galaxies quiescent during the late stages of their evolution, as the feedback from their old stellar population (the only plausible alternative mechanism for maintaining the quiescence in isolation) is too weak to have any significant effect on the thermal state of the ISM at these galactic mass ranges.
How the AGN feedback is implemented has a major influence on its effect on the galactic gas:
While radiative feedback is very efficient in maintaining a low star formation rate through heating the central ISM, it is incapable of driving significant outflows by itself and does not affect the growth of the SMBH much, while, on the other hand, the kinetic momentum feedback resulting from broad-line region winds drives large-scale outflows that can enrich the nearby CGM of the galaxy, but is less effective at keeping the galaxy quiescent.
Both forms of feedback need to act together to maintain quiescence, prevent excessive SMBH growth, and distribute metals through the galactic gas via an AGN-driven galactic fountain, and thereby produce galactic properties in accordance with the observational constraints.

It is worth noting that, while the AGN feedback in our simulations is perfectly capable of keeping the ETG quiescent when we assume 20\% of the cosmological baryon fraction for the initial mass of the hot halo, if we instead take an initial condition with 50\% of the cosmological baryon fraction, corresponding to an about three times more massive gaseous halo, the central gas density becomes too high for the AGN feedback to have much of an effect, resulting in a constantly star-forming galaxy even when the full feedback model is enabled.
This implies (for our assumed gas density profile) that local quiescent ETGs must have lost much of their gas at higher redshifts, likely during the time when they were quenched in the first place, possibly because it was ejected by still more effective (due to shallower potential wells and much higher quasar activity) AGN-driven winds.
Alternatively, the hot gas could have been not fully ejected, but instead distributed much more evenly throughout the CGM of the galaxy, leading to a much shallower, and hence less X-ray bright and less efficiently cooling gas density profile.

While our isolated initial conditions allow us to study the effects of AGN feedback in a controlled way, free from degeneracies with effects from intergalactic interactions, this of course also means that these interactions---and more generally, the effects of the galactic environment on the ETG's evolution and the AGN feedback's efficiency---are not included.
These isolated simulations also do not provide any statistics: we look at only one possible ETG out of a range of possible ones that are consistent with observational constraints on their masses, etc.
To address these issues, cosmological (zoom-in) simulations are necessary.
For example, \citet{2015MNRAS.449.4105C} did such a study for massive ETGs, and their results in regard to the necessity of AGN feedback are consistent with ours.

Another weakness of our approach in this work is the limited resolution:
While it is good enough to investigate the effects of the AGN feedback on the galactic gas on large, galaxy-wide scales, it limits our ability to resolve the central dense gas.
This makes it impossible to attempt more detailed comparisons of AGN-driven outflows of different phases in the simulation with observed ones, as, for example, warm-ionized outflows of $\sim 10^5 M_\odot$ \citep[e.g. those observed by][]{2016Natur.533..504C} can simply not be well resolved with gas particles of $10^5 M_\odot$.
To investigate the effect of AGN feedback on the multi-phase ISM in the centre of ETGs in detail, it will therefore be necessary (at least with current computing technology) to focus simulations on smaller scales than a whole galaxy.

\section*{Acknowledgements}
We thank Eugene Churazov for assistance with the calculation of the X-ray luminosities as well as for helpful comments and suggestions, and Veronica Biffi for a lot of useful advice regarding the X-ray calculations.
We are grateful to Ben Moster for providing the program used to generate the initial conditions.
We also thank Andreas Schmidt for useful conversations, especially about technical details of the code.
Additional thanks goes to the anonymous referee for constructive suggestions to improve the clarity of the paper.
ME, TN and EE acknowledge support from the DFG priority program 1573 ``The physics of the interstellar medium".
TN acknowledges support from the DFG cluster of excellence ``Origin and Structure of the Universe". 

\bibliographystyle{mn2e}
\bibliography{./references}

\begin{thebibliography}{}
\makeatletter
\relax
\def\mn@urlcharsother{\let\do\@makeother \do\$\do\&\do\#\do\^\do\_\do\%\do\~}
\def\mn@doi{\begingroup\mn@urlcharsother \@ifnextchar [ {\mn@doi@}
  {\mn@doi@[]}}
\def\mn@doi@[#1]#2{\def\@tempa{#1}\ifx\@tempa\@empty \href
  {http://dx.doi.org/#2} {doi:#2}\else \href {http://dx.doi.org/#2} {#1}\fi
  \endgroup}
\def\mn@eprint#1#2{\mn@eprint@#1:#2::\@nil}
\def\mn@eprint@arXiv#1{\href {http://arxiv.org/abs/#1} {{\tt arXiv:#1}}}
\def\mn@eprint@dblp#1{\href {http://dblp.uni-trier.de/rec/bibtex/#1.xml}
  {dblp:#1}}
\def\mn@eprint@#1:#2:#3:#4\@nil{\def\@tempa {#1}\def\@tempb {#2}\def\@tempc
  {#3}\ifx \@tempc \@empty \let \@tempc \@tempb \let \@tempb \@tempa \fi \ifx
  \@tempb \@empty \def\@tempb {arXiv}\fi \@ifundefined
  {mn@eprint@\@tempb}{\@tempb:\@tempc}{\expandafter \expandafter \csname
  mn@eprint@\@tempb\endcsname \expandafter{\@tempc}}}

\bibitem[\protect\citeauthoryear{{Alatalo} et~al.,}{{Alatalo}
  et~al.}{2011}]{2011ApJ...735...88A}
{Alatalo} K.,  et~al., 2011, \mn@doi [\apj] {10.1088/0004-637X/735/2/88}, \href
  {http://adsabs.harvard.edu/abs/2011ApJ...735...88A} {735, 88}

\bibitem[\protect\citeauthoryear{{Alatalo} et~al.,}{{Alatalo}
  et~al.}{2015}]{2015ApJ...798...31A}
{Alatalo} K.,  et~al., 2015, \mn@doi [\apj] {10.1088/0004-637X/798/1/31}, \href
  {http://adsabs.harvard.edu/abs/2015ApJ...798...31A} {798, 31}

\bibitem[\protect\citeauthoryear{{Anderson}, {Gaspari}, {White}, {Wang}  \&
  {Dai}}{{Anderson} et~al.}{2015}]{2015MNRAS.449.3806A}
{Anderson} M.~E.,  {Gaspari} M.,  {White} S.~D.~M.,  {Wang} W.,   {Dai} X.,
  2015, \mn@doi [\mnras] {10.1093/mnras/stv437}, \href
  {http://adsabs.harvard.edu/abs/2015MNRAS.449.3806A} {449, 3806}

\bibitem[\protect\citeauthoryear{{Arnaud}}{{Arnaud}}{1996}]{1996ASPC..101...17A}
{Arnaud} K.~A.,  1996, in {Jacoby} G.~H.,  {Barnes} J.,  eds,  Astronomical
  Society of the Pacific Conference Series Vol. 101, Astronomical Data Analysis
  Software and Systems V. p.~17

\bibitem[\protect\citeauthoryear{{Aumer}, {White}, {Naab}  \&
  {Scannapieco}}{{Aumer} et~al.}{2013}]{2013MNRAS.434.3142A}
{Aumer} M.,  {White} S.~D.~M.,  {Naab} T.,   {Scannapieco} C.,  2013, \mn@doi
  [\mnras] {10.1093/mnras/stt1230}, \href
  {http://adsabs.harvard.edu/abs/2013MNRAS.434.3142A} {434, 3142}

\bibitem[\protect\citeauthoryear{{Barnab{\`e}}, {Czoske}, {Koopmans}, {Treu}
  \& {Bolton}}{{Barnab{\`e}} et~al.}{2011}]{2011MNRAS.415.2215B}
{Barnab{\`e}} M.,  {Czoske} O.,  {Koopmans} L.~V.~E.,  {Treu} T.,   {Bolton}
  A.~S.,  2011, \mn@doi [\mnras] {10.1111/j.1365-2966.2011.18842.x}, \href
  {http://adsabs.harvard.edu/abs/2011MNRAS.415.2215B} {415, 2215}

\bibitem[\protect\citeauthoryear{{Bieri}, {Dubois}, {Rosdahl}, {Wagner}, {Silk}
   \& {Mamon}}{{Bieri} et~al.}{2017}]{2017MNRAS.464.1854B}
{Bieri} R.,  {Dubois} Y.,  {Rosdahl} J.,  {Wagner} A.,  {Silk} J.,   {Mamon}
  G.~A.,  2017, \mn@doi [\mnras] {10.1093/mnras/stw2380}, \href
  {http://esoads.eso.org/abs/2017MNRAS.464.1854B} {464, 1854}

\bibitem[\protect\citeauthoryear{{Binney}}{{Binney}}{1977}]{1977ApJ...215..483B}
{Binney} J.,  1977, \mn@doi [\apj] {10.1086/155378}, \href
  {http://adsabs.harvard.edu/abs/1977ApJ...215..483B} {215, 483}

\bibitem[\protect\citeauthoryear{{B{\^i}rzan}, {Rafferty}, {McNamara}, {Wise}
  \& {Nulsen}}{{B{\^i}rzan} et~al.}{2004}]{2004ApJ...607..800B}
{B{\^i}rzan} L.,  {Rafferty} D.~A.,  {McNamara} B.~R.,  {Wise} M.~W.,
  {Nulsen} P.~E.~J.,  2004, \mn@doi [\apj] {10.1086/383519}, \href
  {http://adsabs.harvard.edu/abs/2004ApJ...607..800B} {607, 800}

\bibitem[\protect\citeauthoryear{{Bondi}}{{Bondi}}{1952}]{1952MNRAS.112..195B}
{Bondi} H.,  1952, \mnras, \href
  {http://adsabs.harvard.edu/abs/1952MNRAS.112..195B} {112, 195}

\bibitem[\protect\citeauthoryear{{Bondi} \& {Hoyle}}{{Bondi} \&
  {Hoyle}}{1944}]{1944MNRAS.104..273B}
{Bondi} H.,  {Hoyle} F.,  1944, \mnras, \href
  {http://adsabs.harvard.edu/abs/1944MNRAS.104..273B} {104, 273}

\bibitem[\protect\citeauthoryear{{Booth} \& {Schaye}}{{Booth} \&
  {Schaye}}{2009}]{2009MNRAS.398...53B}
{Booth} C.~M.,  {Schaye} J.,  2009, \mn@doi [\mnras]
  {10.1111/j.1365-2966.2009.15043.x}, \href
  {http://adsabs.harvard.edu/abs/2009MNRAS.398...53B} {398, 53}

\bibitem[\protect\citeauthoryear{{Boroson}, {Kim}  \& {Fabbiano}}{{Boroson}
  et~al.}{2011}]{2011ApJ...729...12B}
{Boroson} B.,  {Kim} D.-W.,   {Fabbiano} G.,  2011, \mn@doi [\apj]
  {10.1088/0004-637X/729/1/12}, \href
  {http://adsabs.harvard.edu/abs/2011ApJ...729...12B} {729, 12}

\bibitem[\protect\citeauthoryear{{Bower}, {Benson}, {Malbon}, {Helly}, {Frenk},
  {Baugh}, {Cole}  \& {Lacey}}{{Bower} et~al.}{2006}]{2006MNRAS.370..645B}
{Bower} R.~G.,  {Benson} A.~J.,  {Malbon} R.,  {Helly} J.~C.,  {Frenk} C.~S.,
  {Baugh} C.~M.,  {Cole} S.,   {Lacey} C.~G.,  2006, \mn@doi [\mnras]
  {10.1111/j.1365-2966.2006.10519.x}, \href
  {http://adsabs.harvard.edu/abs/2006MNRAS.370..645B} {370, 645}

\bibitem[\protect\citeauthoryear{{Brighenti} \& {Mathews}}{{Brighenti} \&
  {Mathews}}{2006}]{2006ApJ...643..120B}
{Brighenti} F.,  {Mathews} W.~G.,  2006, \mn@doi [\apj] {10.1086/502645}, \href
  {http://adsabs.harvard.edu/abs/2006ApJ...643..120B} {643, 120}

\bibitem[\protect\citeauthoryear{{Brusa} et~al.,}{{Brusa}
  et~al.}{2015}]{2015A&A...578A..11B}
{Brusa} M.,  et~al., 2015, \mn@doi [\aap] {10.1051/0004-6361/201425491}, \href
  {http://adsabs.harvard.edu/abs/2015A%26A...578A..11B} {578, A11}

\bibitem[\protect\citeauthoryear{{Cano-D{\'{\i}}az}, {Maiolino}, {Marconi},
  {Netzer}, {Shemmer}  \& {Cresci}}{{Cano-D{\'{\i}}az}
  et~al.}{2012}]{2012A&A...537L...8C}
{Cano-D{\'{\i}}az} M.,  {Maiolino} R.,  {Marconi} A.,  {Netzer} H.,  {Shemmer}
  O.,   {Cresci} G.,  2012, \mn@doi [\aap] {10.1051/0004-6361/201118358}, \href
  {http://adsabs.harvard.edu/abs/2012A%26A...537L...8C} {537, L8}

\bibitem[\protect\citeauthoryear{{Carniani} et~al.,}{{Carniani}
  et~al.}{2016}]{2016A&A...591A..28C}
{Carniani} S.,  et~al., 2016, \mn@doi [\aap] {10.1051/0004-6361/201528037},
  \href {http://adsabs.harvard.edu/abs/2016A%26A...591A..28C} {591, A28}

\bibitem[\protect\citeauthoryear{{Cavaliere} \& {Fusco-Femiano}}{{Cavaliere} \&
  {Fusco-Femiano}}{1976}]{1976A&A....49..137C}
{Cavaliere} A.,  {Fusco-Femiano} R.,  1976, \aap, \href
  {http://adsabs.harvard.edu/abs/1976A%26A....49..137C} {49, 137}

\bibitem[\protect\citeauthoryear{{Chang}, {van der Wel}, {da Cunha}  \&
  {Rix}}{{Chang} et~al.}{2015}]{2015ApJS..219....8C}
{Chang} Y.-Y.,  {van der Wel} A.,  {da Cunha} E.,   {Rix} H.-W.,  2015, \mn@doi
  [\apjs] {10.1088/0067-0049/219/1/8}, \href
  {http://adsabs.harvard.edu/abs/2015ApJS..219....8C} {219, 8}

\bibitem[\protect\citeauthoryear{{Cheung} et~al.,}{{Cheung}
  et~al.}{2016}]{2016Natur.533..504C}
{Cheung} E.,  et~al., 2016, \mn@doi [\nat] {10.1038/nature18006}, \href
  {http://adsabs.harvard.edu/abs/2016Natur.533..504C} {533, 504}

\bibitem[\protect\citeauthoryear{{Choi}, {Ostriker}, {Naab}  \&
  {Johansson}}{{Choi} et~al.}{2012}]{2012ApJ...754..125C}
{Choi} E.,  {Ostriker} J.~P.,  {Naab} T.,   {Johansson} P.~H.,  2012, \mn@doi
  [\apj] {10.1088/0004-637X/754/2/125}, \href
  {http://adsabs.harvard.edu/abs/2012ApJ...754..125C} {754, 125}

\bibitem[\protect\citeauthoryear{{Choi}, {Naab}, {Ostriker}, {Johansson}  \&
  {Moster}}{{Choi} et~al.}{2014}]{2014MNRAS.442..440C}
{Choi} E.,  {Naab} T.,  {Ostriker} J.~P.,  {Johansson} P.~H.,   {Moster} B.~P.,
   2014, \mn@doi [\mnras] {10.1093/mnras/stu874}, \href
  {http://adsabs.harvard.edu/abs/2014MNRAS.442..440C} {442, 440}

\bibitem[\protect\citeauthoryear{{Choi}, {Ostriker}, {Naab}, {Oser}  \&
  {Moster}}{{Choi} et~al.}{2015}]{2015MNRAS.449.4105C}
{Choi} E.,  {Ostriker} J.~P.,  {Naab} T.,  {Oser} L.,   {Moster} B.~P.,  2015,
  \mn@doi [\mnras] {10.1093/mnras/stv575}, \href
  {http://adsabs.harvard.edu/abs/2015MNRAS.449.4105C} {449, 4105}

\bibitem[\protect\citeauthoryear{{Choi}, {Ostriker}, {Naab}, {Somerville},
  {Hirschmann}, {N{\'u}{\~n}ez}, {Hu}  \& {Oser}}{{Choi}
  et~al.}{2016}]{2016arXiv161009389C}
{Choi} E.,  {Ostriker} J.~P.,  {Naab} T.,  {Somerville} R.~S.,  {Hirschmann}
  M.,  {N{\'u}{\~n}ez} A.,  {Hu} C.-Y.,   {Oser} L.,  2016, preprint, \href
  {http://adsabs.harvard.edu/abs/2016arXiv161009389C} {} (\mn@eprint {arXiv}
  {1610.09389})

\bibitem[\protect\citeauthoryear{{Churazov}, {Forman}, {Jones}  \&
  {B{\"o}hringer}}{{Churazov} et~al.}{2000}]{2000A&A...356..788C}
{Churazov} E.,  {Forman} W.,  {Jones} C.,   {B{\"o}hringer} H.,  2000, \aap,
  \href {http://adsabs.harvard.edu/abs/2000A%26A...356..788C} {356, 788}

\bibitem[\protect\citeauthoryear{{Churazov}, {Br{\"u}ggen}, {Kaiser},
  {B{\"o}hringer}  \& {Forman}}{{Churazov} et~al.}{2001}]{2001ApJ...554..261C}
{Churazov} E.,  {Br{\"u}ggen} M.,  {Kaiser} C.~R.,  {B{\"o}hringer} H.,
  {Forman} W.,  2001, \mn@doi [\apj] {10.1086/321357}, \href
  {http://adsabs.harvard.edu/abs/2001ApJ...554..261C} {554, 261}

\bibitem[\protect\citeauthoryear{{Churazov}, {Sunyaev}, {Forman}  \&
  {B{\"o}hringer}}{{Churazov} et~al.}{2002}]{2002MNRAS.332..729C}
{Churazov} E.,  {Sunyaev} R.,  {Forman} W.,   {B{\"o}hringer} H.,  2002,
  \mn@doi [\mnras] {10.1046/j.1365-8711.2002.05332.x}, \href
  {http://adsabs.harvard.edu/abs/2002MNRAS.332..729C} {332, 729}

\bibitem[\protect\citeauthoryear{{Churazov}, {Sazonov}, {Sunyaev}, {Forman},
  {Jones}  \& {B{\"o}hringer}}{{Churazov} et~al.}{2005}]{2005MNRAS.363L..91C}
{Churazov} E.,  {Sazonov} S.,  {Sunyaev} R.,  {Forman} W.,  {Jones} C.,
  {B{\"o}hringer} H.,  2005, \mn@doi [\mnras]
  {10.1111/j.1745-3933.2005.00093.x}, \href
  {http://adsabs.harvard.edu/abs/2005MNRAS.363L..91C} {363, L91}

\bibitem[\protect\citeauthoryear{{Cicone} et~al.,}{{Cicone}
  et~al.}{2014}]{2014A&A...562A..21C}
{Cicone} C.,  et~al., 2014, \mn@doi [\aap] {10.1051/0004-6361/201322464}, \href
  {http://adsabs.harvard.edu/abs/2014A%26A...562A..21C} {562, A21}

\bibitem[\protect\citeauthoryear{{Ciotti} \& {Ostriker}}{{Ciotti} \&
  {Ostriker}}{1997}]{1997ApJ...487L.105C}
{Ciotti} L.,  {Ostriker} J.~P.,  1997, \mn@doi [\apjl] {10.1086/310902}, \href
  {http://adsabs.harvard.edu/abs/1997ApJ...487L.105C} {487, L105}

\bibitem[\protect\citeauthoryear{{Ciotti} \& {Ostriker}}{{Ciotti} \&
  {Ostriker}}{2001}]{2001ApJ...551..131C}
{Ciotti} L.,  {Ostriker} J.~P.,  2001, \mn@doi [\apj] {10.1086/320053}, \href
  {http://adsabs.harvard.edu/abs/2001ApJ...551..131C} {551, 131}

\bibitem[\protect\citeauthoryear{{Ciotti} \& {Ostriker}}{{Ciotti} \&
  {Ostriker}}{2007}]{2007ApJ...665.1038C}
{Ciotti} L.,  {Ostriker} J.~P.,  2007, \mn@doi [\apj] {10.1086/519833}, \href
  {http://adsabs.harvard.edu/abs/2007ApJ...665.1038C} {665, 1038}

\bibitem[\protect\citeauthoryear{{Ciotti}, {Ostriker}  \& {Proga}}{{Ciotti}
  et~al.}{2009}]{2009ApJ...699...89C}
{Ciotti} L.,  {Ostriker} J.~P.,   {Proga} D.,  2009, \mn@doi [\apj]
  {10.1088/0004-637X/699/1/89}, \href
  {http://adsabs.harvard.edu/abs/2009ApJ...699...89C} {699, 89}

\bibitem[\protect\citeauthoryear{{Ciotti}, {Ostriker}  \& {Proga}}{{Ciotti}
  et~al.}{2010}]{2010ApJ...717..708C}
{Ciotti} L.,  {Ostriker} J.~P.,   {Proga} D.,  2010, \mn@doi [\apj]
  {10.1088/0004-637X/717/2/708}, \href
  {http://adsabs.harvard.edu/abs/2010ApJ...717..708C} {717, 708}

\bibitem[\protect\citeauthoryear{{Ciotti}, {Pellegrini}, {Negri}  \&
  {Ostriker}}{{Ciotti} et~al.}{2017}]{2017ApJ...835...15C}
{Ciotti} L.,  {Pellegrini} S.,  {Negri} A.,   {Ostriker} J.~P.,  2017, \mn@doi
  [\apj] {10.3847/1538-4357/835/1/15}, \href
  {http://adsabs.harvard.edu/abs/2017ApJ...835...15C} {835, 15}

\bibitem[\protect\citeauthoryear{{Crenshaw}, {Kraemer}  \& {George}}{{Crenshaw}
  et~al.}{2003}]{2003ARA&A..41..117C}
{Crenshaw} D.~M.,  {Kraemer} S.~B.,   {George} I.~M.,  2003, \mn@doi [\araa]
  {10.1146/annurev.astro.41.082801.100328}, \href
  {http://adsabs.harvard.edu/abs/2003ARA%26A..41..117C} {41, 117}

\bibitem[\protect\citeauthoryear{{Cresci} et~al.,}{{Cresci}
  et~al.}{2015a}]{2015A&A...582A..63C}
{Cresci} G.,  et~al., 2015a, \mn@doi [\aap] {10.1051/0004-6361/201526581},
  \href {http://adsabs.harvard.edu/abs/2015A%26A...582A..63C} {582, A63}

\bibitem[\protect\citeauthoryear{{Cresci} et~al.,}{{Cresci}
  et~al.}{2015b}]{2015ApJ...799...82C}
{Cresci} G.,  et~al., 2015b, \mn@doi [\apj] {10.1088/0004-637X/799/1/82}, \href
  {http://adsabs.harvard.edu/abs/2015ApJ...799...82C} {799, 82}

\bibitem[\protect\citeauthoryear{{Croton} et~al.,}{{Croton}
  et~al.}{2006}]{2006MNRAS.365...11C}
{Croton} D.~J.,  et~al., 2006, \mn@doi [\mnras]
  {10.1111/j.1365-2966.2005.09675.x}, \href
  {http://adsabs.harvard.edu/abs/2006MNRAS.365...11C} {365, 11}

\bibitem[\protect\citeauthoryear{{Dai}, {Bregman}, {Kochanek}  \&
  {Rasia}}{{Dai} et~al.}{2010}]{2010ApJ...719..119D}
{Dai} X.,  {Bregman} J.~N.,  {Kochanek} C.~S.,   {Rasia} E.,  2010, \mn@doi
  [\apj] {10.1088/0004-637X/719/1/119}, \href
  {http://adsabs.harvard.edu/abs/2010ApJ...719..119D} {719, 119}

\bibitem[\protect\citeauthoryear{{Danielson}, {Lehmer}, {Alexander}, {Brandt},
  {Luo}, {Miller}, {Xue}  \& {Stott}}{{Danielson}
  et~al.}{2012}]{2012MNRAS.422..494D}
{Danielson} A.~L.~R.,  {Lehmer} B.~D.,  {Alexander} D.~M.,  {Brandt} W.~N.,
  {Luo} B.,  {Miller} N.,  {Xue} Y.~Q.,   {Stott} J.~P.,  2012, \mn@doi
  [\mnras] {10.1111/j.1365-2966.2012.20626.x}, \href
  {http://adsabs.harvard.edu/abs/2012MNRAS.422..494D} {422, 494}

\bibitem[\protect\citeauthoryear{{Dasyra}, {Bostrom}, {Combes}  \&
  {Vlahakis}}{{Dasyra} et~al.}{2015}]{2015ApJ...815...34D}
{Dasyra} K.~M.,  {Bostrom} A.~C.,  {Combes} F.,   {Vlahakis} N.,  2015, \mn@doi
  [\apj] {10.1088/0004-637X/815/1/34}, \href
  {http://adsabs.harvard.edu/abs/2015ApJ...815...34D} {815, 34}

\bibitem[\protect\citeauthoryear{{Debuhr}, {Quataert}  \& {Ma}}{{Debuhr}
  et~al.}{2011}]{2011MNRAS.412.1341D}
{Debuhr} J.,  {Quataert} E.,   {Ma} C.-P.,  2011, \mn@doi [\mnras]
  {10.1111/j.1365-2966.2010.17992.x}, \href
  {http://adsabs.harvard.edu/abs/2011MNRAS.412.1341D} {412, 1341}

\bibitem[\protect\citeauthoryear{{Debuhr}, {Quataert}  \& {Ma}}{{Debuhr}
  et~al.}{2012}]{2012MNRAS.420.2221D}
{Debuhr} J.,  {Quataert} E.,   {Ma} C.-P.,  2012, \mn@doi [\mnras]
  {10.1111/j.1365-2966.2011.20187.x}, \href
  {http://adsabs.harvard.edu/abs/2012MNRAS.420.2221D} {420, 2221}

\bibitem[\protect\citeauthoryear{{Delvecchio} et~al.,}{{Delvecchio}
  et~al.}{2015}]{2015MNRAS.449..373D}
{Delvecchio} I.,  et~al., 2015, \mn@doi [\mnras] {10.1093/mnras/stv213}, \href
  {http://adsabs.harvard.edu/abs/2015MNRAS.449..373D} {449, 373}

\bibitem[\protect\citeauthoryear{{Di Matteo}, {Springel}  \& {Hernquist}}{{Di
  Matteo} et~al.}{2005}]{2005Natur.433..604D}
{Di Matteo} T.,  {Springel} V.,   {Hernquist} L.,  2005, \mn@doi [\nat]
  {10.1038/nature03335}, \href
  {http://adsabs.harvard.edu/abs/2005Natur.433..604D} {433, 604}

\bibitem[\protect\citeauthoryear{{Diehl} \& {Statler}}{{Diehl} \&
  {Statler}}{2007}]{2007ApJ...668..150D}
{Diehl} S.,  {Statler} T.~S.,  2007, \mn@doi [\apj] {10.1086/521009}, \href
  {http://adsabs.harvard.edu/abs/2007ApJ...668..150D} {668, 150}

\bibitem[\protect\citeauthoryear{{Dubois}, {Devriendt}, {Slyz}  \&
  {Teyssier}}{{Dubois} et~al.}{2010}]{2010MNRAS.409..985D}
{Dubois} Y.,  {Devriendt} J.,  {Slyz} A.,   {Teyssier} R.,  2010, \mn@doi
  [\mnras] {10.1111/j.1365-2966.2010.17338.x}, \href
  {http://adsabs.harvard.edu/abs/2010MNRAS.409..985D} {409, 985}

\bibitem[\protect\citeauthoryear{{Dubois}, {Devriendt}, {Slyz}  \&
  {Teyssier}}{{Dubois} et~al.}{2012}]{2012MNRAS.420.2662D}
{Dubois} Y.,  {Devriendt} J.,  {Slyz} A.,   {Teyssier} R.,  2012, \mn@doi
  [\mnras] {10.1111/j.1365-2966.2011.20236.x}, \href
  {http://adsabs.harvard.edu/abs/2012MNRAS.420.2662D} {420, 2662}

\bibitem[\protect\citeauthoryear{{Dubois}, {Gavazzi}, {Peirani}  \&
  {Silk}}{{Dubois} et~al.}{2013}]{2013MNRAS.433.3297D}
{Dubois} Y.,  {Gavazzi} R.,  {Peirani} S.,   {Silk} J.,  2013, \mn@doi [\mnras]
  {10.1093/mnras/stt997}, \href
  {http://adsabs.harvard.edu/abs/2013MNRAS.433.3297D} {433, 3297}

\bibitem[\protect\citeauthoryear{{Dubois}, {Peirani}, {Pichon}, {Devriendt},
  {Gavazzi}, {Welker}  \& {Volonteri}}{{Dubois}
  et~al.}{2016}]{2016MNRAS.463.3948D}
{Dubois} Y.,  {Peirani} S.,  {Pichon} C.,  {Devriendt} J.,  {Gavazzi} R.,
  {Welker} C.,   {Volonteri} M.,  2016, \mn@doi [\mnras]
  {10.1093/mnras/stw2265}, \href
  {http://adsabs.harvard.edu/abs/2016MNRAS.463.3948D} {463, 3948}

\bibitem[\protect\citeauthoryear{{Dunn} et~al.,}{{Dunn}
  et~al.}{2010}]{2010ApJ...709..611D}
{Dunn} J.~P.,  et~al., 2010, \mn@doi [\apj] {10.1088/0004-637X/709/2/611},
  \href {http://adsabs.harvard.edu/abs/2010ApJ...709..611D} {709, 611}

\bibitem[\protect\citeauthoryear{{Eddington}}{{Eddington}}{1916}]{1916MNRAS..77...16E}
{Eddington} A.~S.,  1916, \mnras, \href
  {http://adsabs.harvard.edu/abs/1916MNRAS..77...16E} {77, 16}

\bibitem[\protect\citeauthoryear{{Eke}, {Navarro}  \& {Frenk}}{{Eke}
  et~al.}{1998}]{1998ApJ...503..569E}
{Eke} V.~R.,  {Navarro} J.~F.,   {Frenk} C.~S.,  1998, \mn@doi [\apj]
  {10.1086/306008}, \href {http://adsabs.harvard.edu/abs/1998ApJ...503..569E}
  {503, 569}

\bibitem[\protect\citeauthoryear{{Fabian}}{{Fabian}}{2012}]{2012ARA&A..50..455F}
{Fabian} A.~C.,  2012, \mn@doi [\araa] {10.1146/annurev-astro-081811-125521},
  \href {http://adsabs.harvard.edu/abs/2012ARA%26A..50..455F} {50, 455}

\bibitem[\protect\citeauthoryear{{Forman}, {Schwarz}, {Jones}, {Liller}  \&
  {Fabian}}{{Forman} et~al.}{1979}]{1979ApJ...234L..27F}
{Forman} W.,  {Schwarz} J.,  {Jones} C.,  {Liller} W.,   {Fabian} A.~C.,  1979,
  \mn@doi [\apjl] {10.1086/183103}, \href
  {http://adsabs.harvard.edu/abs/1979ApJ...234L..27F} {234, L27}

\bibitem[\protect\citeauthoryear{{Forman} et~al.,}{{Forman}
  et~al.}{2007}]{2007ApJ...665.1057F}
{Forman} W.,  et~al., 2007, \mn@doi [\apj] {10.1086/519480}, \href
  {http://adsabs.harvard.edu/abs/2007ApJ...665.1057F} {665, 1057}

\bibitem[\protect\citeauthoryear{{Franx}, {van Dokkum}, {Schreiber}, {Wuyts},
  {Labb{\'e}}  \& {Toft}}{{Franx} et~al.}{2008}]{2008ApJ...688..770F}
{Franx} M.,  {van Dokkum} P.~G.,  {Schreiber} N.~M.~F.,  {Wuyts} S.,
  {Labb{\'e}} I.,   {Toft} S.,  2008, \mn@doi [\apj] {10.1086/592431}, \href
  {http://adsabs.harvard.edu/abs/2008ApJ...688..770F} {688, 770}

\bibitem[\protect\citeauthoryear{{Gaibler}, {Khochfar}, {Krause}  \&
  {Silk}}{{Gaibler} et~al.}{2012}]{2012MNRAS.425..438G}
{Gaibler} V.,  {Khochfar} S.,  {Krause} M.,   {Silk} J.,  2012, \mn@doi
  [\mnras] {10.1111/j.1365-2966.2012.21479.x}, \href
  {http://adsabs.harvard.edu/abs/2012MNRAS.425..438G} {425, 438}

\bibitem[\protect\citeauthoryear{{Gan}, {Yuan}, {Ostriker}, {Ciotti}  \&
  {Novak}}{{Gan} et~al.}{2014}]{2014ApJ...789..150G}
{Gan} Z.,  {Yuan} F.,  {Ostriker} J.~P.,  {Ciotti} L.,   {Novak} G.~S.,  2014,
  \mn@doi [\apj] {10.1088/0004-637X/789/2/150}, \href
  {http://adsabs.harvard.edu/abs/2014ApJ...789..150G} {789, 150}

\bibitem[\protect\citeauthoryear{{Genel} et~al.,}{{Genel}
  et~al.}{2014}]{2014MNRAS.445..175G}
{Genel} S.,  et~al., 2014, \mn@doi [\mnras] {10.1093/mnras/stu1654}, \href
  {http://adsabs.harvard.edu/abs/2014MNRAS.445..175G} {445, 175}

\bibitem[\protect\citeauthoryear{{Gomes} et~al.,}{{Gomes}
  et~al.}{2016}]{2016A&A...588A..68G}
{Gomes} J.~M.,  et~al., 2016, \mn@doi [\aap] {10.1051/0004-6361/201525976},
  \href {http://adsabs.harvard.edu/abs/2016A%26A...588A..68G} {588, A68}

\bibitem[\protect\citeauthoryear{{Grange}, {de Plaa}, {Kaastra}, {Werner},
  {Verbunt}, {Paerels}  \& {de Vries}}{{Grange}
  et~al.}{2011}]{2011A&A...531A..15G}
{Grange} Y.~G.,  {de Plaa} J.,  {Kaastra} J.~S.,  {Werner} N.,  {Verbunt} F.,
  {Paerels} F.,   {de Vries} C.~P.,  2011, \mn@doi [\aap]
  {10.1051/0004-6361/201016187}, \href
  {http://adsabs.harvard.edu/abs/2011A%26A...531A..15G} {531, A15}

\bibitem[\protect\citeauthoryear{{Greene} \& {Ho}}{{Greene} \&
  {Ho}}{2007}]{2007ApJ...667..131G}
{Greene} J.~E.,  {Ho} L.~C.,  2007, \mn@doi [\apj] {10.1086/520497}, \href
  {http://adsabs.harvard.edu/abs/2007ApJ...667..131G} {667, 131}

\bibitem[\protect\citeauthoryear{{Greene}, {Murphy}, {Graves}, {Gunn},
  {Raskutti}, {Comerford}  \& {Gebhardt}}{{Greene}
  et~al.}{2013}]{2013ApJ...776...64G}
{Greene} J.~E.,  {Murphy} J.~D.,  {Graves} G.~J.,  {Gunn} J.~E.,  {Raskutti}
  S.,  {Comerford} J.~M.,   {Gebhardt} K.,  2013, \mn@doi [\apj]
  {10.1088/0004-637X/776/2/64}, \href
  {http://adsabs.harvard.edu/abs/2013ApJ...776...64G} {776, 64}

\bibitem[\protect\citeauthoryear{{Heckman}, {Kauffmann}, {Brinchmann},
  {Charlot}, {Tremonti}  \& {White}}{{Heckman}
  et~al.}{2004}]{2004ApJ...613..109H}
{Heckman} T.~M.,  {Kauffmann} G.,  {Brinchmann} J.,  {Charlot} S.,  {Tremonti}
  C.,   {White} S.~D.~M.,  2004, \mn@doi [\apj] {10.1086/422872}, \href
  {http://adsabs.harvard.edu/abs/2004ApJ...613..109H} {613, 109}

\bibitem[\protect\citeauthoryear{{Hernquist}}{{Hernquist}}{1990}]{1990ApJ...356..359H}
{Hernquist} L.,  1990, \mn@doi [\apj] {10.1086/168845}, \href
  {http://adsabs.harvard.edu/abs/1990ApJ...356..359H} {356, 359}

\bibitem[\protect\citeauthoryear{{Ho}}{{Ho}}{2009}]{2009ApJ...699..626H}
{Ho} L.~C.,  2009, \mn@doi [\apj] {10.1088/0004-637X/699/1/626}, \href
  {http://adsabs.harvard.edu/abs/2009ApJ...699..626H} {699, 626}

\bibitem[\protect\citeauthoryear{{Hopkins}, {Hernquist}, {Cox}, {Di Matteo},
  {Martini}, {Robertson}  \& {Springel}}{{Hopkins}
  et~al.}{2005}]{2005ApJ...630..705H}
{Hopkins} P.~F.,  {Hernquist} L.,  {Cox} T.~J.,  {Di Matteo} T.,  {Martini} P.,
   {Robertson} B.,   {Springel} V.,  2005, \mn@doi [\apj] {10.1086/432438},
  \href {http://adsabs.harvard.edu/abs/2005ApJ...630..705H} {630, 705}

\bibitem[\protect\citeauthoryear{{Hopkins}, {Cox}, {Kere{\v s}}  \&
  {Hernquist}}{{Hopkins} et~al.}{2008}]{2008ApJS..175..390H}
{Hopkins} P.~F.,  {Cox} T.~J.,  {Kere{\v s}} D.,   {Hernquist} L.,  2008,
  \mn@doi [\apjs] {10.1086/524363}, \href
  {http://adsabs.harvard.edu/abs/2008ApJS..175..390H} {175, 390}

\bibitem[\protect\citeauthoryear{{Hopkins}, {Torrey}, {Faucher-Gigu{\`e}re},
  {Quataert}  \& {Murray}}{{Hopkins} et~al.}{2016}]{2016MNRAS.458..816H}
{Hopkins} P.~F.,  {Torrey} P.,  {Faucher-Gigu{\`e}re} C.-A.,  {Quataert} E.,
  {Murray} N.,  2016, \mn@doi [\mnras] {10.1093/mnras/stw289}, \href
  {http://esoads.eso.org/abs/2016MNRAS.458..816H} {458, 816}

\bibitem[\protect\citeauthoryear{{Hoyle} \& {Lyttleton}}{{Hoyle} \&
  {Lyttleton}}{1939}]{1939PCPS...35..405H}
{Hoyle} F.,  {Lyttleton} R.~A.,  1939, \mn@doi [Proceedings of the Cambridge
  Philosophical Society] {10.1017/S0305004100021150}, \href
  {http://adsabs.harvard.edu/abs/1939PCPS...35..405H} {35, 405}

\bibitem[\protect\citeauthoryear{{Hu}, {Naab}, {Walch}, {Moster}  \&
  {Oser}}{{Hu} et~al.}{2014}]{2014MNRAS.443.1173H}
{Hu} C.-Y.,  {Naab} T.,  {Walch} S.,  {Moster} B.~P.,   {Oser} L.,  2014,
  \mn@doi [\mnras] {10.1093/mnras/stu1187}, \href
  {http://adsabs.harvard.edu/abs/2014MNRAS.443.1173H} {443, 1173}

\bibitem[\protect\citeauthoryear{{Humphrey}, {Buote}, {Gastaldello},
  {Zappacosta}, {Bullock}, {Brighenti}  \& {Mathews}}{{Humphrey}
  et~al.}{2006}]{2006ApJ...646..899H}
{Humphrey} P.~J.,  {Buote} D.~A.,  {Gastaldello} F.,  {Zappacosta} L.,
  {Bullock} J.~S.,  {Brighenti} F.,   {Mathews} W.~G.,  2006, \mn@doi [\apj]
  {10.1086/505019}, \href {http://adsabs.harvard.edu/abs/2006ApJ...646..899H}
  {646, 899}

\bibitem[\protect\citeauthoryear{{Iwamoto}, {Brachwitz}, {Nomoto}, {Kishimoto},
  {Umeda}, {Hix}  \& {Thielemann}}{{Iwamoto}
  et~al.}{1999}]{1999ApJS..125..439I}
{Iwamoto} K.,  {Brachwitz} F.,  {Nomoto} K.,  {Kishimoto} N.,  {Umeda} H.,
  {Hix} W.~R.,   {Thielemann} F.-K.,  1999, \mn@doi [\apjs] {10.1086/313278},
  \href {http://adsabs.harvard.edu/abs/1999ApJS..125..439I} {125, 439}

\bibitem[\protect\citeauthoryear{{Johansson}, {Naab}  \&
  {Ostriker}}{{Johansson} et~al.}{2009}]{2009ApJ...697L..38J}
{Johansson} P.~H.,  {Naab} T.,   {Ostriker} J.~P.,  2009, \mn@doi [\apjl]
  {10.1088/0004-637X/697/1/L38}, \href
  {http://adsabs.harvard.edu/abs/2009ApJ...697L..38J} {697, L38}

\bibitem[\protect\citeauthoryear{{Jones} \& {Forman}}{{Jones} \&
  {Forman}}{1984}]{1984ApJ...276...38J}
{Jones} C.,  {Forman} W.,  1984, \mn@doi [\apj] {10.1086/161591}, \href
  {http://adsabs.harvard.edu/abs/1984ApJ...276...38J} {276, 38}

\bibitem[\protect\citeauthoryear{{Kauffmann} \& {Heckman}}{{Kauffmann} \&
  {Heckman}}{2009}]{2009MNRAS.397..135K}
{Kauffmann} G.,  {Heckman} T.~M.,  2009, \mn@doi [\mnras]
  {10.1111/j.1365-2966.2009.14960.x}, \href
  {http://adsabs.harvard.edu/abs/2009MNRAS.397..135K} {397, 135}

\bibitem[\protect\citeauthoryear{{Kauffmann} et~al.,}{{Kauffmann}
  et~al.}{2003a}]{2003MNRAS.341...33K}
{Kauffmann} G.,  et~al., 2003a, \mn@doi [\mnras]
  {10.1046/j.1365-8711.2003.06291.x}, \href
  {http://adsabs.harvard.edu/abs/2003MNRAS.341...33K} {341, 33}

\bibitem[\protect\citeauthoryear{{Kauffmann} et~al.,}{{Kauffmann}
  et~al.}{2003b}]{2003MNRAS.346.1055K}
{Kauffmann} G.,  et~al., 2003b, \mn@doi [\mnras]
  {10.1111/j.1365-2966.2003.07154.x}, \href
  {http://adsabs.harvard.edu/abs/2003MNRAS.346.1055K} {346, 1055}

\bibitem[\protect\citeauthoryear{{Kehrig} et~al.,}{{Kehrig}
  et~al.}{2012}]{2012A&A...540A..11K}
{Kehrig} C.,  et~al., 2012, \mn@doi [\aap] {10.1051/0004-6361/201118357}, \href
  {http://adsabs.harvard.edu/abs/2012A%26A...540A..11K} {540, A11}

\bibitem[\protect\citeauthoryear{{Kim} \& {Fabbiano}}{{Kim} \&
  {Fabbiano}}{2013}]{2013ApJ...776..116K}
{Kim} D.-W.,  {Fabbiano} G.,  2013, \mn@doi [\apj]
  {10.1088/0004-637X/776/2/116}, \href
  {http://adsabs.harvard.edu/abs/2013ApJ...776..116K} {776, 116}

\bibitem[\protect\citeauthoryear{{Kim} \& {Fabbiano}}{{Kim} \&
  {Fabbiano}}{2015}]{2015ApJ...812..127K}
{Kim} D.-W.,  {Fabbiano} G.,  2015, \mn@doi [\apj]
  {10.1088/0004-637X/812/2/127}, \href
  {http://adsabs.harvard.edu/abs/2015ApJ...812..127K} {812, 127}

\bibitem[\protect\citeauthoryear{{Kim}, {Wise}, {Alvarez}  \& {Abel}}{{Kim}
  et~al.}{2011}]{2011ApJ...738...54K}
{Kim} J.-h.,  {Wise} J.~H.,  {Alvarez} M.~A.,   {Abel} T.,  2011, \mn@doi
  [\apj] {10.1088/0004-637X/738/1/54}, \href
  {http://adsabs.harvard.edu/abs/2011ApJ...738...54K} {738, 54}

\bibitem[\protect\citeauthoryear{{Konami}, {Matsushita}, {Nagino}  \&
  {Tamagawa}}{{Konami} et~al.}{2014}]{2014ApJ...783....8K}
{Konami} S.,  {Matsushita} K.,  {Nagino} R.,   {Tamagawa} T.,  2014, \mn@doi
  [\apj] {10.1088/0004-637X/783/1/8}, \href
  {http://adsabs.harvard.edu/abs/2014ApJ...783....8K} {783, 8}

\bibitem[\protect\citeauthoryear{{Kormendy} \& {Ho}}{{Kormendy} \&
  {Ho}}{2013}]{2013ARA&A..51..511K}
{Kormendy} J.,  {Ho} L.~C.,  2013, \mn@doi [\araa]
  {10.1146/annurev-astro-082708-101811}, \href
  {http://adsabs.harvard.edu/abs/2013ARA%26A..51..511K} {51, 511}

\bibitem[\protect\citeauthoryear{{Kravtsov}, {Vikhlinin}  \&
  {Meshscheryakov}}{{Kravtsov} et~al.}{2014}]{2014arXiv1401.7329K}
{Kravtsov} A.,  {Vikhlinin} A.,   {Meshscheryakov} A.,  2014, preprint, \href
  {http://adsabs.harvard.edu/abs/2014arXiv1401.7329K} {} (\mn@eprint {arXiv}
  {1401.7329})

\bibitem[\protect\citeauthoryear{{Lagan{\'a}}, {Martinet}, {Durret}, {Lima
  Neto}, {Maughan}  \& {Zhang}}{{Lagan{\'a}}
  et~al.}{2013}]{2013A&A...555A..66L}
{Lagan{\'a}} T.~F.,  {Martinet} N.,  {Durret} F.,  {Lima Neto} G.~B.,
  {Maughan} B.,   {Zhang} Y.-Y.,  2013, \mn@doi [\aap]
  {10.1051/0004-6361/201220423}, \href
  {http://adsabs.harvard.edu/abs/2013A%26A...555A..66L} {555, A66}

\bibitem[\protect\citeauthoryear{{Le Brun}, {McCarthy}, {Schaye}  \&
  {Ponman}}{{Le Brun} et~al.}{2014}]{2014MNRAS.441.1270L}
{Le Brun} A.~M.~C.,  {McCarthy} I.~G.,  {Schaye} J.,   {Ponman} T.~J.,  2014,
  \mn@doi [\mnras] {10.1093/mnras/stu608}, \href
  {http://adsabs.harvard.edu/abs/2014MNRAS.441.1270L} {441, 1270}

\bibitem[\protect\citeauthoryear{{Makino}, {Sasaki}  \& {Suto}}{{Makino}
  et~al.}{1998}]{1998ApJ...497..555M}
{Makino} N.,  {Sasaki} S.,   {Suto} Y.,  1998, \mn@doi [\apj] {10.1086/305507},
  \href {http://adsabs.harvard.edu/abs/1998ApJ...497..555M} {497, 555}

\bibitem[\protect\citeauthoryear{{Maoz} \& {Mannucci}}{{Maoz} \&
  {Mannucci}}{2012}]{2012PASA...29..447M}
{Maoz} D.,  {Mannucci} F.,  2012, \mn@doi [\pasa] {10.1071/AS11052}, \href
  {http://adsabs.harvard.edu/abs/2012PASA...29..447M} {29, 447}

\bibitem[\protect\citeauthoryear{{Martizzi}, {Teyssier}  \& {Moore}}{{Martizzi}
  et~al.}{2012}]{2012MNRAS.420.2859M}
{Martizzi} D.,  {Teyssier} R.,   {Moore} B.,  2012, \mn@doi [\mnras]
  {10.1111/j.1365-2966.2011.19950.x}, \href
  {http://adsabs.harvard.edu/abs/2012MNRAS.420.2859M} {420, 2859}

\bibitem[\protect\citeauthoryear{{Mathews} \& {Brighenti}}{{Mathews} \&
  {Brighenti}}{2003}]{2003ARA&A..41..191M}
{Mathews} W.~G.,  {Brighenti} F.,  2003, \mn@doi [\araa]
  {10.1146/annurev.astro.41.090401.094542}, \href
  {http://adsabs.harvard.edu/abs/2003ARA%26A..41..191M} {41, 191}

\bibitem[\protect\citeauthoryear{{McCarthy} et~al.,}{{McCarthy}
  et~al.}{2010}]{2010MNRAS.406..822M}
{McCarthy} I.~G.,  et~al., 2010, \mn@doi [\mnras]
  {10.1111/j.1365-2966.2010.16750.x}, \href
  {http://adsabs.harvard.edu/abs/2010MNRAS.406..822M} {406, 822}

\bibitem[\protect\citeauthoryear{{McCarthy}, {Schaye}, {Bower}, {Ponman},
  {Booth}, {Dalla Vecchia}  \& {Springel}}{{McCarthy}
  et~al.}{2011}]{2011MNRAS.412.1965M}
{McCarthy} I.~G.,  {Schaye} J.,  {Bower} R.~G.,  {Ponman} T.~J.,  {Booth}
  C.~M.,  {Dalla Vecchia} C.,   {Springel} V.,  2011, \mn@doi [\mnras]
  {10.1111/j.1365-2966.2010.18033.x}, \href
  {http://adsabs.harvard.edu/abs/2011MNRAS.412.1965M} {412, 1965}

\bibitem[\protect\citeauthoryear{{McDermid} et~al.,}{{McDermid}
  et~al.}{2015}]{2015MNRAS.448.3484M}
{McDermid} R.~M.,  et~al., 2015, \mn@doi [\mnras] {10.1093/mnras/stv105}, \href
  {http://adsabs.harvard.edu/abs/2015MNRAS.448.3484M} {448, 3484}

\bibitem[\protect\citeauthoryear{{McNamara} et~al.,}{{McNamara}
  et~al.}{2000}]{2000ApJ...534L.135M}
{McNamara} B.~R.,  et~al., 2000, \mn@doi [\apjl] {10.1086/312662}, \href
  {http://adsabs.harvard.edu/abs/2000ApJ...534L.135M} {534, L135}

\bibitem[\protect\citeauthoryear{{Moe}, {Arav}, {Bautista}  \& {Korista}}{{Moe}
  et~al.}{2009}]{2009ApJ...706..525M}
{Moe} M.,  {Arav} N.,  {Bautista} M.~A.,   {Korista} K.~T.,  2009, \mn@doi
  [\apj] {10.1088/0004-637X/706/1/525}, \href
  {http://adsabs.harvard.edu/abs/2009ApJ...706..525M} {706, 525}

\bibitem[\protect\citeauthoryear{{Morganti}, {Oosterloo}, {Oonk}, {Frieswijk}
  \& {Tadhunter}}{{Morganti} et~al.}{2015}]{2015A&A...580A...1M}
{Morganti} R.,  {Oosterloo} T.,  {Oonk} J.~B.~R.,  {Frieswijk} W.,
  {Tadhunter} C.,  2015, \mn@doi [\aap] {10.1051/0004-6361/201525860}, \href
  {http://adsabs.harvard.edu/abs/2015A%26A...580A...1M} {580, A1}

\bibitem[\protect\citeauthoryear{{Moster}, {Macci{\`o}}, {Somerville}, {Naab}
  \& {Cox}}{{Moster} et~al.}{2011}]{2011MNRAS.415.3750M}
{Moster} B.~P.,  {Macci{\`o}} A.~V.,  {Somerville} R.~S.,  {Naab} T.,   {Cox}
  T.~J.,  2011, \mn@doi [\mnras] {10.1111/j.1365-2966.2011.18984.x}, \href
  {http://adsabs.harvard.edu/abs/2011MNRAS.415.3750M} {415, 3750}

\bibitem[\protect\citeauthoryear{{Moster}, {Naab}  \& {White}}{{Moster}
  et~al.}{2013}]{2013MNRAS.428.3121M}
{Moster} B.~P.,  {Naab} T.,   {White} S.~D.~M.,  2013, \mn@doi [\mnras]
  {10.1093/mnras/sts261}, \href
  {http://adsabs.harvard.edu/abs/2013MNRAS.428.3121M} {428, 3121}

\bibitem[\protect\citeauthoryear{{Mulchaey} \& {Jeltema}}{{Mulchaey} \&
  {Jeltema}}{2010}]{2010ApJ...715L...1M}
{Mulchaey} J.~S.,  {Jeltema} T.~E.,  2010, \mn@doi [\apjl]
  {10.1088/2041-8205/715/1/L1}, \href
  {http://adsabs.harvard.edu/abs/2010ApJ...715L...1M} {715, L1}

\bibitem[\protect\citeauthoryear{{Naab} \& {Ostriker}}{{Naab} \&
  {Ostriker}}{2016}]{2016arXiv161206891N}
{Naab} T.,  {Ostriker} J.~P.,  2016, preprint, \href
  {http://adsabs.harvard.edu/abs/2016arXiv161206891N} {} (\mn@eprint {arXiv}
  {1612.06891})

\bibitem[\protect\citeauthoryear{{Nagai}, {Kravtsov}  \& {Vikhlinin}}{{Nagai}
  et~al.}{2007}]{2007ApJ...668....1N}
{Nagai} D.,  {Kravtsov} A.~V.,   {Vikhlinin} A.,  2007, \mn@doi [\apj]
  {10.1086/521328}, \href {http://adsabs.harvard.edu/abs/2007ApJ...668....1N}
  {668, 1}

\bibitem[\protect\citeauthoryear{{Navarro}, {Frenk}  \& {White}}{{Navarro}
  et~al.}{1996}]{1996ApJ...462..563N}
{Navarro} J.~F.,  {Frenk} C.~S.,   {White} S.~D.~M.,  1996, \mn@doi [\apj]
  {10.1086/177173}, \href {http://adsabs.harvard.edu/abs/1996ApJ...462..563N}
  {462, 563}

\bibitem[\protect\citeauthoryear{{Novak}, {Ostriker}  \& {Ciotti}}{{Novak}
  et~al.}{2011}]{2011ApJ...737...26N}
{Novak} G.~S.,  {Ostriker} J.~P.,   {Ciotti} L.,  2011, \mn@doi [\apj]
  {10.1088/0004-637X/737/1/26}, \href
  {http://adsabs.harvard.edu/abs/2011ApJ...737...26N} {737, 26}

\bibitem[\protect\citeauthoryear{{N{\'u}{\~n}ez}, {Ostriker}, {Naab}, {Oser},
  {Hu}  \& {Choi}}{{N{\'u}{\~n}ez} et~al.}{2017}]{2017arXiv170101082N}
{N{\'u}{\~n}ez} A.,  {Ostriker} J.~P.,  {Naab} T.,  {Oser} L.,  {Hu} C.-Y.,
  {Choi} E.,  2017, preprint, \href
  {http://adsabs.harvard.edu/abs/2017arXiv170101082N} {} (\mn@eprint {arXiv}
  {1701.01082})

\bibitem[\protect\citeauthoryear{{O'Sullivan}, {Sanderson}  \&
  {Ponman}}{{O'Sullivan} et~al.}{2007}]{2007MNRAS.380.1409O}
{O'Sullivan} E.,  {Sanderson} A.~J.~R.,   {Ponman} T.~J.,  2007, \mn@doi
  [\mnras] {10.1111/j.1365-2966.2007.12229.x}, \href
  {http://adsabs.harvard.edu/abs/2007MNRAS.380.1409O} {380, 1409}

\bibitem[\protect\citeauthoryear{{Omma}, {Binney}, {Bryan}  \& {Slyz}}{{Omma}
  et~al.}{2004}]{2004MNRAS.348.1105O}
{Omma} H.,  {Binney} J.,  {Bryan} G.,   {Slyz} A.,  2004, \mn@doi [\mnras]
  {10.1111/j.1365-2966.2004.07382.x}, \href
  {http://adsabs.harvard.edu/abs/2004MNRAS.348.1105O} {348, 1105}

\bibitem[\protect\citeauthoryear{{Ostriker}, {Choi}, {Ciotti}, {Novak}  \&
  {Proga}}{{Ostriker} et~al.}{2010}]{2010ApJ...722..642O}
{Ostriker} J.~P.,  {Choi} E.,  {Ciotti} L.,  {Novak} G.~S.,   {Proga} D.,
  2010, \mn@doi [\apj] {10.1088/0004-637X/722/1/642}, \href
  {http://adsabs.harvard.edu/abs/2010ApJ...722..642O} {722, 642}

\bibitem[\protect\citeauthoryear{{Page} et~al.,}{{Page}
  et~al.}{2012}]{2012Natur.485..213P}
{Page} M.~J.,  et~al., 2012, \mn@doi [\nat] {10.1038/nature11096}, \href
  {http://adsabs.harvard.edu/abs/2012Natur.485..213P} {485, 213}

\bibitem[\protect\citeauthoryear{{Peng} et~al.,}{{Peng}
  et~al.}{2010}]{2010ApJ...721..193P}
{Peng} Y.-j.,  et~al., 2010, \mn@doi [\apj] {10.1088/0004-637X/721/1/193},
  \href {http://adsabs.harvard.edu/abs/2010ApJ...721..193P} {721, 193}

\bibitem[\protect\citeauthoryear{{Pinto} et~al.,}{{Pinto}
  et~al.}{2014}]{2014A&A...572L...8P}
{Pinto} C.,  et~al., 2014, \mn@doi [\aap] {10.1051/0004-6361/201425270}, \href
  {http://adsabs.harvard.edu/abs/2014A%26A...572L...8P} {572, L8}

\bibitem[\protect\citeauthoryear{{Planck Collaboration} et~al.,}{{Planck
  Collaboration} et~al.}{2013}]{2013A&A...557A..52P}
{Planck Collaboration} et~al., 2013, \mn@doi [\aap]
  {10.1051/0004-6361/201220941}, \href
  {http://adsabs.harvard.edu/abs/2013A%26A...557A..52P} {557, A52}

\bibitem[\protect\citeauthoryear{{Planck Collaboration} et~al.,}{{Planck
  Collaboration} et~al.}{2014}]{2014A&A...571A..16P}
{Planck Collaboration} et~al., 2014, \mn@doi [\aap]
  {10.1051/0004-6361/201321591}, \href
  {http://adsabs.harvard.edu/abs/2014A%26A...571A..16P} {571, A16}

\bibitem[\protect\citeauthoryear{{Proga} \& {Kallman}}{{Proga} \&
  {Kallman}}{2004}]{2004ApJ...616..688P}
{Proga} D.,  {Kallman} T.~R.,  2004, \mn@doi [\apj] {10.1086/425117}, \href
  {http://adsabs.harvard.edu/abs/2004ApJ...616..688P} {616, 688}

\bibitem[\protect\citeauthoryear{{Rees} \& {Ostriker}}{{Rees} \&
  {Ostriker}}{1977}]{1977MNRAS.179..541R}
{Rees} M.~J.,  {Ostriker} J.~P.,  1977, \mn@doi [\mnras]
  {10.1093/mnras/179.4.541}, \href
  {http://adsabs.harvard.edu/abs/1977MNRAS.179..541R} {179, 541}

\bibitem[\protect\citeauthoryear{{Renzini} \& {Peng}}{{Renzini} \&
  {Peng}}{2015}]{2015ApJ...801L..29R}
{Renzini} A.,  {Peng} Y.-j.,  2015, \mn@doi [\apjl]
  {10.1088/2041-8205/801/2/L29}, \href
  {http://adsabs.harvard.edu/abs/2015ApJ...801L..29R} {801, L29}

\bibitem[\protect\citeauthoryear{{Renzini}, {Ciotti}, {D'Ercole}  \&
  {Pellegrini}}{{Renzini} et~al.}{1993}]{1993ApJ...419...52R}
{Renzini} A.,  {Ciotti} L.,  {D'Ercole} A.,   {Pellegrini} S.,  1993, \mn@doi
  [\apj] {10.1086/173458}, \href
  {http://adsabs.harvard.edu/abs/1993ApJ...419...52R} {419, 52}

\bibitem[\protect\citeauthoryear{{Sarzi} et~al.,}{{Sarzi}
  et~al.}{2013}]{2013MNRAS.432.1845S}
{Sarzi} M.,  et~al., 2013, \mn@doi [\mnras] {10.1093/mnras/stt062}, \href
  {http://adsabs.harvard.edu/abs/2013MNRAS.432.1845S} {432, 1845}

\bibitem[\protect\citeauthoryear{{Sazonov}, {Ostriker}, {Ciotti}  \&
  {Sunyaev}}{{Sazonov} et~al.}{2005}]{2005MNRAS.358..168S}
{Sazonov} S.~Y.,  {Ostriker} J.~P.,  {Ciotti} L.,   {Sunyaev} R.~A.,  2005,
  \mn@doi [\mnras] {10.1111/j.1365-2966.2005.08763.x}, \href
  {http://adsabs.harvard.edu/abs/2005MNRAS.358..168S} {358, 168}

\bibitem[\protect\citeauthoryear{{Scannapieco}, {Tissera}, {White}  \&
  {Springel}}{{Scannapieco} et~al.}{2005}]{2005MNRAS.364..552S}
{Scannapieco} C.,  {Tissera} P.~B.,  {White} S.~D.~M.,   {Springel} V.,  2005,
  \mn@doi [\mnras] {10.1111/j.1365-2966.2005.09574.x}, \href
  {http://adsabs.harvard.edu/abs/2005MNRAS.364..552S} {364, 552}

\bibitem[\protect\citeauthoryear{{Scannapieco}, {Tissera}, {White}  \&
  {Springel}}{{Scannapieco} et~al.}{2006}]{2006MNRAS.371.1125S}
{Scannapieco} C.,  {Tissera} P.~B.,  {White} S.~D.~M.,   {Springel} V.,  2006,
  \mn@doi [\mnras] {10.1111/j.1365-2966.2006.10785.x}, \href
  {http://adsabs.harvard.edu/abs/2006MNRAS.371.1125S} {371, 1125}

\bibitem[\protect\citeauthoryear{{Schaye} et~al.,}{{Schaye}
  et~al.}{2015}]{2015MNRAS.446..521S}
{Schaye} J.,  et~al., 2015, \mn@doi [\mnras] {10.1093/mnras/stu2058}, \href
  {http://adsabs.harvard.edu/abs/2015MNRAS.446..521S} {446, 521}

\bibitem[\protect\citeauthoryear{{Sedov}}{{Sedov}}{1959}]{1959sdmm.book.....S}
{Sedov} L.~I.,  1959, {Similarity and Dimensional Methods in Mechanics}.
Academic Press, New York

\bibitem[\protect\citeauthoryear{{Shakura} \& {Sunyaev}}{{Shakura} \&
  {Sunyaev}}{1973}]{1973A&A....24..337S}
{Shakura} N.~I.,  {Sunyaev} R.~A.,  1973, \aap, \href
  {http://adsabs.harvard.edu/abs/1973A%26A....24..337S} {24, 337}

\bibitem[\protect\citeauthoryear{{Shin}, {Ostriker}  \& {Ciotti}}{{Shin}
  et~al.}{2010}]{2010ApJ...711..268S}
{Shin} M.-S.,  {Ostriker} J.~P.,   {Ciotti} L.,  2010, \mn@doi [\apj]
  {10.1088/0004-637X/711/1/268}, \href
  {http://adsabs.harvard.edu/abs/2010ApJ...711..268S} {711, 268}

\bibitem[\protect\citeauthoryear{{Sijacki}, {Springel}, {Di Matteo}  \&
  {Hernquist}}{{Sijacki} et~al.}{2007}]{2007MNRAS.380..877S}
{Sijacki} D.,  {Springel} V.,  {Di Matteo} T.,   {Hernquist} L.,  2007, \mn@doi
  [\mnras] {10.1111/j.1365-2966.2007.12153.x}, \href
  {http://adsabs.harvard.edu/abs/2007MNRAS.380..877S} {380, 877}

\bibitem[\protect\citeauthoryear{{Sijacki}, {Vogelsberger}, {Genel},
  {Springel}, {Torrey}, {Snyder}, {Nelson}  \& {Hernquist}}{{Sijacki}
  et~al.}{2015}]{2015MNRAS.452..575S}
{Sijacki} D.,  {Vogelsberger} M.,  {Genel} S.,  {Springel} V.,  {Torrey} P.,
  {Snyder} G.~F.,  {Nelson} D.,   {Hernquist} L.,  2015, \mn@doi [\mnras]
  {10.1093/mnras/stv1340}, \href
  {http://adsabs.harvard.edu/abs/2015MNRAS.452..575S} {452, 575}

\bibitem[\protect\citeauthoryear{{Silk}}{{Silk}}{1977}]{1977ApJ...211..638S}
{Silk} J.,  1977, \mn@doi [\apj] {10.1086/154972}, \href
  {http://adsabs.harvard.edu/abs/1977ApJ...211..638S} {211, 638}

\bibitem[\protect\citeauthoryear{{Silk}}{{Silk}}{2013}]{2013ApJ...772..112S}
{Silk} J.,  2013, \mn@doi [\apj] {10.1088/0004-637X/772/2/112}, \href
  {http://adsabs.harvard.edu/abs/2013ApJ...772..112S} {772, 112}

\bibitem[\protect\citeauthoryear{{Silk} \& {Rees}}{{Silk} \&
  {Rees}}{1998}]{1998A&A...331L...1S}
{Silk} J.,  {Rees} M.~J.,  1998, \aap, \href
  {http://adsabs.harvard.edu/abs/1998A%26A...331L...1S} {331, L1}

\bibitem[\protect\citeauthoryear{{Smith}, {Lucey}, {Price}, {Hudson}  \&
  {Phillipps}}{{Smith} et~al.}{2012}]{2012MNRAS.419.3167S}
{Smith} R.~J.,  {Lucey} J.~R.,  {Price} J.,  {Hudson} M.~J.,   {Phillipps} S.,
  2012, \mn@doi [\mnras] {10.1111/j.1365-2966.2011.19956.x}, \href
  {http://adsabs.harvard.edu/abs/2012MNRAS.419.3167S} {419, 3167}

\bibitem[\protect\citeauthoryear{{Somerville} \& {Dav{\'e}}}{{Somerville} \&
  {Dav{\'e}}}{2015}]{2015ARA&A..53...51S}
{Somerville} R.~S.,  {Dav{\'e}} R.,  2015, \mn@doi [\araa]
  {10.1146/annurev-astro-082812-140951}, \href
  {http://adsabs.harvard.edu/abs/2015ARA%26A..53...51S} {53, 51}

\bibitem[\protect\citeauthoryear{{Springel}}{{Springel}}{2005}]{2005MNRAS.364.1105S}
{Springel} V.,  2005, \mn@doi [\mnras] {10.1111/j.1365-2966.2005.09655.x},
  \href {http://adsabs.harvard.edu/abs/2005MNRAS.364.1105S} {364, 1105}

\bibitem[\protect\citeauthoryear{{Springel} \& {Hernquist}}{{Springel} \&
  {Hernquist}}{2003}]{2003MNRAS.339..289S}
{Springel} V.,  {Hernquist} L.,  2003, \mn@doi [\mnras]
  {10.1046/j.1365-8711.2003.06206.x}, \href
  {http://adsabs.harvard.edu/abs/2003MNRAS.339..289S} {339, 289}

\bibitem[\protect\citeauthoryear{{Springel}, {Di Matteo}  \&
  {Hernquist}}{{Springel} et~al.}{2005}]{2005MNRAS.361..776S}
{Springel} V.,  {Di Matteo} T.,   {Hernquist} L.,  2005, \mn@doi [\mnras]
  {10.1111/j.1365-2966.2005.09238.x}, \href
  {http://adsabs.harvard.edu/abs/2005MNRAS.361..776S} {361, 776}

\bibitem[\protect\citeauthoryear{{Steinborn}, {Dolag}, {Hirschmann}, {Prieto}
  \& {Remus}}{{Steinborn} et~al.}{2015}]{2015MNRAS.448.1504S}
{Steinborn} L.~K.,  {Dolag} K.,  {Hirschmann} M.,  {Prieto} M.~A.,   {Remus}
  R.-S.,  2015, \mn@doi [\mnras] {10.1093/mnras/stv072}, \href
  {http://adsabs.harvard.edu/abs/2015MNRAS.448.1504S} {448, 1504}

\bibitem[\protect\citeauthoryear{{Thomas}, {Maraston}, {Bender}  \& {Mendes de
  Oliveira}}{{Thomas} et~al.}{2005}]{2005ApJ...621..673T}
{Thomas} D.,  {Maraston} C.,  {Bender} R.,   {Mendes de Oliveira} C.,  2005,
  \mn@doi [\apj] {10.1086/426932}, \href
  {http://adsabs.harvard.edu/abs/2005ApJ...621..673T} {621, 673}

\bibitem[\protect\citeauthoryear{{Vogelsberger} et~al.,}{{Vogelsberger}
  et~al.}{2014}]{2014MNRAS.444.1518V}
{Vogelsberger} M.,  et~al., 2014, \mn@doi [\mnras] {10.1093/mnras/stu1536},
  \href {http://adsabs.harvard.edu/abs/2014MNRAS.444.1518V} {444, 1518}

\bibitem[\protect\citeauthoryear{{Weinberger} et~al.,}{{Weinberger}
  et~al.}{2017}]{2017MNRAS.465.3291W}
{Weinberger} R.,  et~al., 2017, \mn@doi [\mnras] {10.1093/mnras/stw2944}, \href
  {http://adsabs.harvard.edu/abs/2017MNRAS.465.3291W} {465, 3291}

\bibitem[\protect\citeauthoryear{{Williams}, {Quadri}, {Franx}, {van Dokkum},
  {Toft}, {Kriek}  \& {Labb{\'e}}}{{Williams}
  et~al.}{2010}]{2010ApJ...713..738W}
{Williams} R.~J.,  {Quadri} R.~F.,  {Franx} M.,  {van Dokkum} P.,  {Toft} S.,
  {Kriek} M.,   {Labb{\'e}} I.,  2010, \mn@doi [\apj]
  {10.1088/0004-637X/713/2/738}, \href
  {http://adsabs.harvard.edu/abs/2010ApJ...713..738W} {713, 738}

\bibitem[\protect\citeauthoryear{{Woosley} \& {Weaver}}{{Woosley} \&
  {Weaver}}{1995}]{1995ApJS..101..181W}
{Woosley} S.~E.,  {Weaver} T.~A.,  1995, \mn@doi [\apjs] {10.1086/192237},
  \href {http://adsabs.harvard.edu/abs/1995ApJS..101..181W} {101, 181}

\bibitem[\protect\citeauthoryear{{Wylezalek} \& {Zakamska}}{{Wylezalek} \&
  {Zakamska}}{2016}]{2016arXiv160608442W}
{Wylezalek} D.,  {Zakamska} N.~L.,  2016, preprint, \href
  {http://adsabs.harvard.edu/abs/2016arXiv160608442W} {} (\mn@eprint {arXiv}
  {1606.08442})

\bibitem[\protect\citeauthoryear{{Zubovas}, {Nayakshin}, {King}  \&
  {Wilkinson}}{{Zubovas} et~al.}{2013}]{2013MNRAS.433.3079Z}
{Zubovas} K.,  {Nayakshin} S.,  {King} A.,   {Wilkinson} M.,  2013, \mn@doi
  [\mnras] {10.1093/mnras/stt952}, \href
  {http://adsabs.harvard.edu/abs/2013MNRAS.433.3079Z} {433, 3079}

\makeatother
\end{thebibliography}

\label{lastpage}
\end{document}